\documentclass[12pt,preprint]{emulateapj}
\usepackage{epsfig}

\def\arcsec{$^{\prime\prime}$}
\def\arcmin{$^{\prime}$}

\shorttitle{Kinematics of Globular Clusters in NGC 5128}
\shortauthors{Woodley et al.}

\begin{document}

\title{The Kinematics of the Globular Cluster System of NGC 5128 with a New, Large Sample of Radial Velocity Measurements}

\author{Kristin A.~Woodley\altaffilmark{1}}
\affil{Department of Physics \& Astronomy, McMaster University,
  Hamilton ON  L8S 4M1, Canada}

\author{Mat{\'i}as G{\'o}mez}
\affil{Departamento de Ciencias Fisicas, Facultad de Ingenieria, Universidad Andres Bello, Chile}
\email{matiasgomez@unab.cl}

\author{William E.~Harris}
\affil{Department of Physics \& Astronomy, McMaster University,
  Hamilton ON  L8S 4M1, Canada}
\email{harris@physics.mcmaster.ca}

\author{Doug Geisler}
\affil{Departamento de Astronom\'ia, Universidad de
  Concepci\'on, Chile}
\email{dgeisler@astro-udec.cl}

\author{Gretchen L.~H.~Harris}
\affil{Department of Physics and Astronomy, University of
  Waterloo, Waterloo, ON, Canada}
\email{glharris@astro.uwaterloo.ca}

\altaffiltext{1} {Current Address: Department of Physics \& Astronomy,
  University of British Columbia, Vancouver BC V6T 1Z1, Canada; kwoodley@phas.ubc.ca}

\begin{abstract}
New radial velocity measurements for  previously known and
newly confirmed  globular clusters in  the nearby massive  galaxy NGC
5128 are presented.   We  have obtained  spectroscopy  from  LDSS-2/Magellan,
VIMOS/VLT,  and Hydra/CTIO from which we have measured the radial  velocities  of 218
known, and  identified 155 new, globular clusters.  The  current sample of confirmed
globular clusters in NGC 5128 is now 605 with 564 of these having radial velocity measurements, 
the second largest kinematic database for any galaxy.
We have performed  a new kinematic analysis of the globular cluster system that
extends  out to 45\arcmin\ in  galactocentric radius.   We
have examined  the  systemic  velocity, projected
rotation amplitude and axis, and the projected velocity dispersion of 
the globular clusters as functions of galactocentric distance and metallicity.
Our results indicate that the  metal-poor  globular clusters  have a very mild  
rotation signature of $26\pm15$ km s$^{-1}$.  The
metal-rich globular clusters are rotating with a higher, though still
small signature of  $43\pm15$ km s$^{-1}$ around the
isophotal  major axis of  NGC 5128  within 15\arcmin.   Their velocity
dispersions are consistent within the uncertainties and the profiles 
appear flat or declining  within 20\arcmin.  We note
the small  sample of metal-rich globular clusters with  ages less than 5  Gyr in the
literature  appear to  have different kinematic properties than the
old, metal-rich  globular cluster subpopulation.   The
mass and  mass-to-light ratios have also been estimated using  the globular clusters  as tracer
particles for NGC  5128.  Out to a distance of  20\arcmin, we have obtained a
mass of $(5.9\pm2.0)\times 10^{11}$  M$_{\sun}$ and a mass-to-light ratio in
the B-band  of 16 M$_{\sun}/$L$_{B\sun}$.  Combined with previous work
on the ages and metallicities of its globular clusters, as well as 
properties of its stellar halo, our findings suggest NGC 5128 formed via hierarchical
merging over other methods of
formation, such as major merging at late times.
\end{abstract}


\section{Introduction}
\label{sec:intro}

The modern  idea of  galaxy formation involves  hierarchically merging
smaller galaxies or  protogalaxies \citep{toomre77,white78,peebles80,whitefrenk91,kauffmann93,baugh98,cole00,somerville01} that contain both baryonic
and  dark matter.   The merging  of these  smaller clumps  into larger
systems  results in a  bound system  embedded in  a large  dark matter
halo.   

Globular clusters (GCs) are great tracers  of the formation and evolution of
their host galaxy  as there is significant evidence  that they trace
episodes        of         star        formation        \citep[][among
others]{holtzman92,schweizer93,whitmore93,whitmore95,zepf95,schweizer96,miller97,carlson98,schweizer98,whitmore99,zepf99,chien07,goudfrooij07,trancho07},
and generally  have ages $> 10$  Gyr, making these some of the oldest structures in the Universe       \citep[][as
examples]{kisslerpatig98a,forbes01,puzia02,hempel03,kundu05,puzia05,strader05,sharina06,beasley08,puzia08,woodley09}.
GCs are  also an excellent way  to study massive  galaxies because  they are
luminous, compact and
can be found in large  quantities out to large galactocentric radii in
relatively spherical  distributions.  They are also expected to retain
a memory  of their formation  conditions \citep[e.g.][]{puzia06}, so
by understanding their formation we  may be able to constrain  formation
scenarios of their host galaxy.  Finally, each cluster has a unique 
age and abundance which are relatively straightforward to determine,
providing us 
with the key data we need to trace their formation and chemical evolution. 

The color bimodality, consisting of GCs having two distinct peaks in
their broadband color histograms, corresponding to red  and blue GCs,
is a common  feature in all  early-type galaxies 
such as NGC 4472
\citep{cohen03,strader07}, M87 \citep{kundu07}, 
and NGC 5128 \citep{peng04b,beasley08,woodley09}, as well as late-type galaxies,
such as  the Milky Way and M31
\citep{zinn85,elson88,armandroff89,brodie90,zinn90,harris91,ashman98}. 
The separate
populations of GCs have led to the idea that they form via different
mechanisms/conditions and perhaps at different times.
\cite{ashman92}  proposed  that red, or metal-rich,  GCs  can  result from  the
merging  events of  gaseous  disk galaxies.   This  type of  formation
suggests an age spread in the metal-rich GCs as disk mergers can continue to low redshifts. 
\cite{beasley02} and \cite{strader05} have proposed an interesting formation scenario
that   combines   two   scenarios proposed   early:  multiphase   collapse
\citep{forbes97} which  suggested that most  (if not all) GCs  form in
their parent galaxy, but that  the galaxy undergoes more than one phase of
collapse  and  hence  star   formation;  and  the  accretion  scenario
\citep{cote98,cote00,cote02}   suggesting  that  blue, or metal-poor,   GCs  are
accreted into the massive  galaxy via merging of surrounding satellite
galaxies. 
Within this scenario,  the  metal-poor GCs could form in
small proto-galactic clouds in the early universe.  As first suggested
by \cite{santos03},  their formation could have been  truncated by the
reionization  process.  It  is also  possible  that the  star formation  was
truncated  by feedback  processes  such  as the  removal  of gas  from
supernovae  driven  winds \citep[e.g.][]{dekel03,kauffmann03}.   These
protogalaxies would merge and  form a massive early-type galaxy with
the field stars  and metal-rich GCs forming in  the second collapse of
star formation.  

In any of the above, it is
clear that age is a distinguishing factor.  However, the metallicity
bimodality also suggests the blue and red components may have different kinematical signatures.  Kinematics
can  be used  to test  whether red  and blue  GCs share  signatures of
rotation  and  velocity  dispersion,   and  how  those signatures  can  vary  with
galactocentric distance.   We already see in  many elliptical galaxies
that  the  red GCs  are  more  centrally  concentrated than  the  blue
\citep[e.g.][]{geisler96,forbes98,rhode01,dirsch03,forbes04,peng04b,tamura06a,woodley07},
so  we  question  whether  their  kinematics  also  change with radius.

NGC 5128 (Centaurus A)  is a moderately luminous \citep[M$_B= -21.1$
mag,][]{dufour79} elliptical galaxy in the nearby Centaurus group, at
a  distance of  $3.8\pm 0.1$  Mpc  \citep{harris09}.   It has evidence  for a
relatively  recent merger  event such  as  a warped disk and faint shells \citep{malin78}, 
recent star formation \citep{graham98,rejkuba01,rejkuba02}, and a small blue elliptical arc that appears
to be a tidally disrupted  stellar stream, which could be a recently
accreted  satellite galaxy \citep{peng02}.   The close  proximity of
this  galaxy allows  a  detailed study  of  its stellar  population,
including the GCs, that is not yet possible for other giant elliptical galaxies.
However, this close proximity also means the galaxy is spread across
a very wide area in the projected  sky.  NGC 5128 also has a low  galactic latitude  of
$b=19^o$ so its GCs need to be securely confirmed via
their radial  velocities, which are generally distinct from the two  forms of
contamination, Milky Way  foreground stars and background galaxies.
Currently,  over 400   GCs  have  been  confirmed  with  this
methodology  in the  past  \citep{vandenbergh81,hesser84,hesser86,harris92,peng04b,woodley05,rejkuba07,beasley08,woodley09},  with a  large number of these compiled in the catalog of \cite{woodley07}.

In this paper, we confirm a number of new GCs as well as
remeasure previously known GCs in NGC 5128.  The
observations and data reduction are presented in
Section~\ref{sec:observations}, followed by our presentation of radial
velocity measurements in Section~\ref{sec:rv}.  In
Section~\ref{sec:kin_dyn}, we present a new kinematic analysis of 
the GCs in NGC 5128 as well as estimate a new mass and mass-to-light ratio
of NGC 5128.  We then discuss our findings in
Section~\ref{sec:discussion} and conclude in Section~\ref{sec:conclusions}.

\section{Observations and Data Reduction}
\label{sec:observations}

Our spectroscopy involves three  independent observing runs, each with
a primary objective to obtain radial velocity measurements of GC 
candidates to confirm new membership.  Table~\ref{tab:fields} lists the
center of field  location in R.  A. and  Decl. (in J2000 coordinates) and 
the  total exposure  times  for  each field.

\subsection{LDSS-2/Baade Dataset}
\label{sec:ldss2}

We obtained multi-object slit spectroscopy with the now decommissioned
Low Dispersion Survey Spectrograph-2 (LDSS-2) on the
Baade/Magellan 6.5 m telescope at the Las Campanas Observatory in
Chile.  We observed 12 fields scattered around the central regions of NGC 5128 on May 9-10, 2002
(PI: D. Geisler) with the 6\arcmin\ field of view of LDSS-2 (at the
distance of NGC 5128, 1\arcmin\ is approximately 1.1 kpc).  
Our slit sizes easily encompassed both our target objects as well
as the sky immediately surrounding each object.  
A gain of 1.0  e$^-$/ADU and a readout noise of 5.0
e$^-$ were used in this study.

Our  general reduction  was  performed with  the  Image Reduction  and
Analysis Facility (IRAF)\footnote{IRAF  is distributed by the National
  Optical   Astronomy  Observatories,  which   are  operated   by  the
  Association of  Universities for Research in  Astronomy, Inc., under
  cooperative  agreement   with  the  National   Science  Foundation.}
software.  Each  of our target  fields was taken with  exposure times
between  1000-1800  seconds.  We  obtained  and  combined zero  (bias)
frames ({\it  zerocombine}) for each  night and removed the  bias from
our  science exposures.   The  overscan  region of  the  CCD was  also
removed from our science target exposures ({\em ccdproc}), followed by
the   removal   of   cosmic    rays   with   the   European   Southern
Observatory-Munich  Image Data Analysis  System (ESO-MIDAS)  task {\em
  filter/cosmic}.  The flat fields  were combined for each field ({\em
  flatcombine}  in  IRAF) and  applied  to  our  science fields  ({\em
  ccdproc}).   Within each  slit we  fit, traced,  and  extracted each
object  and  sky  spectrum  ({\em  apall}).   Comparison  arcs,  taken
adjacent to  field exposures, were  used to generate  a transformation
from  pixel  to  wavelength  for  each of  our  target  objects  ({\em
  identify})  and  applied with  {\em  dispcor}.   We  used the
neutral oxygen $5577
\rm{\AA}$ sky  line to correct  for small shifts (typically  less than
$0.5  \rm{\AA}$)  in  our  spectroscopy  ({\em  specshift})  and  then
subtracted the  sky from the  targets ({\em sarith}) and  then cleaned
the spectra  ({\em lineclean}) using a  low rejection of 8  and a high
rejection of  2.5 below  and above the  residual sigma with  a spline3
function.  Each  field was  observed two times  and then  sum combined
({\em scombine}).  Our radial  velocities were then measured with {\em
  fxcor} using  GC0106 as a  template spectrum.  This object is a
bright cluster with previously well determined radial velocity from the literature.

\subsection{VIMOS MOS/VLT Dataset}
\label{sec:vimos}

We obtained spectroscopy with the VIMOS MOS/VLT 8 m telescope in 2007 (PI:
M. G{\'o}mez, Program ID 079.D-0539), on Paranal, Chile.  We obtained 5 fields located around
the inner regions of NGC 5128, each with 4 CCD quadrants of 7\arcmin\
$\times$ 8\arcmin\ separated by  2\arcmin\ gaps.  We targetted 166 
previously known GCs and the remaining slits were placed on
candidate GCs.  Our slit width was 1.0\arcsec, emcompassing both
our target objects as well as surrounding sky.  

Our target data fields were taken in $\sim1000$ second exposures with the
high-resolution (HR) blue grism with no filter and approximate
wavelength coverage of $4100-6300\rm{\AA}$.  The spectral
resolution was 2050 and the dispersion 0.51 $\rm{\AA}$/px for a 1\arcsec\
slit.   We observed with a binning of 1 by 1
and a gain of 1.96 e$^-$/ADU and a readout noise of 3.1 e$^{-}$. 

Our general reduction performed with IRAF involved combining  ({\em zerocombine}) and removing our zero
(bias) exposures from the science targets along with subtracting
the overscan region ({\em ccdproc}).  We then subtracted the
cosmic rays in our target fields with the ESO-MIDAS task {\em
  filter/cosmic}.  Following the cosmic ray removal, we returned to IRAF
to identify and fit the sky
regions as well as trace and fit each target spectrum with a legendre function of order 5.  
We then extracted and removed the sky spectrum from our target spectrum ({\em apall}).
A comparison arc,  taken with each target field, was used to 
identify lines ({\em identify}) from which we 
applied the dispersion correction to the data ({\em dispcor}).  The different 
exposures for each field were combined  ({\em
  scombine}) and the radial velocities were measured ({\em fxcor}) using 
the template GC 158-213 \citep{puzia02}.  

\subsection{Hydra/CTIO Dataset}
\label{sec:hydra}

We obtained spectroscopy from  the Hydra Spectrograph on the Cerro
Tololo  Interamerican  Observatory (CTIO)  Blanco  4  m telescope,  in
Chile (PI: W. Harris, Program ID 2007A-0070).  We observed from April 7  to April 10, 2007 and had 2.5 usable
nights  of  data  out  of  4, with lost time due almost  entirely  to  poor  weather
conditions.  We observed 4 target fields, each with a 40 \arcmin\
field.  The fiber spectrograph  has 2 \arcsec\ diameter fibers, easily
capturing the light of a GC at the distance of NGC 5128.   We placed
available  fibers on  both candidate and known GCs  as well as approximately  25 fibers placed on
random sky positions in each field.   Our sky
fibers were  scattered throughout the  full radial range and within  the full  concentric range  of the  field.  

All of our data  was obtained in the red regime with  a range of $\sim
2300  \rm{\AA}$ centered  on $7600  \rm{\AA}$, capturing  H$\alpha$ and  the two
bluest of the Calcium triplet lines.  We used the KPGLF grating with a
dispersion of  $0.57 \rm{\AA} /$px  and no filter.   Our binning was $1\times  2$
($2048\times 4096$ pixels) and we  used a gain setting of 0.84 e$^{-}/$ADU
and a readout noise of 3 e$^{-}$.

We reduced  our data  with the IRAF software.  We removed the combined {\em zerocombine} bias in  IRAF  as well as subtracting the overscan region of the
CCD from the  exposures ({\em ccdproc}).
The cosmic rays were removed using the L.A. Cosmic task
\citep{vandokkum01}.  We created our flat field using
Milky  flats taken on the  first day of  observations by combining
our 5 exposures (each of  300 seconds) using {\em flatcombine}.    
The  task  {\em  dohydra}  was then  run  on  each  target
exposure to define the  apertures, fit a flat field to our combined milky flat,
as well as  define a pixel-to-wavelength transformation with our
comparison arc spectrum, which was then applied to our target data.   Next,
we combined  our twilight exposures  for each night  ({\em flatcombine}) and corrected for the fibre throughput 
differences.  During  our pixel-to-wavelength correction,  we also
trimmed the exposures to include $6500-8740 {\rm{\AA}}$ with 2000 pixels.  We
then extracted the spectra  from multi- to one-dimensional fits file format
with {\em  scopy}.  
  
In order to remove the sky contribution, we
scaled the averaged sky spectrum for all fibers in each field to match the intensity of the spectrum of each 
individual object, which was then subtracted from the target
spectrum using {\em sarith}.  
We removed the  continuum from each exposure ({\em continuum})
with  a 10th order  legendre function,  rejecting the  6 lowest  and 3
highest pixels.   Each object was  then combined with the  same target
from all exposures with {\em scombine} by taking the average.

We measured  radial  velocities of  our  target objects by
creating  a  template  GC  using  two  bright  GCs  with
previously measured radial  velocity measurements from high-resolution
spectroscopy.  We chose GC0330 and GC0365 from
field 1 with weighted velocity measurements of $673\pm1$ km
s$^{-1}$ and $594\pm1$ km s$^{-1}$, respectively.  We combined the two
fully reduced, sky subtracted spectra by averaging with {\em scombine}
after each spectrum had been doppler corrected to 0 km s$^{-1}$.  With
our template in  hand, we used {\em fxcor} to
cross-correlate  each  object  with  our template  M31 cluster GC 158-213. 

\section{Globular Cluster System of NGC 5128}
\label{sec:rv}

\subsection{Newly Confirmed Globular Clusters}
\label{sec:newGCs}

\begin{figure}
\plotone{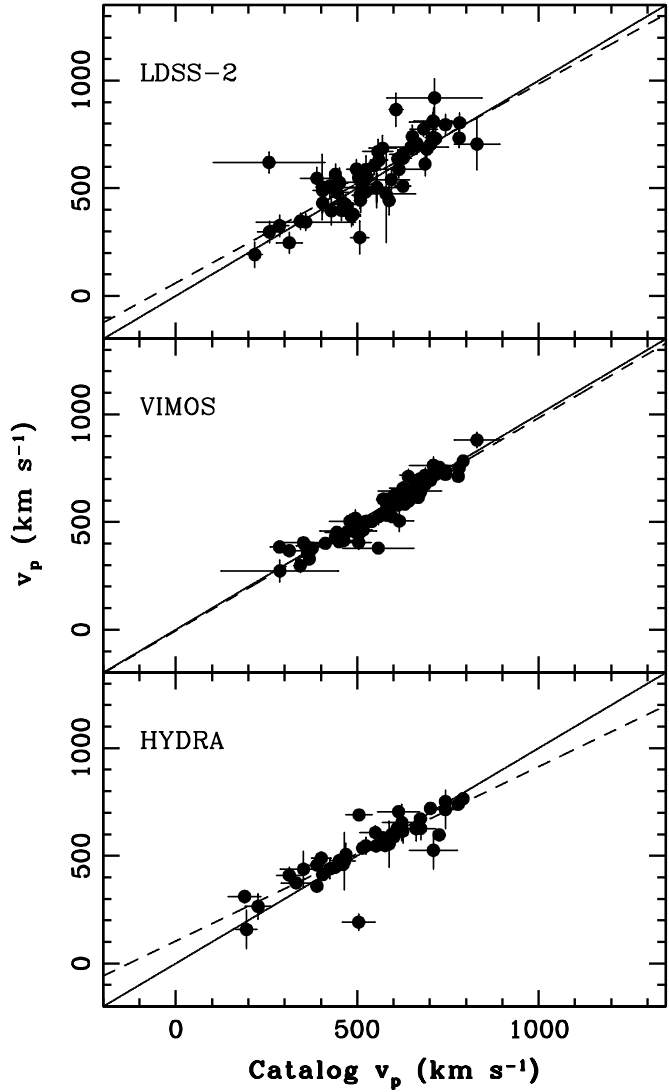}
\caption{We compare radial velocities measured from our three surveys,
LDSS-2 ({\it top panel}), VIMOS ({\it middle panel}), and HYDRA ({\it
  bottom panel}) to the weighted radial velocity measurements in the
\cite{woodley07} catalog.  The solid line is a 1:1 fit and the dashed
line is the best least squares fit between the two sets of
measurements.  The least square fits are (slope,
intercept)=$(0.92,60.2)_{LDSS-2}$, $(0.99,-4.71)_{VIMOS}$, and $
0.81,104.1)_{HYDRA}$.  We find good agreement in the LDSS-2 and VIMOS
studies with the previously cataloged values. The HYDRA study has one
significant outlier, which is GC0445 with cataloged velocity
$504\pm45$ km s$^{-1}$ and our
new measurement $192\pm39$ km s$^{-1}$.  GC0445 has been measured only
one previous time in \cite{woodley09}.} 
\label{fig:vel_match}
\end{figure}

We have remeasured radial
velocities for 218 previously known GCs, listed in
Table~\ref{tab:vel_knowns}.   This table lists the
known GC ID, the R.A. and Decl. (J2000), and the velocity measurements
presented here.  It also lists the {\it new} weighted velocity
measurement of all GCs in the past literature \citep{vandenbergh81,hesser84,hesser86,harris92,peng04c,woodley05,rejkuba07,beasley08,woodley09} {\it including} the new
measurements of this study.
We have remeasured 64 GCs with LDSS-2, 154 GCs with
VIMOS, and 57 GCs
with Hydra.  Least square fits are shown for the comparison of our measured velocities with known values in the
literature in Figure~\ref{fig:vel_match}.  The comparisons for nearly
all GCs are well within the measurement uncertainties.  In the rare
cases of outlying measurements, the most obvious being GC0445 in the
Hydra study, there is only one previous radial velocity measurement
from \cite{woodley09}, but both measurements are still within the bounds of
accepted radial velocity confirmation of $200-1000$ km s$^{-1}$.  We have higher confidence in the radial velocity measured with GMOS because the spectrum had a higher S/N than the Hydra data.

Table~\ref{tab:vel_new} lists our 155 newly confirmed GCs from our radial velocity
measurements.  Table~\ref{tab:vel_new} lists the GC ID, the R.A. and
Decl. (J2000), the C, M, and T$_1$ magnitudes from
\cite{harris04a}, and the measured radial velocity.  We identified 29 new GCs with LDSS-2
(GC0416-GC0444), 80 new GCs\footnote{Four of these newly identified
  GCs may be stars based on their structural paramter measurements,
  but have velocities within the accepted range.  These objects,
  GC0485, GC0508, GC0523, and GC0526, are included here as new GCs but
  with caution.} with VIMOS (GC0480-GC0559), and 46 new GCs with Hydra
(GC0560-GC0605).  The final column in the table, lists the weighted
average 
of all radial velocity measurements of these new GCs.

\begin{figure}
\plotone{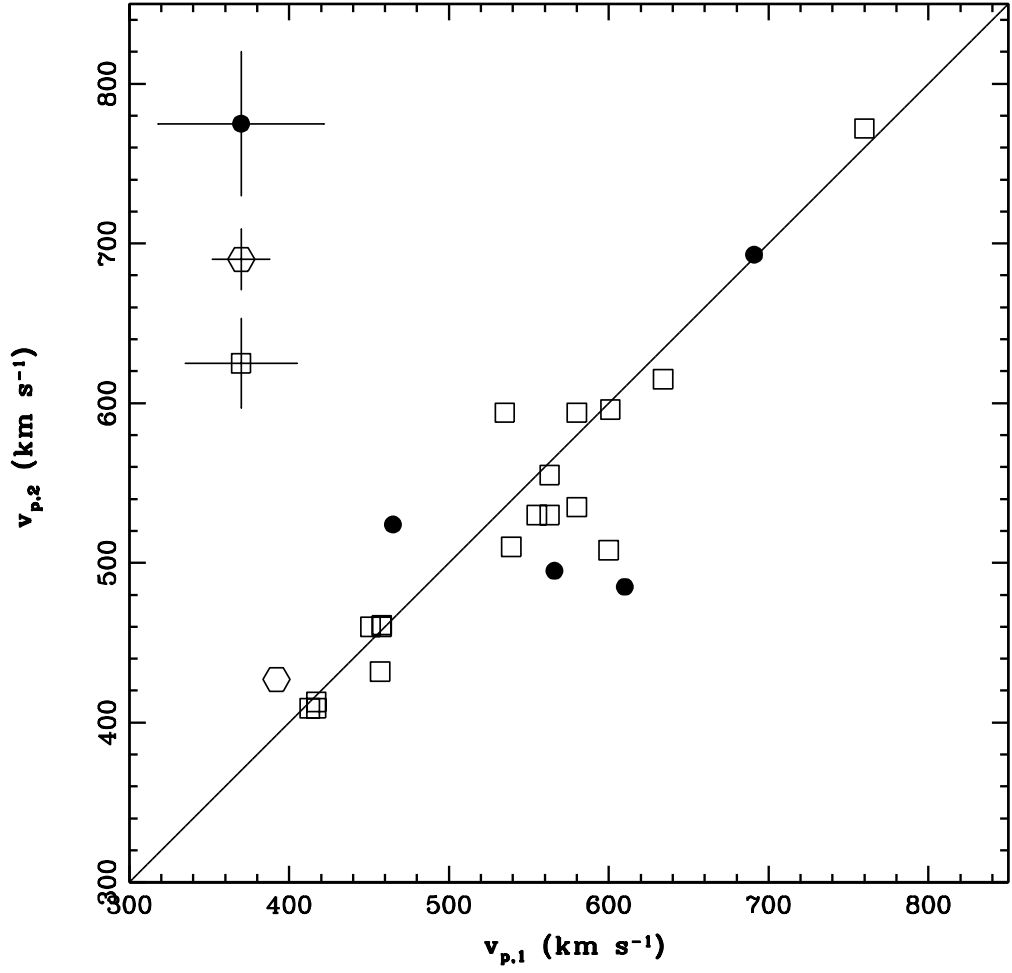}
\caption{A comparison of GCs that were measured more than once within
  each survey.  The solid line is a 1:1 fit.  LDSS-2 had 4
  ({\it solid circles}), VIMOS had 1 ({\it hexagon}), and HYDRA had 17
  multiple measurements ({\it squares}).  The least squares fit for
  the Hydra data has a slope of 0.96 and an intercept of 13.1,
  indicating decent consistency of multiple measurements within the
  different HYDRA fields.}
\label{fig:vel_mult}
\end{figure}

We compare all multiple measurements of the same GC within each radial
velocity study to test for consistent measurements.  A comparison fit
is shown in Figure~\ref{fig:vel_mult} for the 4, 1, and 17 multiple
measurements for the LDSS-2, VIMOS, and Hydra studies, respectively.  An
average uncertainty for the studies is also indicated in the figure.
We find the multiple measurements match a 1:1 fit within the
measurement uncertainties of each study.  The mean rms difference
between multiple measurements are 64 km s$^{-1}$ for LDSS-2, 22 km
s$^{-1}$ for Hydra, and 35 km s$^{-1}$ for VIMOS (recall there is only
1 GC measured in the VIMOS dataset that was measured in multiple fields).

Of our measured  GC candidates there are a small number with velocities  between $150-250$ km
s$^{-1}$ which may be either GCs or foreground Milky Way stars.   This
velocity domain is the major region of contamination of the GC catalog
in NGC 5128. 
To distinguish between these two objects, we measured their structural
parameters  using our superb $1.2 \times
1.2$ deg$^2$  Magellan/Inamori Magellan Areal  Camera and Spectrograph
(IMACS)  images   ({\it  R}   filter)  taken  in   0.5\arcsec\
seeing.
The structural parameters were measured with
ISHAPE  \citep{larsen99,larsen01}  which  convolves  the  stellar  point
spread (PSF) function with  an
analytical King  profile \citep{king62}  and compares the  result with
the  input candidate  image achieving  a best  match. 
The structural
parameters,  measured from  the  models,  that we  used  are the  core
radius, $r_c$,  the tidal radius, $r_t$,  the concentration parameter,
$c=r_t/r_c$,  and  ellipticity.   The  half-light radii  can  also  be
obtained  from  the  transformation, $r_e/r_c  \simeq  0.547c^{0.486}$
which  is  good  to  $\pm2\%$  for $c>4$  \citep{larsen01},  which  is
satisfied for  GCs in NGC 5128 \cite[see][]{gomez07}.  
We compared the measured structural parameters of the GC candidates to normal GCs  in  NGC  5128
\citep{harris06,gomez06,gomez07,mclaughlin08},
as well as to a delta distribution light profile, which describes the PSF
of a star.  Based on their
structural  parameters, we 
classified a small handful of candidates  as GCs with low velocity
measurements.  With our preliminary look at the structural parameters
of GCs in NGC 5128, we note that five GCs are well fit by a delta
distribution light profile.  Four of these are newly identified here (GC0485, GC0508, GC0523, and GC0526), 
while the other is GC0472.  A closer look at the structural
parameters of these objects is underway, however, we have included
them here because they satisfy the velocity criteria for the GCs in
NGC 5128.  We have determined the kinematics below including and
excluding these 5 GCs and found that there is no kinematic difference
between the two solution sets within the uncertainties.  The results
presented thus include the objects.

\subsection{Radial Velocity Distributions}
\label{sec:rvdist}

We have divided our globular cluster system (GCS) into either metal-poor or metal-rich GCs based
on their color information.  
We calculated a  metallicity [Fe/H]$_{C-T_1}$  obtained from a  transformation of
dereddened (C-T$_1$) derived by \cite{harris02} calibrated through the
Milky Way GC data.  The transformation, calibrated in the range of (C-T$_1$)$_o = 0.9$ to $1.9$ mag and [Fe/H]$=-2.3$ to $-0.1$, is
\begin{equation}
\label{eqn:ct1_feh}
[Fe/H]_{C-T_1} = -6.037\times(1-0.82 \times (C-T_1)_o) + (0.162 \times (C-T_1)_o^2)
\end{equation}
where
\begin{equation}
(C-T_1)_o = C-T_1 - 1.966\times E(B -V)
\end{equation}
We use a  foreground reddening value of E(B - V) =
0.11  for NGC  5128, corresponding  to  E(C-T$_1$) =  0.22 for  the
transformation.  The uncertainties
for  a typical  color  are  $\pm0.07$  dex in  the
metal-rich  and   $\pm0.2$  dex  in  the   metal-poor  regimes.
If there is no C-T$_1$ color for the GC, we examine their U-V
color and followed the transformation of \cite{reed94} 
\begin{equation}
\label{eqn:uv_feh}
[Fe/H]_{U-V} = -3.061 + (2.015 \times ((U - V) - E(U-V)))
\end{equation}
with a  foreground reddening value of $E(U-V)= E(U-B) + E(B-V) = 0.2$.  
The GCs were classified as metal-poor ([Fe/H]$_{C-T_1}< -1.0$) or as metal-rich ([Fe/H]$_{C-T_1} \geq -1.0$)
\citep{harris04b,woodley05,woodley07}.
When neither C, T$_1$, U, or V information is
available, we divided the GCs as metal-rich if $(B-I)\geq 2.072$
and as metal-poor if $(B-I)< 2.072$ following \cite{peng04b}.
Our final GC sample of 605 has 291 metal-poor, 292 metal-rich, and 22
GCs with insufficient photometry for any of the above
transformations.  

\begin{figure}
\plotone{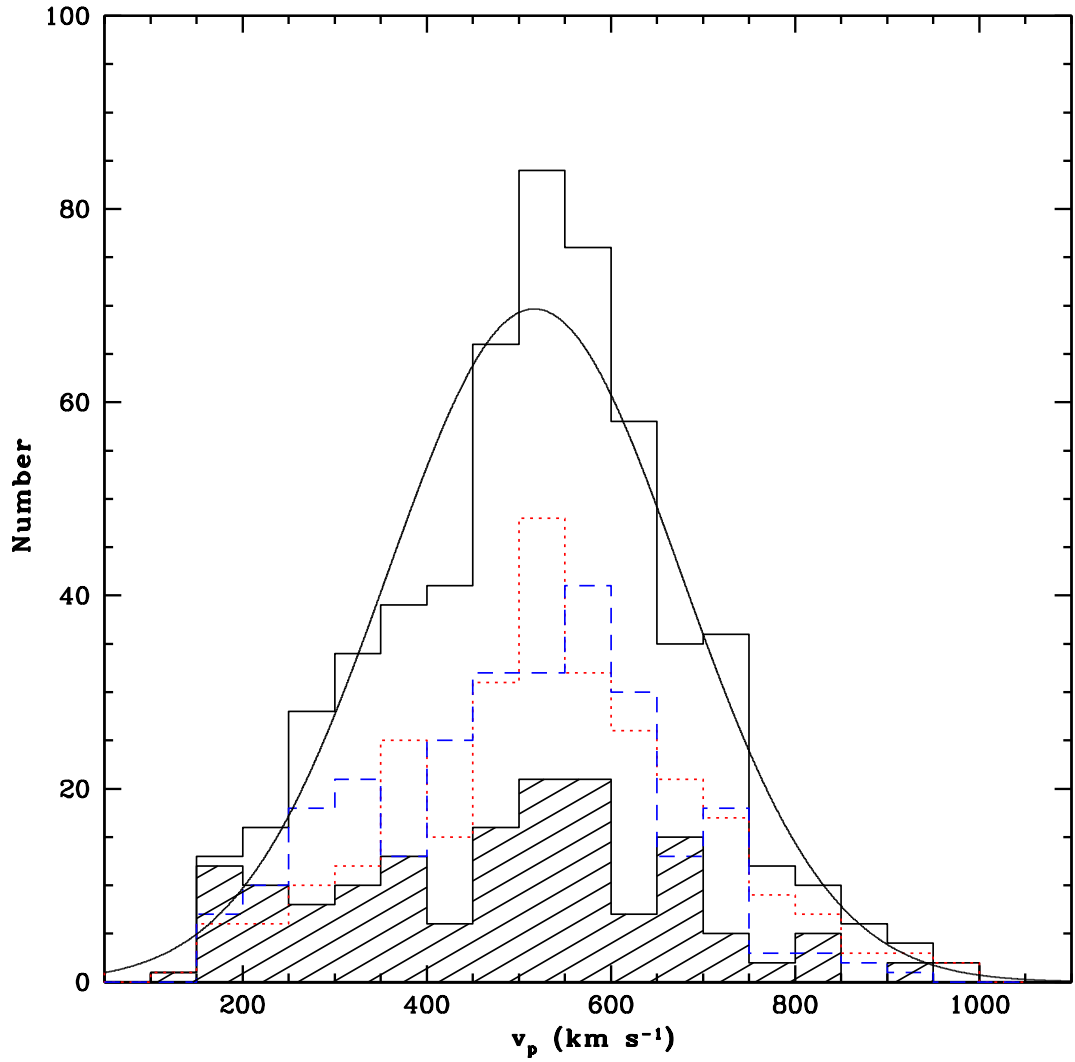}
\caption{The radial velocity distribution of the entire GCS ({\it open
  histogram}), the metal-rich GC subpopulation ({\it red dotted
  histogram}), the metal-poor GC subpopulation ({\it blue long dashed
  histogram}), and the newly confirmed GCs from this study ({\it
  hatched histogram}). The entire GCS is fit with a Gaussian function
using Rmix (mean = $516.7\pm6.8$ km s$^{-1}$ and sigma = $159.5\pm4.8$ km s$^{-1}$). {\it See the electronic version for
    the color figure}. }
\label{fig:vel_N}
\end{figure}

The   radial   velocity    distribution   functions   are   shown   in
Figure~\ref{fig:vel_N} for the entire  GC population, the red GCs, the
blue  GCs, and  the newly  confirmed GCs  from this  study.  We  fit a
Gaussian function  to the entire GCS  using RMIX\footnote{The complete
  code, available  for a variety  of platforms, is  publicly available
  from        Peter        MacDonald's        Web       site        at
  http://www.math.mcmaster.ca/peter/mix/mix.html.}.  We find  the best
Gaussian fit  has a mean of  $517\pm7$ km s$^{-1}$ and  a sigma of
$160\pm5$ km s$^{-1}$.  The median  is $v_p = 526$ km s$^{-1}$ and
the standard Pearson's approximation for the mode is $v_p = 3 \times \rm{median} - 2 \times \rm{mean} = 545$ km
s$^{-1}$, which  is closer to  the accepted systemic velocity  of the
galaxy of $541$  km s$^{-1}$.  The shift  to lower mean
velocity than the galaxy's systemic velocity may suggest that our catalog of GCs may
still contain  contamination at the  low velocity end  from foreground
blue Milky  Way  halo  stars.   We  expect  to  find  the  higher  velocity
counterparts  to these  lower  velocity  GCs, or  hope  to remove  the
possible low velocity contamination in future work.  
We do not find any bias in previous radial velocity searches 
which would hinder finding these higher velocity GCs.

\begin{figure}
\plotone{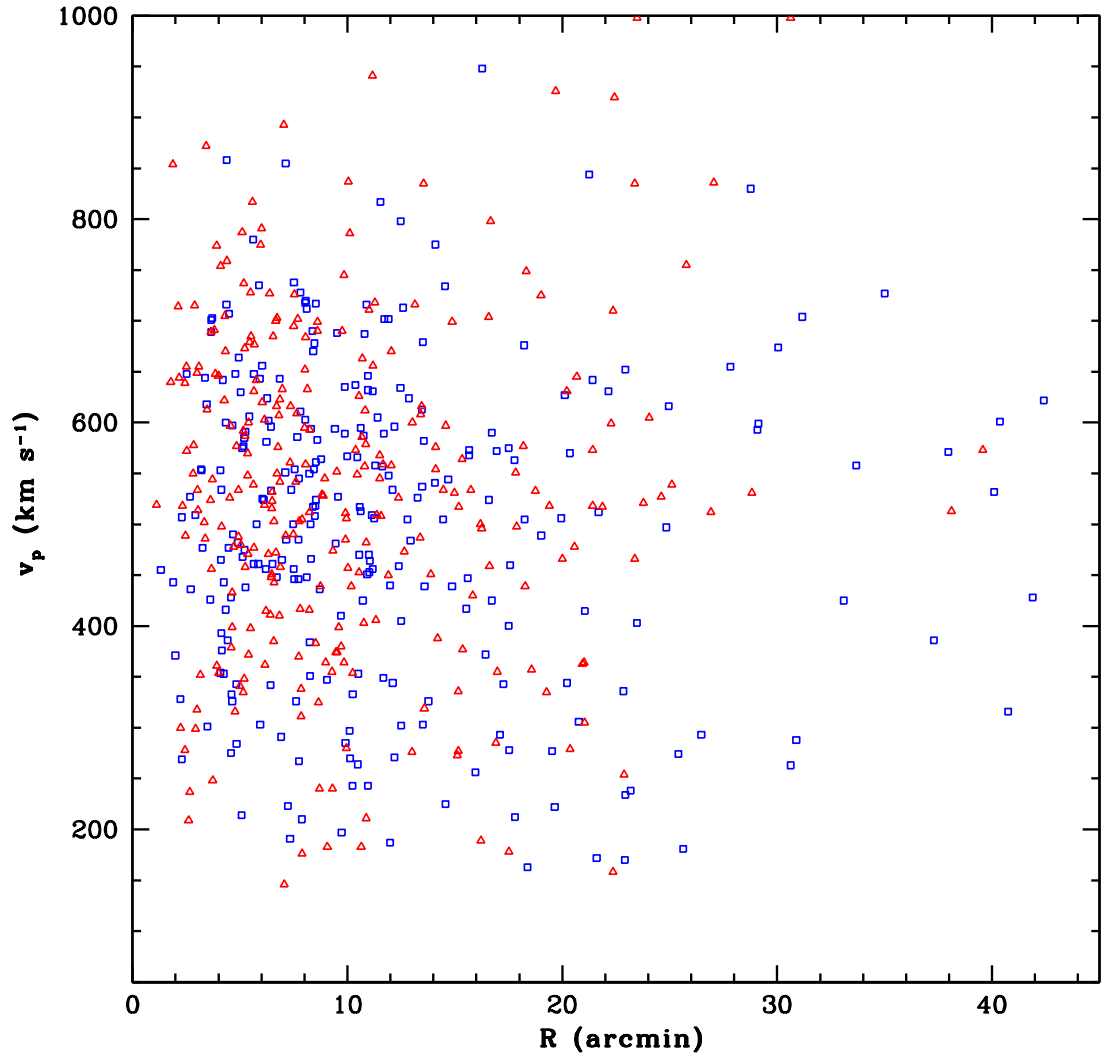}
\caption{Radial velocity measurements for all metal-rich ({\it red
    triangles}) and metal-poor ({\it blue squares}) GCs as a function of
  projected galactocentric radius. {\it See the electronic version for
    the color figure}.} 
\label{fig:vel_R}
\end{figure}

We have analyzed the radial velocity
measurements as a function of galactocentric distance, shown in
Figure~\ref{fig:vel_R}.  If a large number of low-velocity objects
are Milky Way foreground star contamination, we would expect to see a
large number of blue objects covering this velocity range.  Looking at
Fig.~\ref{fig:vel_R}, this is not clear, but in the
15\arcmin-20\arcmin\ range, we can see there are more GCs with lower
radial velocities than high radial velocities.  

To examine this more closely, we have created velocity distributions as a function of
galactocentric position, shown in Figure~\ref{fig:velrad_histo}.
We have binned our GCs in units of effective radius of the galaxy
light (R$_{eff} \sim 5$\arcmin), from
0\arcmin-5\arcmin, 5\arcmin-10\arcmin, 10\arcmin-15\arcmin,
15\arcmin-20\arcmin, with the outer GCs in the last bin of
20\arcmin-45\arcmin.  We fit the distributions with unimodal
Gaussians with RMIX, overplotted in the figure, with mean, sigma, and
reduced $\chi^2$ 
values listed in Table~\ref{tab:vel_fits}.

While it is clear that the
metal-poor GCs on average lie at lower radial velocities than the metal-rich GCs, only
the 10\arcmin-15\arcmin and 20\arcmin-45\arcmin\ bins show the metal-poor GCs
lie at significantly lower velocities. They may well be affected 
by contamination of foreground field stars in
these outer regions, but the number of known GCs here is quite low, 
and the asymmetries may be purely low number statistics. 

\subsection{Photometry and Projected Distribution}
\label{sec:phot_rad}

\begin{figure}
\plotone{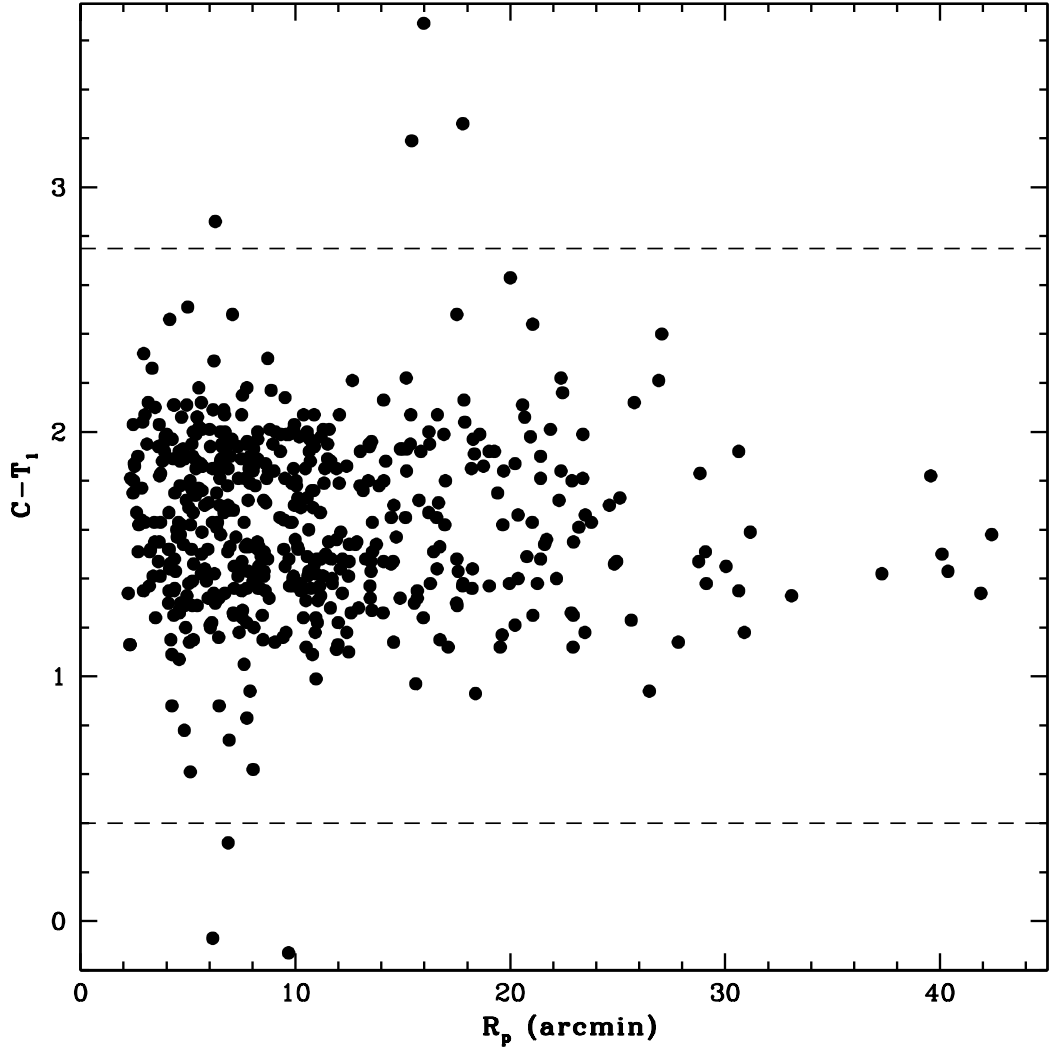}
\caption{The C-T$_1$ color of all GCs in NGC 5128 as a function of
  galactocentric radius. The dashed lines indicate the bounds of the GCs
  included in the kinematical and subsequent dynamical analysis.
  Excluded are the 4 reddest (GC0078, GC0408, GC0411, and GC0552) and the
  3 bluest (GC0084, GC0282, GC0432) GCs. } 
\label{fig:rad_ct1}
\end{figure}

With a total sample of 605 GCs in NGC 5128, we briefly examine their
photometric properties and projected distribution.  In
Figure~\ref{fig:rad_ct1} we show the color, C-T$_1$, as a function of the
projected galactocentric radial distance in the galaxy. 
There are 4 very red GCs (GC0078, GC0408, GC0411, and GC0552) which by
their colors may be red background galaxies, but their radial velocity measurements
are $568\pm28$, $838\pm63$, $548\pm33$, and $811\pm32$ km s$^{-1}$,
respectively.   The 3 very blue GCs (GC0084, GC0282 and GC0432) 
have radial velocities of $348\pm31$, $613\pm6$, and $396\pm84$ km
s$^{-1}$, respectively.  These blue GCs could
either be Milky Way foreground stars or they could
genuinely be very young GCs.  Ages of these blue GCs determined
spectroscopically are needed to securely distinguish the difference.   
We exclude these 7
objects from our kinematical analysis below in an attempt to examine
the kinematics of the normal GC population.  We have also analyzed the
kinematics after removal of the 5 GCs that may be stars (discussed in
Section~\ref{sec:newGCs}), and find there is no difference of any
determined parameter within uncertainties.

\begin{figure}
\centering
\begin{tabular}{cc}
\epsfig{file=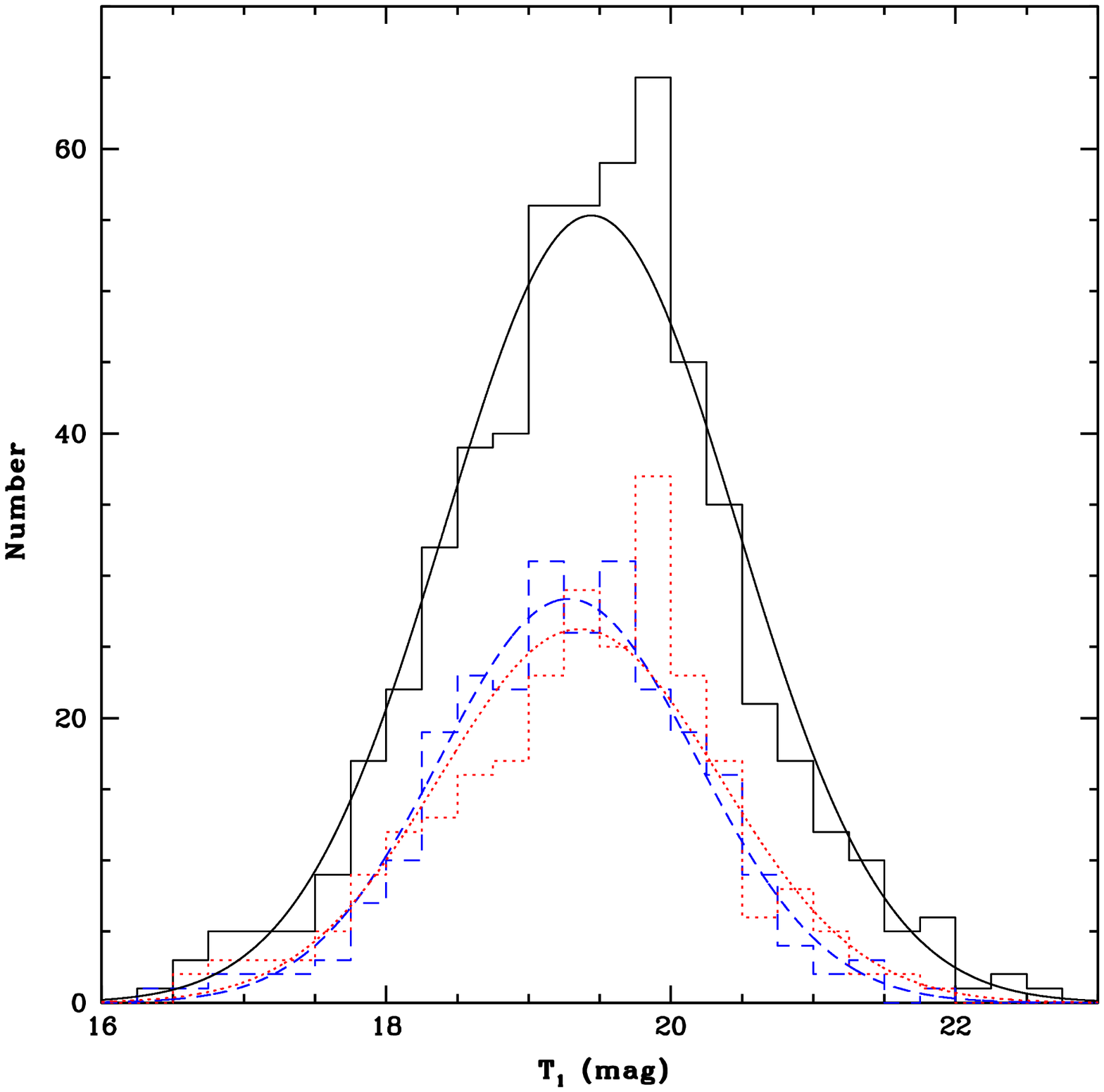,width=0.4\linewidth,clip=} &
\epsfig{file=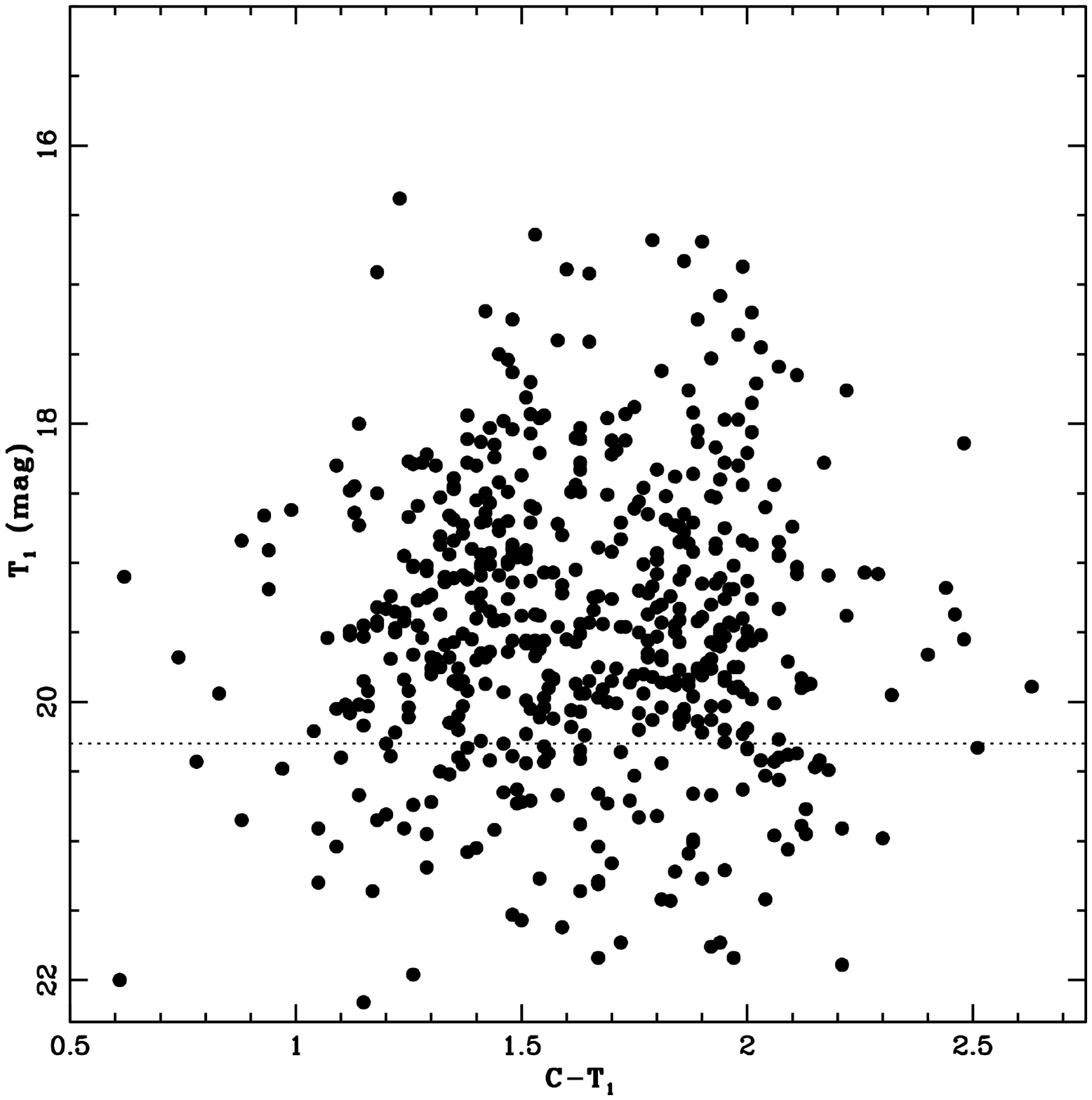,width=0.4\linewidth,clip=} \\
\end{tabular}
\caption{{\it Left:} The luminosity function of all GCs in NGC 5128 with available
  T$_1$ magnitude (N$=569$) is fit by a Gaussian function with a mean
  of $19.44\pm0.04$ mag and sigma $= 1.03\pm0.03$ mag. The luminosity functions for the metal-poor ({\it blue}) and
  metal-rich ({\it red}) GCs are also fit by Gaussian functions with means of $19.28\pm0.06$ mag and $19.36\pm0.06$ mag and
  sigmas of $ 0.90\pm0.04$ mag and  $ 0.98\pm0.04$ mag, respectively.
  {\it Right:}  The color-magnitude diagram of the entire GCS in NGC 5128 exhibits
the color-bimodality of blue and red GCs.  The {\it dotted line}
represents the expected turnover magnitude.  {\it See the electronic version for
    the color figure}.} 
\label{fig:Tmag}
\end{figure}

The GC luminosity function is considered a {\it standard candle}
distance indicator.  
The turn-over magnitude of
the GC luminosity function of the Milky Way and M31 GCSs has a mean of
$V = -7.4\pm0.2$ mag \citep{harris91,ashman98}.  With our {\it a priori}
knowledge of a distance, we can therefore estimate the turnover
magnitude of the GCS in NGC 5128 to be $V = 20.8$ mag, corresponding to T$_1 = 20.3$ mag.
The left hand side of
Figure~\ref{fig:Tmag} shows the luminosity function for 569 GCs with the necessary photometry.  
The distribution appears Gaussian with a mean of $19.44\pm0.04$ mag and
a sigma of $1.03\pm0.03$ mag.  Therefore we are clearly missing the
faintest GCs in the system.  
We have also fit the luminosity
functions of the metal-rich and metal-poor GC subpopulations and found
a mean of $19.28\pm0.06$ mag and a sigma of $0.90\pm0.04$ mag for the
metal-poor GCs and a mean of $19.36\pm0.06$ mag and a sigma of $0.98\pm0.04$ mag for the
metal-rich GCs.  The metal-rich GCs have a slightly fainter peak luminosity than the metal-poor GCs, which has also been
seen in other GCSs \citep{larsen01,tamura06a}.  Assuming the mass
distribution between the metal-rich and metal-poor GCs is the same
(which may not be true if the GC subpopulations formed at different
times), the difference in the turn-over magnitude would be expected if the
color of a GC reflects its metallicity, which is a good approximation for GCs older than a
few Gyrs \citep{worthey94}.  The integrated light of a metal-poor GC would appear
brighter than a metal-rich GC, at a given age, because the giant stars would be
brighter for a metal-poor GC \citep{elson96}.  

On the right side of Fig.~\ref{fig:Tmag}, the color-magnitude diagram of
the GCS shows a clear bimodality, also seen in
Fig.~\ref{fig:rad_ct1}.  The photometric uncertainties for the bright objects 
from \cite{harris04a,harris04b} range typically from $0.05-0.09$ mag, so much of the observed scatter
in both sequences in Fig.~\ref{fig:Tmag} represents real
cluster-to-cluster metallicity differences.

\begin{figure}
\plotone{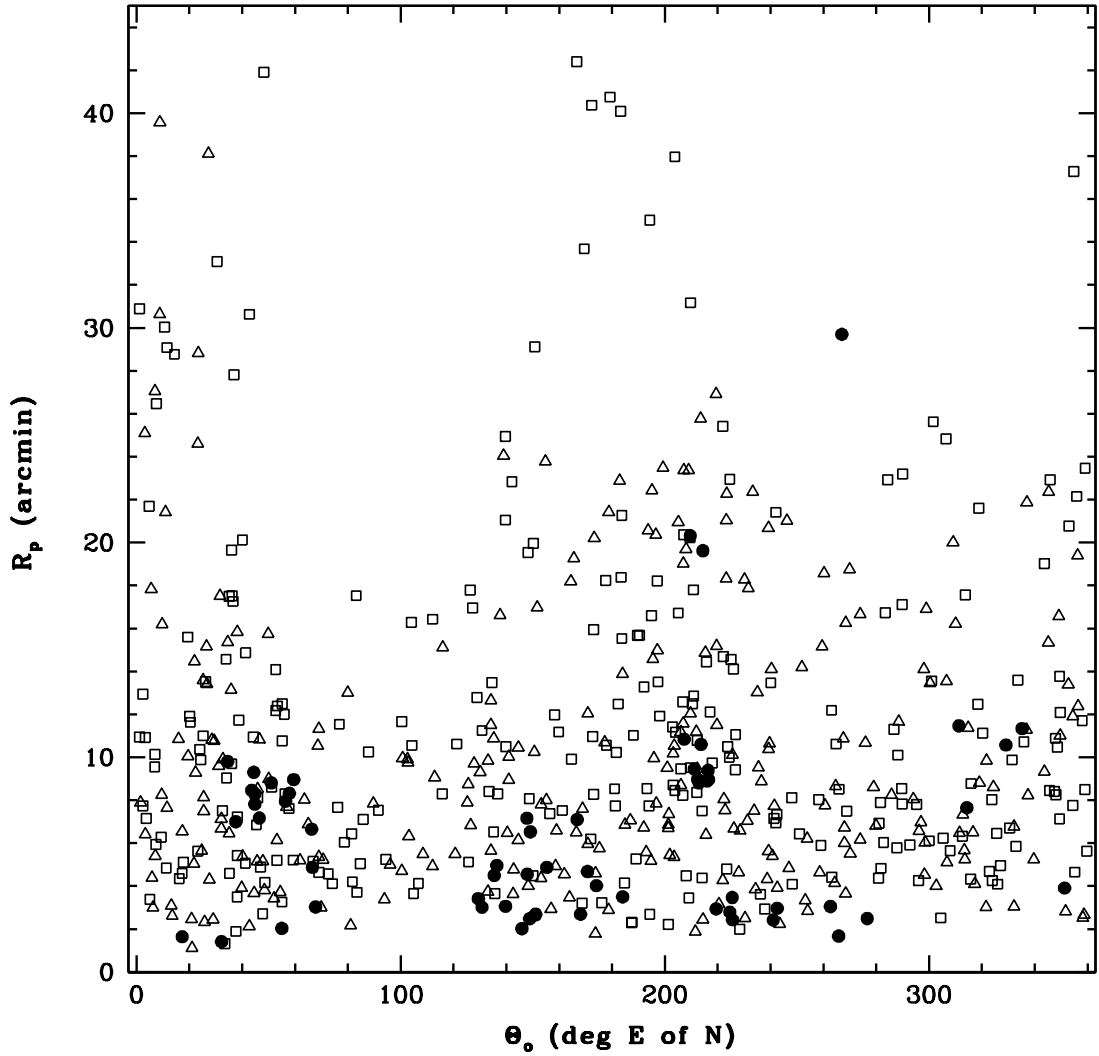}
\caption{The GCS of NGC 5128 is shown as their galactocentric radial
  position as a function of their azimuthal position measured in
  degrees East of North on the projected sky.  The {\it squares} are
  the metal-poor GCs with a measured velocity, the {\it triangles} are
metal-rich GCs with a measured velocity, and the {\it solid circles}
are GCs that either have measured velocities but no color information,
or GCs with no measured radial velocities.  These latter objects were
confirmed from resolved {\it Hubble Space Telescope} images
\citep{harris06}. Clearly, the outer regions ($>10$\arcmin) of NGC 5128 need to be
searched for GCs to remove any potential spatial bias in this region.} 
\label{fig:Theta_R}
\end{figure}

The projected radial distribution of the 605 GCs in NGC 5128 is shown
as a function of projected azimuthal angle in the sky (measured in
degrees east of north) in Figure~\ref{fig:Theta_R}.
The GCS of NGC 5128 has been spatially searched 
extensively out to 10\arcmin, while only a few searches, including 
the work presented here, have tried to search thoroughly out to
20\arcmin.  
Beyond this distance, only \cite{peng04c} has
made an attempt to extensively search along the isophotal major axis
of the galaxy of $35^o$ E of N \citep{dufour79}. 
Within 20\arcmin, we do see a lack of GCs along the
isophotal minor axis \citep[$119^o$ E of N][]{dufour79}.  However,
with upcoming searches for GCs beyond 20\arcmin, we will be able to test if the
GCS follows the ellipticity of the underlying galaxy light.

\section{The Kinematics and Dynamics of the Globular Cluster System}
\label{sec:kin_dyn}

\subsection{Mass and Mass-to-Light Determinations}
\label{sec:mass}

In the  current paradigm  of cold dark  matter (CDM)  and hierarchical
merging, we expect to find luminous, baryonic matter at the centers
of  much larger  dark matter  halos.  Determining  the full  extent of
these dark matter  halos and the total mass of a  galaxy
has proven to  be quite challenging.  

The technique  of measuring  integrated stellar
light spectroscopy is useful  for obtaining a velocity dispersion from
high S/N data  in which one can calculate a  mass.  However, the stellar
halo gets fainter with galactocentric distance, limiting studies to a
few inner effective radii \citep{kronawitter00,gerhard01}.  HI gas has
also been successfully used to  determine mass content, but is
inapplicable for early-type galaxies without large amounts of gas.  
Extended X-ray  halos   are    found   in   massive   early-type   galaxies
\citep{mathews03,humphrey06} which may  be used to determine the mass out to large  distances,  where dark
matter  is expected to  dominate.  This has been done in   luminous  
ellipticals  such   as  in  NGC   4486  (M87)
\citep{mclaughlin99},   and  NGC   4649  (M60)   \citep{bridges06,hwang08}.   

In NGC 5128, tracer particles have been used  to estimate the mass
via  both  PNe  \citep{peng04a,woodley07}  and  GCs 
\citep{peng04b,woodley06,woodley07}.   With our larger sample of over 
560 GCs with radial velocity measurements, we can derive an 
improved estimate of the mass and  mass-to-light ratio of NGC 5128.  

\subsubsection{Pressure and Rotation Supported Mass}
\label{sec:tme}

We calculate the mass of NGC 5128 using its GC
population by adding together the component of mass
supported by pressure, $M_p$, and that supported by rotation, $M_r$
with
\begin{equation}
\label{eqn:mtot}
M_{tot} = M_p + M_r
\end{equation}
where $M_p$ is determined by the Tracer Mass estimator, discussed by
\cite{evans03} as 
\begin{equation}
\label{eqn:tme}
M_{p} = \frac{C}{GN} \sum_{i}(v_{f_i} - v_{sys})^2 R_i
\end{equation}
where $N$ is the number of objects in the sample, $R_i$ is the
projected galactocentric radius of the tracer object, and $v_{f_i}$ is the
radial velocity of the tracer object {\it with the rotation
component removed}.  We have determined the rotational component for
the GCs using the kinematic solution described in
Section~\ref{sec:vaa}.    The  Tracer Mass  estimator is  an advantageous
mass estimator over the projected and virial mass estimators described
in  \cite{bahcall81} and \cite{heisler85}  because  it does  not assume  that
the distribution of tracer objects follows the underlying dark matter halo.

We assume the GCs
are an isotropic population, so the value of C is
\begin{equation}
\label{eqn:C}
C = \frac{4(\alpha + \gamma)(4 - \alpha -\gamma)(1-(\frac{r_{in}}{r_{out}})^{(3-\gamma)})}{\pi(3-\gamma)(1-(\frac{r_{in}}{r_{out}})^{(4-\alpha-\gamma)})}
\end{equation}
where $r_{in}$ and $r_{out}$ are the 3-dimensional radii corresponding
to the  2-dimensional projected radii of the  innermost, $R_{in}$, and
outermost, $R_{out}$,  tracers in the sample.   The parameter $\alpha$
is set  to zero for an  isothermal halo potential in  which the system
has a flat rotation curve at large distances.  The slope of the volume
density  distribution is $\gamma$  which is  found by  determining the
surface  density slope  of the  sample and  deprojecting the  slope to
three-dimensions,  written as  $r^{-\gamma}$.   Although the  Tracer
Mass  estimator uses  a sample  of GCs  defined between
$r_{in}$ and  $r_{out}$, it determines the total  enclosed mass within
the  outermost point.

The mass component supported by rotation 
in Eqn.~\ref{eqn:mtot} is determined from the rotational
component of the Jeans Equation,
\begin{equation}
\label{eqn:rje}
M_{r} = \frac{R_{out}v^{2}_{max}}{G}
\end{equation}
where $R_{out}$ is the outermost tracer projected radius in the sample and
$v_{max}$ is the rotation amplitude.

\subsubsection{Systemic Velocity, Rotation Amplitude, and Rotation Axis}
\label{sec:vaa}

To obtain the amount of rotation that needs to be removed from each GC
in Equation~\ref{eqn:tme}, we use the equation
\begin{equation}
\label{eqn:kin}
v_p(\Theta) = v_{sys} + \Omega R  sin(\Theta - \Theta_o)
\end{equation}
described in \cite{cote01}.  With our known quantities, the measured radial  velocity,  $v_p$, and  the
angular  position of  each GC  measured on  the projected sky  in
degrees East of North, $\Theta$,   we fit our GC dataset
with  a weighted least  squares fit with the  non-linear function.  From the fit, we  
extract the systemic
velocity,  $v_{sys}$,  the  rotation   amplitude  (a  product  of  the
projected angular velocity and  the projected galactocentric radius of
the  GC) $\Omega  R$, and  the projected  rotation axis,
$\Theta_o$, also measured  in degrees East of North in  the projected sky.   
Eqn.~\ref{eqn:kin}  assumes spherical  symmetry,
that $\Omega$ is only a function of the projected radius, and that the
rotation axis  lies in the plane  of the sky. 

\subsubsection{Surface Density Profiles}
\label{sec:surfacedensity}

\begin{figure}
\plotone{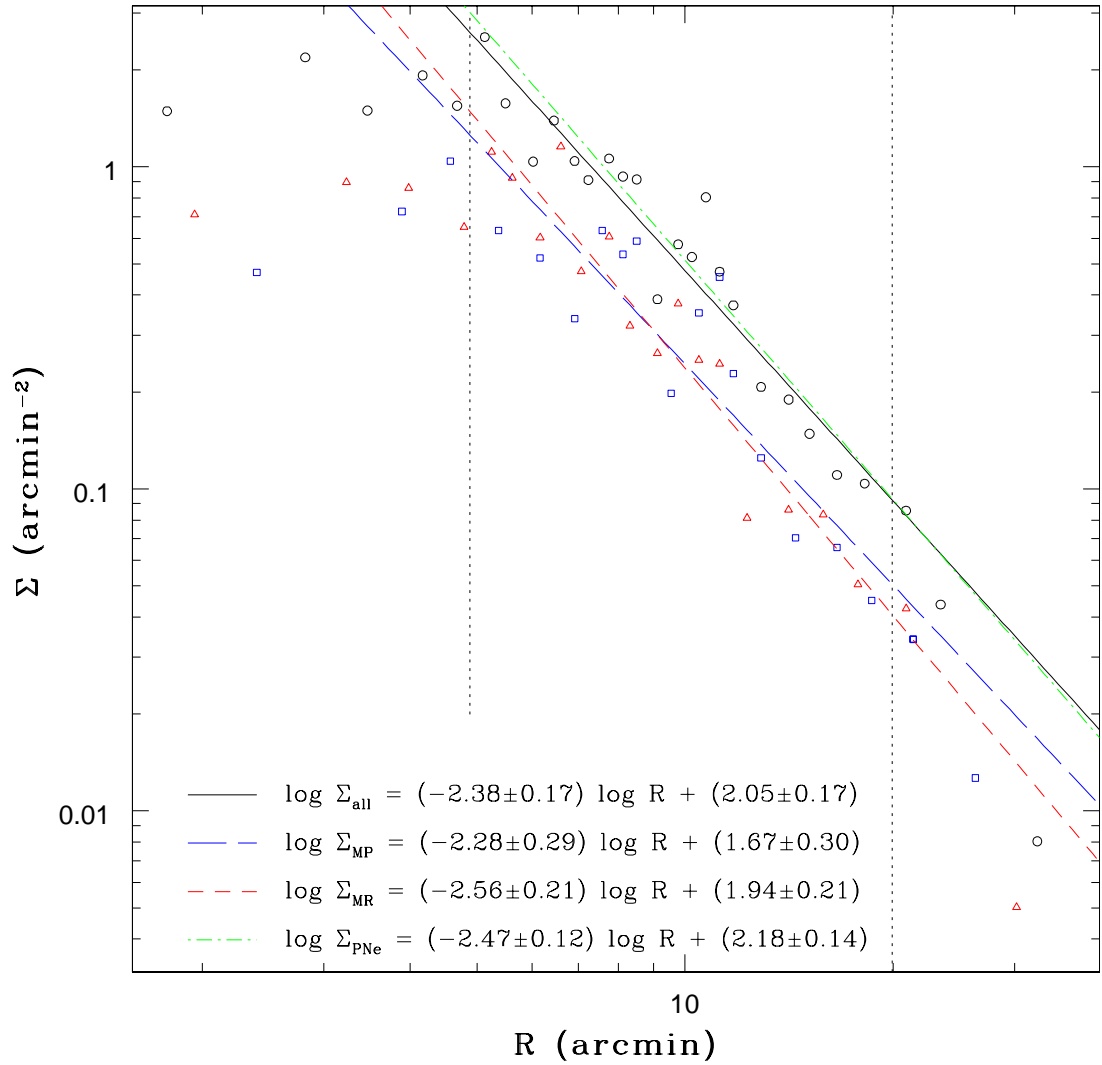} 
\caption{The surface density profiles of the GCS is shown binned in
  circular annuli of equal numbers for the entire GCS ({\it black
    circles}), the MP GCs ({\it blue squares}), and the MR GCs ({\it
    red triangles}).  These three populations were fit with a power
  law function between the regions of azimuthal coverage between
  5\arcmin-20\arcmin\ indicated by the horizontal dashed lines.  The
  best fits are labelled,
  and shown for the entire GCS ({\it black line}), MP GCs
  ({\it long dashed blue line}), and MR GCs ({\it short dashed red
    line}), respectively.  The PNe power law fit between
  5\arcmin-20\arcmin\ is also shown ({\it green dashed dotted line}).
  {\it See the electronic version for
    the color figure}.}
\label{fig:surdensity}
\end{figure}

In order to calculate Equation~\ref{eqn:C}, we require the surface
density profile of the GCS.  
GCSs typically have radial surface density profiles that fall
off with a power law, with an exponent that varies between $1.0-2.5$
\citep{ashman98}.   
To calculate the surface density slope of the GCS in NGC 5128, shown in
Figure~\ref{fig:surdensity}, we bin the GCs into
circular annuli with equal number of GCs in each to allow near equal
statistical weighting \citep{maiz05}.  We then fit the data with a
power-law within the region of approximate azimuthal completeness between
5\arcmin-20\arcmin\ for the entire GCS, the metal-rich
and metal-poor subpopulations of GCs.  Within 5\arcmin, a flattening of the
GCS is evident in Fig.~\ref{fig:surdensity} due to the lack of GC
candidates in the dustlane of the galaxy.  Also evident is a drop-off in
the GCS beyond 20\arcmin\ where there are few to no known GCs along
the isophotal minor axis.  Our best fits within these bounds yield a
{\it volume} density slope of $3.38\pm0.17$, $3.26\pm0.29$, and $3.56\pm0.21$ for the GCS, the metal-poor GCs,
and the metal-rich GCs, respectively, which are used in Equation~\ref{eqn:C}
for all subsequent mass determinations.  We also determine the volume 
density profile of the PNe within the 5\arcmin-20\arcmin\ bounds to
yield a slope $3.47\pm0.12$ \citep[obtained similarly in][who 
determined the surface density profile excluding only PNe within 5\arcmin]{woodley07}.

\subsubsection{Mass and Mass-to-Light Results}
\label{sec:mass_results}

\begin{figure}
\plotone{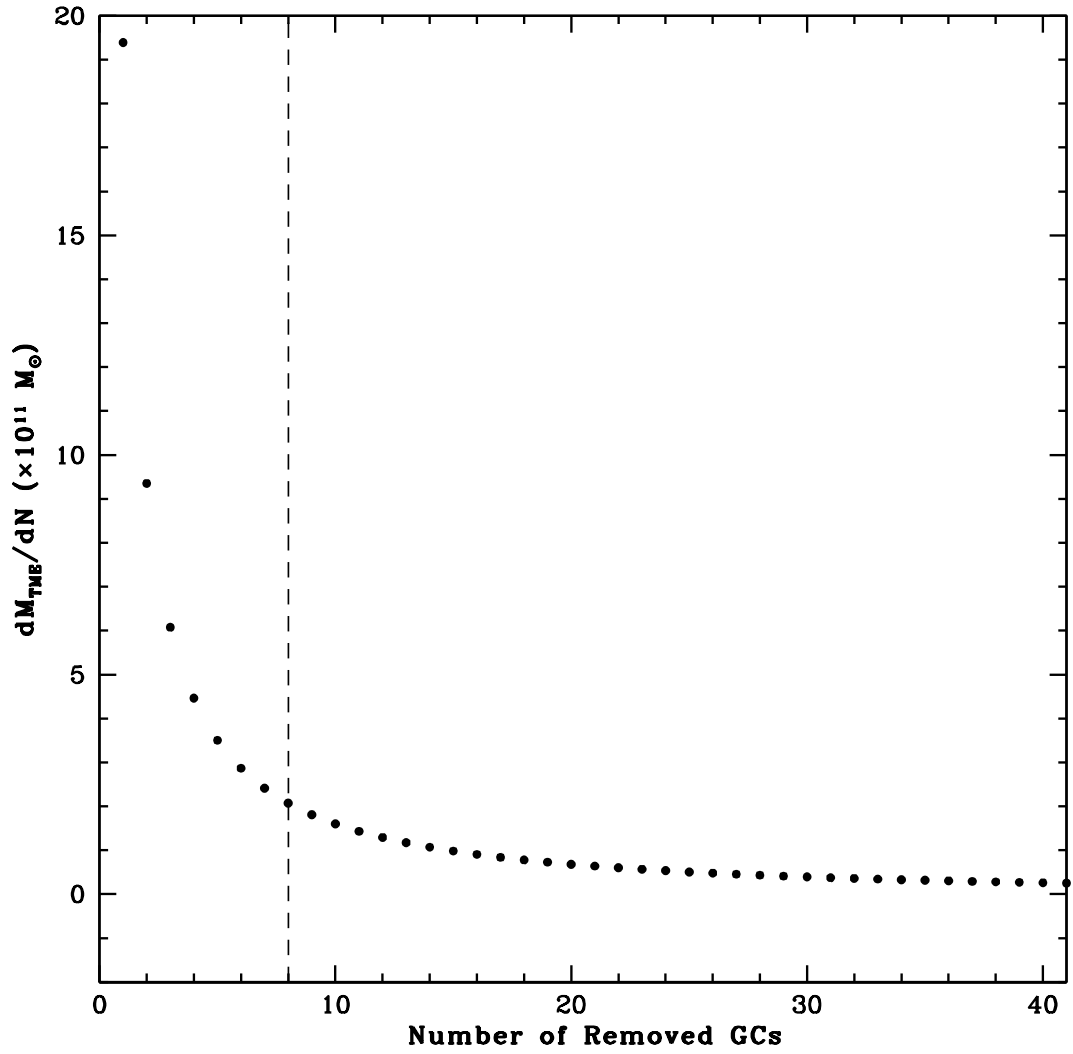}
\caption{The total change in mass normalized by the number
  of GCs subsequently removed with decreasing $v^2R$ contribution in the
  Tracer Mass Estimator, is shown. We selectively remove 8 GCs with
  the largest $v^2R$ contribution indicated by the {\it dashed line}, which inflate the total mass
  estimate of NGC 5128. } 
\label{fig:mass_removed}
\end{figure}

Following a similar technique to \cite{schuberth09}, we have used the
Tracer Mass Estimator as a way to remove GCs that have extreme radial
velocities at a large projected distances which would artificially inflate
the total mass estimate.  This can be seen by their contribution to
the $v^2R$ term in Equation~\ref{eqn:tme}.
First, we determine the total mass of the
NGC 5128 using all of the GCs in our sample.  Then, we remove the GC 
with the highest $v^2R$ contribution from our sample and recalculate
the total mass.  We
continue to calculate the total mass while removing the subsequent largest $v^2R$
contributor in each iteration.  For each step, we determine the change
in total mass from the total mass determined using the entire GCS,
divided by the total number of GCs removed.  Our results are shown in
Figure~\ref{fig:mass_removed}.  We have removed the 8 largest $v^2R$
contributors from the total mass estimate, as well as in the kinematic
analysis that follows. These 8 GCs, 4 of which are metal-rich,  are
all found at a galactocentric radius greater than 16\arcmin, and
lead to an overestimate in the mass of NGC 5128.  

To generate a mass profile within NGC 5128 using its GCS, we determine
the rotation and pressure supported masses in cumulative bins from
0\arcmin-5\arcmin, 0\arcmin-10\arcmin, 0\arcmin-15\arcmin, 
0\arcmin-20\arcmin, 0\arcmin-45\arcmin.  The results are tabulated in
Table~\ref{tab:mass} with columns displaying the GC subgroup, the
radial range of included GCs, the number of GCs per radial bin, the
systemic velocity, the rotation amplitude, the rotation axis, the
pressure supported mass, the rotation supported mass, and the last
column indicates the total mass of the galaxy.  The total mass
determined is the mass enclosed within the outermost tracer point. 

We also calculate the mass of NGC 5128 by excluding the GCs
located within 0-5\arcmin\ projected galactocentric radius listed in Table~\ref{tab:mass}.  
The Tracer Mass estimator
assumes spherical symmetry of the system and the GCs within 5\arcmin\
have severe spatial bias within the dustlane.
While we examine the enclosed mass out to 45\arcmin, our most secure
mass estimate is out to 20\arcmin\ (determined with the GCs between
5\arcmin-20\arcmin) to be $(5.9\pm2.0)\times 10^{11}$ M$_{\sun}$.  We also
determine the total mass using only the metal-poor and metal-rich GCs out to 20\arcmin\
and out to 45\arcmin, also listed in
Table~\ref{tab:mass}.  Both metallicity subpopulations provide
good agreement for the mass estimate to the entire GCS. 

The mass-to-light  ratio of NGC  5128 was also calculated using  the GCS.   We have  calculated the  absolute B-band  magnitude  assuming a
distance of  3.8 Mpc and  an apparent B-band  magnitude of 7.84  and a
galactic extinction  of 0.5  for NGC 5128  \citep{karachentsev02}.  We
determine   the  B-band   mass-to-light  ratio,   M$/$L$_B$,   out  to
20\arcmin\ and  to 45\arcmin\  to be 16 M$_{\sun}/$L$_{B\sun}$ and
30 M$_{\sun}/$L$_{B\sun}$, respectively.

\subsubsection{Comparison to Previous Measurements}

In NGC  5128, the HI  gas shells  were used to  estimate a mass  of $2
\times 10^{11}$ M$_\sun$ within 15 kpc under their assumption that NGC
5128  is  at  a   distance  of  3.5  Mpc  \citep{schiminovich94}.  With
the distance of 3.8 Mpc used in this study, the mass determined by \cite{schiminovich94}
would increase to just less than $2.2\times10^{11}$
M$_{\sun}$ within 15 kpc (or 13.8\arcmin).  Within 13\arcmin\, we re-evaluate our total mass
estimate and find $M_t =(3.5\pm1.2)\times10^{11}$ M$_{\sun}$, in
agreement with the HI gas shells estimate.

\cite{peng04b} have used
215 GCs  extending out to 40 kpc ($\sim36$\arcmin) to estimate a pressure
supported mass of $7.5 \times 10^{11}$ M$_{\sun}$ \citep[see][for this
mass, corrected  from \cite{peng04b}]{woodley07}.  We re-evaulate the
pressure supported mass out to 36\arcmin\ and obtain $M_p =(9.7\pm3.3)\times
10^{11}$  M$_{\sun}$, matching the estimate of \cite{peng04b} within uncertainties.

\cite{woodley07} found  a total  mass, including the contribution from
rotation and pressure,
of $(1.3 \pm0.5) \times 10^{12}$ M$_{\sun}$ from 340 GCs
out to the full extent of the known GCS (45\arcmin). We find
  $M_t =(1.1\pm0.4)\times10^{12}$ M$_{\sun}$ in excellent 
agreement with this result although we use a much larger
sample of 429 GCs (excluding GCs with $R_{gc} < 5$\arcmin). 

Both \cite{peng04b} and \cite{woodley07} have used samples with quite low
number of GCs beyond 10\arcmin\ in galactocentric radius.  
We do, however, find good agreement
with their earlier work, but with better constraints on the
uncertainties associated with the determined mass.

\subsection{Kinematic Analysis}
\label{sec:kin_anal}

The previous kinematic studies of the GCS in NGC 5128  have been done with
smaller samples of GCs, with 215 GCs used in \cite{peng04b} and 340
GCs used in \cite{woodley06} and \cite{woodley07}.  Here, with a larger sample size and 
better spatial coverage, we evaluate the kinematics of the GCS.   

We determine the  kinematics for the GCs using Equation~\ref{eqn:kin} for
the  full sample  or discretely binned  subsamples by weighting  the individual 
GCs according to their individual radial
velocity uncertainties, $v_{p,err}$ and the random velocity component
of  the  system,  $v_{ran}$.   The radial  velocities  and  associated
uncertainties  used in  this study  are the  weighted averages  of all
previous measurements.  The
random velocity  component is the  standard deviation of  the radial
velocities of the GCs from the best fit  rotation curves determined without 
any weighting.   The final uncertainty for each  GC used
in Eqn.~\ref{eqn:kin} is $\sigma = (v_{p,err}^2 + v_{ran}^2)^{1/2}$,
and the weight is $1/\sigma^2$.  The dominant uncertainty is $v_{ran}$
providing all GCs a  near equal weighting in the fitting.

\begin{figure}
\plotone{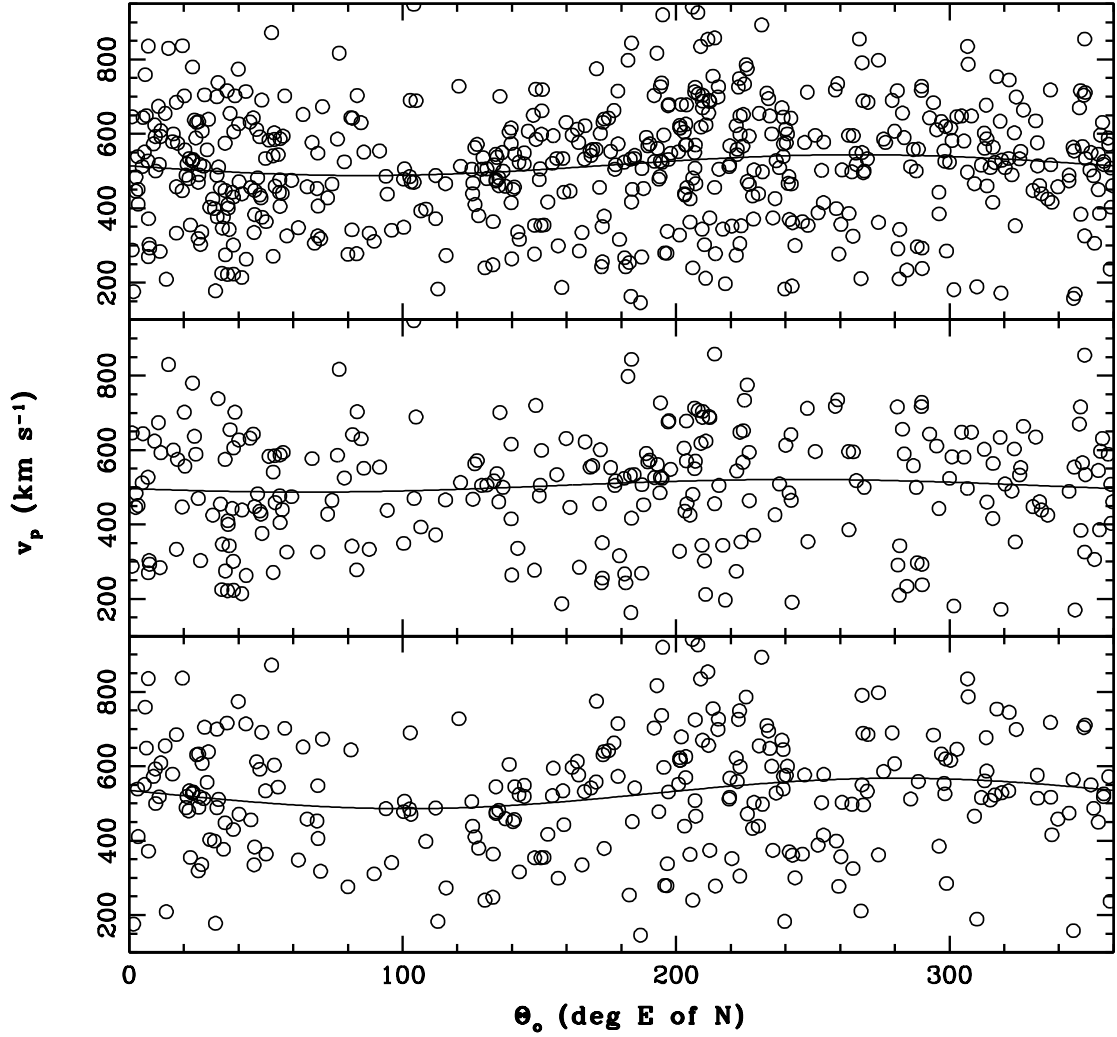}
\caption{The rotation axis and measured radial velocity are shown 
for the entire GC sample ({\it top panel}), the metal-poor sample ({\it
  middle panel}), and the metal-rich sample ({\it bottom panel}).  The
GCs have been fit with Equation~\ref{eqn:kin} which is overplotted as
the solid curve.  The fitted parameters are listed in Table~\ref{tab:kin}.} 
\label{fig:Theta_v}
\end{figure}

Our initial results are shown in Figure~\ref{fig:Theta_v} for the entire GCS,
and for the metal-poor and metal-rich subpopulations of GCs.  The best sine
curve fits clearly show the
large dispersion (or random motion) of the GCS with a small
rotational component. 
Our results are presented in Table~\ref{tab:kin} which lists the GC subgroup, the
radial range of the GC coverage, the average radius,
and number of GCs
within the subgroup, followed by the best fit systemic velocity, rotation
amplitude, rotation axis, the
isophotal major axis of $215 ^o$E of N \citep{dufour79} subtracted from the rotation axis, velocity dispersion, and rotation parameter
(rotation amplitude/velocity dispersion).  The latter two will be
discussed in Section~\ref{sec:veldisp}. 
We note the kinematics
presented in Table~\ref{tab:mass}, while calculated with the same method as
in Table~\ref{tab:kin}, differ by their radial ranges.  In the
mass determination presented in Section~\ref{sec:mass}, we were
developing a mass profile that with the inclusion of more
datapoints reduces the uncertainties on the mass determined using
the enclosed data.  Here, we are searching for the kinematic
signatures at distinct radial positions. 

For the
entire GCs, our best fits are $v_{sys} = 517\pm7$ km s$^{-1}$, $\Omega R
= 33\pm10$ km s$^{-1}$, and $\Theta_o = 185\pm15 ^o$E of N.  Also, our best fits for the overall
kinematics are $v_{sys} = 506\pm9$ km s$^{-1}$, $\Omega R
= 26\pm15$ km s$^{-1}$, and $\Theta_o = 177\pm28 ^o$E of N for the metal-poor
and $v_{sys} = 526\pm10$ km s$^{-1}$, $\Omega R 
= 43\pm15$ km s$^{-1}$, and $\Theta_o = 196\pm17 ^o$E of N for the metal-rich
GC subpopulations.  As discussed in
Section~\ref{sec:rvdist}, the overall velocity of the metal-poor GC
subpopulation is shifted to a lower value than the metal-rich GC
subpopulation, however, both are lower than the systemic
velocity of the galaxy.  The rotation amplitude as first
indicated in Fig.~\ref{fig:Theta_v} is quite low.  The metal-poor GC
subpopulation has very mild rotation and the metal-rich GCs have mild to moderate rotation as a whole.
The metal-rich GC subpopulation appears to rotate around the isophotal major axis of the
galaxy within 15\arcmin, while the metal-poor GC subpopulation does not appear to
follow either the isophotal minor or major axis.  We note
there is no difference in 
the results within uncertainties if  $v_{sys}$ is 
fixed at the systemic velocity of the galaxy or left free in Eqn.~\ref{eqn:kin}.

We create kinematic profiles for the entire GCS as well as the metal-poor and
metal-rich GC subpopulations as functions of galactocentric radius.  We bin the
GCs in galactic effective radial bins of 0\arcmin-5\arcmin,
5\arcmin-10\arcmin, 10\arcmin-15\arcmin, 15\arcmin-20\arcmin, and 
the remaining GCs in the final bin of 20\arcmin-45\arcmin\ for the
entire GCS and the 
metal-poor GC subpopulation.  The metal-rich GC subpopulation has a
more spatially centrally 
concentrated system so we bin in 0\arcmin-5\arcmin,
5\arcmin-10\arcmin, 10\arcmin-15\arcmin\ bins, with the remaining
metal-rich GCs in the final bin of 15\arcmin-45\arcmin.

We list the  results in Table~\ref{tab:kin} for the  best fit systemic
velocity, rotation amplitude, and rotation axis for the entire GCS, as
well  as metal-poor  and  metal-rich GC  subpopulations.  With  radial
distance,  we  see  both  the metal-rich  and  metal-poor  populations
maintain a relatively constant amount of rotation with radial distribution, with the metal-rich
GCs having a  larger rotation amplitude than the  metal-poor GCs.  
Note that the rotation amplitude for the entire metal-poor GC subpopulation
from 0\arcmin-45\arcmin\ is less than most of the values determined in
the specific radial bins,which at first sight is an anomaly.  However,
this curious result is connected with the solutionos for the rotation
axis angles, which differ by large amounts from bin to bin.  When
the rotation amplitude is zero, the rotation axis is ofcourse
undefined, so when we average together the results from the different
radial bins (all with a different axis of rotation), they average down
to zero with no meaningful axis.  We conclude from this that the
rotation in the metal-poor GC subpopulation is very mild.  
There
is an indication  that the metal-poor GCs have  an increased amount of
rotation in  the outer regions  (beyond 20\arcmin), but again,  the GC
population  in this region  clearly violates  the basic  assumption of
spherical symmetry used in Eqn.~\ref{eqn:kin}.  We therefore caution
as to its validity.    If  the metal-poor  GC system rotation does  increase, this  may support the
formation mechanisms of disk-disk  mergers for NGC 5128.  Here, the orbital angular
momenta of the progenitor  galaxies are converted to intrinsic angular
momentum in the remnant  from the merging event \citep{bekki05}.  This
angular momentum  is transferred to  the outer regions of  the galaxy,
and thus  it would be expected  for the rotation amplitude of the GCS to increase at
larger radii. 

The metal-poor GC subpopulation shows a potential 
twist in its rotation axis within the inner 5\arcmin\ from rotation around the
isophotal minor axis to around the isophotal major axis.  Beyond
10\arcmin\ the axis 
may change again.  However, given the near zero rotation
amplitude of the metal-poor GCs, it does not appear that they rotate around 
a well-defined rotation axis.
In the inner 
20\arcmin\ the metal-rich GCs are consistent with rotation around the isophotal
major axis of the galaxy.  In the outer regions, the rotation axis of
the metal-rich GC subpopulation appears to decrease slightly, but again, this is a
region of uncertainty.  

\subsubsection{Velocity Dispersion}
\label{sec:veldisp}

The velocity dispersion of the system is determined using a maximum
likelihood dispersion estimator discussed in \cite{pryor93},
\begin{equation}
\label{eqn:veldisp}
\sum_{i=1}^{N} \frac{(v_i - v_{sys})^2}{(\sigma_{vp}^2 + v_{p,err_i}^2)^2} = \sum_{i=1}^{N} \frac{1}{(\sigma_{vp}^2 + v_{p,err_i}^2)}
\end{equation}
where $N$ is the number of GCs in the sample, $v_i$ is the
radial velocity of the GC after subtraction of the rotational
component determined in Eqn.~\ref{eqn:kin} with $v_{sys}$ held constant at 541 km s$^{-1}$, and $v_{p,err_i}$ is the uncertainty in the velocity
measurement.  The projected velocity dispersion, $\sigma_{vp}$, is 
determined by iteration until satisfying Equation~\ref{eqn:veldisp}.  The
systemic velocity used here was determined from Eqn.~\ref{eqn:kin}.  The
uncertainties in the velocity dispersion are calculated from the
variance of the dispersion, listed in \cite{pryor93}.  This has been shown by \cite{pryor93} to
estimate the uncertainty to $5\%$ with as little as 20 objects in the sample.

\begin{figure}
\plotone{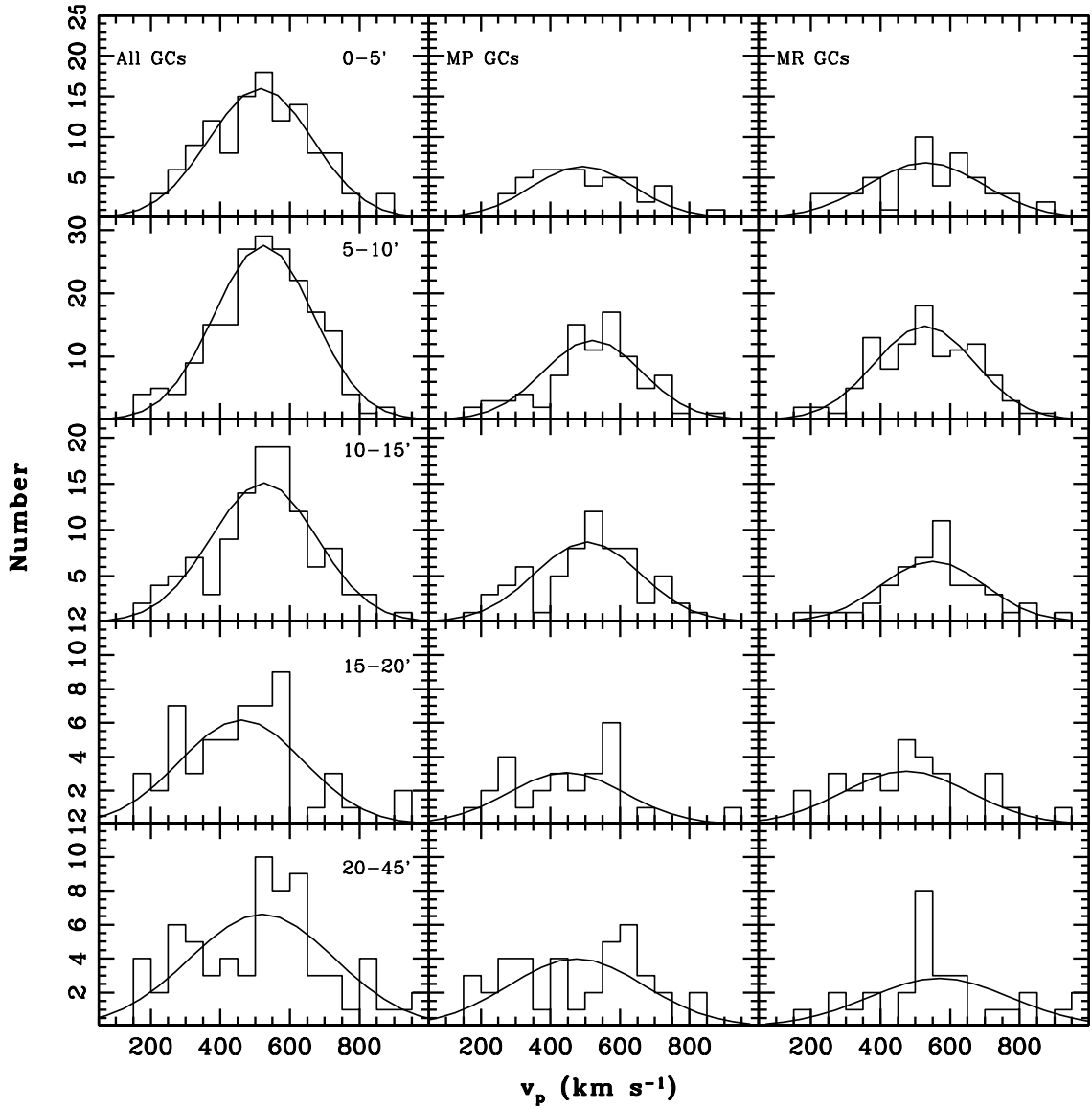}
\caption{Radial velocity histograms for the entire GCS ({\it left
    panels}), the metal-poor ({\it middle panels}) and metal-rich
  ({\it right panels}) GC subpopulations.  Each subpopulation is
  binned radially from 0\arcmin-5\arcmin\ ({\it top panels}),
  5\arcmin-10\arcmin\ ({\it 2nd panels}),10\arcmin-15\arcmin\ ({\it 3rd
    panels}),15\arcmin-20\arcmin\ ({\it 4th panels}), and 20\arcmin-45\arcmin\ ({\it bottom
    panels}). The best fits are listed in Table~\ref{tab:vel_fits}.} 
\label{fig:velrad_histo}
\end{figure}

We list the velocity dispersions in Table~\ref{tab:kin}.  
As a whole the metal-poor and metal-rich GC subpopulations
have identical velocity dispersions within uncertainties of  $149\pm4$ km s$^{-1}$ and
$156\pm4$ km s$^{-1}$, respectively.  However, the behaviour with 
galactocentric radius may be different for the different 
metallicity subpopulations.  The metal-poor GC velocity dispersion is
flat or declining out to 20\arcmin\ beyond which it appears to rise.  The metal-rich GC profile
shows a declining velocity dispersion out to 5\arcmin\ after which it
appears to flatten, which may indicate the metal-rich GCs are more
radially biased than the metal-poor GCs.   A rise in the velocity dispersion in the outer
regions of a galaxy could be an indicator for large amounts of dark
matter or of anisotropic
orbits of the velocity ellipsoid of the GCS itself.  
With an extended X-ray gas profile to trace the mass out to regions
beyond the halo light, 
we could determine the degree of anisotropy of these orbits, if they
exist, 
using a Jeans solution. This has been done for M87 \citep{cote01}, M49
\citep{cote03}, M60 \citep{hwang08}, and NGC 1407 \citep{romanowsky09}
as examples.  
It may be possible to find indications of anisotropy in the GCS by examining its radial
velocity as a function of radius.  To analyse the ellipsoid
distributions, it is more instructive to remove the rotational
component of the GC subpopulations; however, considering the rotation
amplitudes are nearly negligible, we use Fig.~\ref{fig:velrad_histo}
as a first glance into possible anisotropy of the GCS.  

If the GCs are moving on very
radial orbits, we expect the distribution to be strongly peaked at the
systemic velocity, while if they are
moving on very circular orbits, the distribution would have a more flattened shape.
Isotropic orbital motion would have a Gaussian distribution.  The radial velocity distributions do
not appear to be tightly peaked in any histogram in
Fig.~\ref{fig:velrad_histo}, but there is a slight tendency for the GC
velocity histograms to have a wider distribution at larger galactocentric
radius, thus a flattening effect.  In Table~\ref{tab:vel_fits} we
quote the reduced $\chi^2$ values for these Gaussian fits.  
We note a general trend for larger
$\chi^2_{red}$ values for the outer bins, indicating a poorer fit of
the model to the data, perhaps by the lower number of GCs in the outermost regions.  
A flattened distribution may also be caused by the contribution of GCs
from accreted satellite galaxies in the outer regions of the galaxy
\citep{bekki03}.  Evidence for potential substructure has been presented in M31
\citep{merrett03}, M87 \citep{doherty09}, NGC 1399 \citep{schuberth09}, and perhaps in
NGC 1407 \citep{romanowsky09}.  
Interestingly, numerical simulations have suggested that tracer
populations should have isotropic orbits in the inner regions of their
host galaxy, while in the outer regions, their orbits should become
anisotropic \citep{dekel05,diemand05,abadi06,gnedin06}.  In relation
to the GC results presented here, the presence of substructure of
accreted satellites or interlopers will bias the velocity dispersion
to higher values. This is seen in the outer-most bin in the metal-poor GC
subpopulations, possibly indicating substructure from an accreted
satellite or Milky Way halo star contamination.

The rotation parameter is a measure of the maximum rotation divided by
the velocity dispersion in a system.  
We obtain rotational parameters of $0.22\pm0.07$,
$0.17\pm0.09$, and $0.28\pm0.10$ for the entire GCS, metal-poor, and metal-rich
subpopulations of GCs, respectively (listed in Table~\ref{tab:kin}).  
These values reinforce what we
have already discovered about the GCS.  While the velocity dispersion of 
the entire metal-poor and entire metal-rich GC subpopulations is
consistent within uncertainties, the metal-rich
GCS, as a whole, has more rotation.

\subsubsection{Kinematics for Globular Clusters with Ages}
\label{sec:ages}

\cite{woodley09} have estimated ages, metallicities, and $\alpha$-to-Fe 
abundance ratios for 72 GCs in NGC 5128.    By calibrating their high signal-to-noise spectroscopy  
taken with Gemini-S/GMOS to the Lick index system \citep{burstein84,worthey94,worthey97,trager98}, 
they extracted their estimates by comparing their measured indices 
to the simple stellar population models of \cite{tmb03,tmk04}.   
Using Eqn.~\ref{eqn:kin}, we have estimated the kinematic properties of 
these 72 GCs,  as well as
subsets of this group, such as their GCs estimated to be old (ages $\geq 8$ Gyr), intermediate 
($5  - 8$  Gyr), and  young (age $ < 5 $ Gyr).   We also
analyse the old  GCs in subgroups of  metal-rich 
and metal-poor. 
Our results are presented in  Table~\ref{tab:age_kin},  which  lists  our 
subgroup of GCs, the number of GCs in each
subgroup,  the  systemic velocity,  the  rotation  amplitude, and  the
rotation axis. 
The kinematics for the GCs subdivided by
metallicity and age have quite large uncertainties because they are small samples.  They do,
however, 
indicate general properties of these small subsets of GCs.   
As expected, the old metal-poor GCs and old metal-rich GCs generally match the kinematic properties
of the entire metal-poor GCs and metal-rich GCs, respectively. 
The rotation axis of the young GCs, which are
all metal-rich, is quite different than that of the entire metal-rich GC population.
With a rotation axis of $80\pm84^o$ E of N at the average projected
radius of 5.7\arcmin\, this is significantly different from the metal-rich GC
rotation axis of  $244\pm27^o$ E of N at an average radius of 3.49\arcmin\ or
$178\pm22^o$ E of N at an average 7.23\arcmin\ distance.  The young metal-rich
GCs appear to be rotating either along the isophotal major axis, but
in the opposite direction to the metal-rich GC subpopulation, or along the
isophotal minor axis, or anywhere in between.  In any case, their
rotation axis is quite different from the bulk of the metal-rich GCs.

\subsubsection{Kinematics for Globular Clusters versus Magnitude}
\label{sec:kinmag}

\begin{figure}
\centering
\begin{tabular}{cc}
\epsfig{file=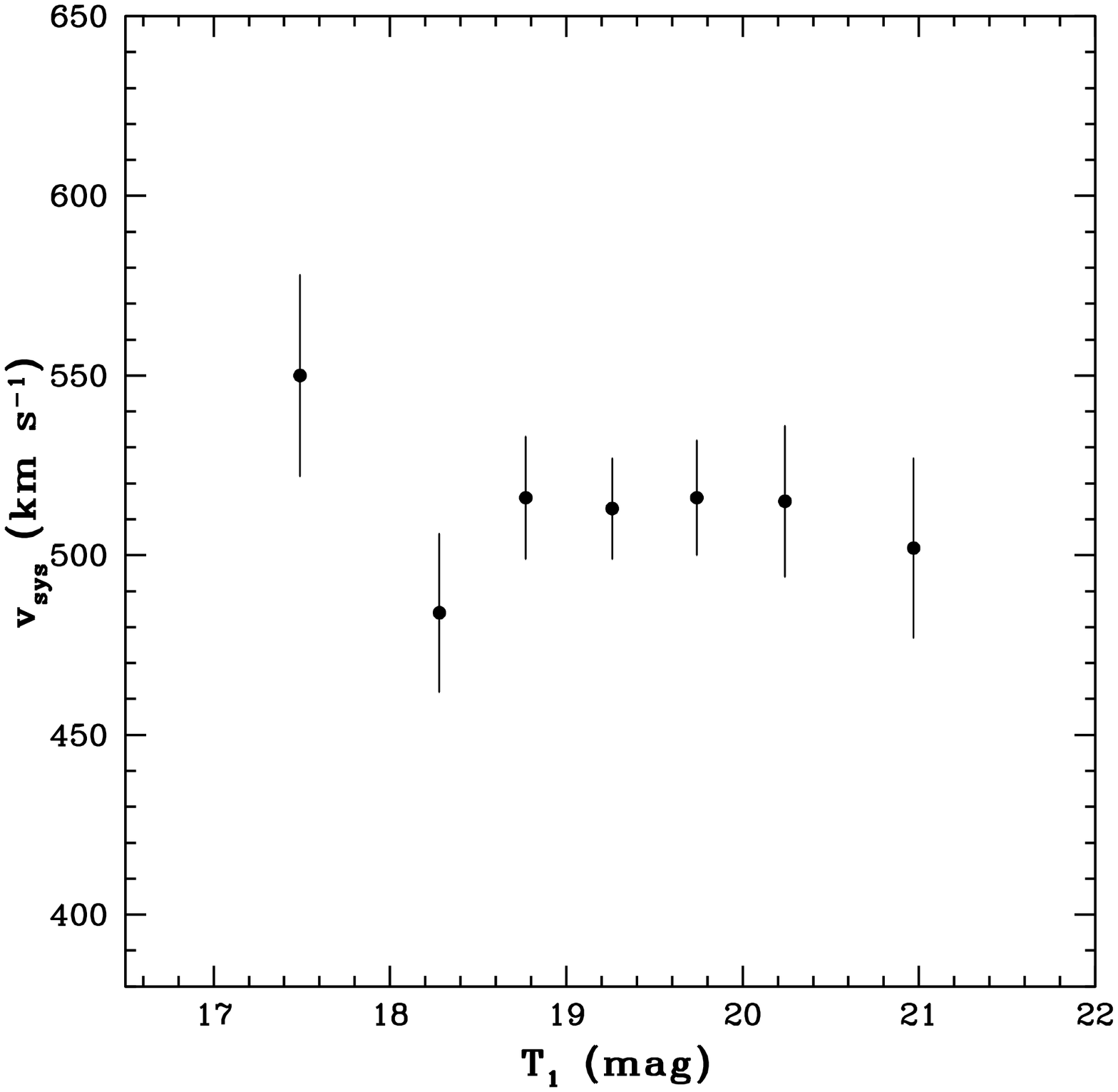,width=0.3\linewidth,clip=} &
\epsfig{file=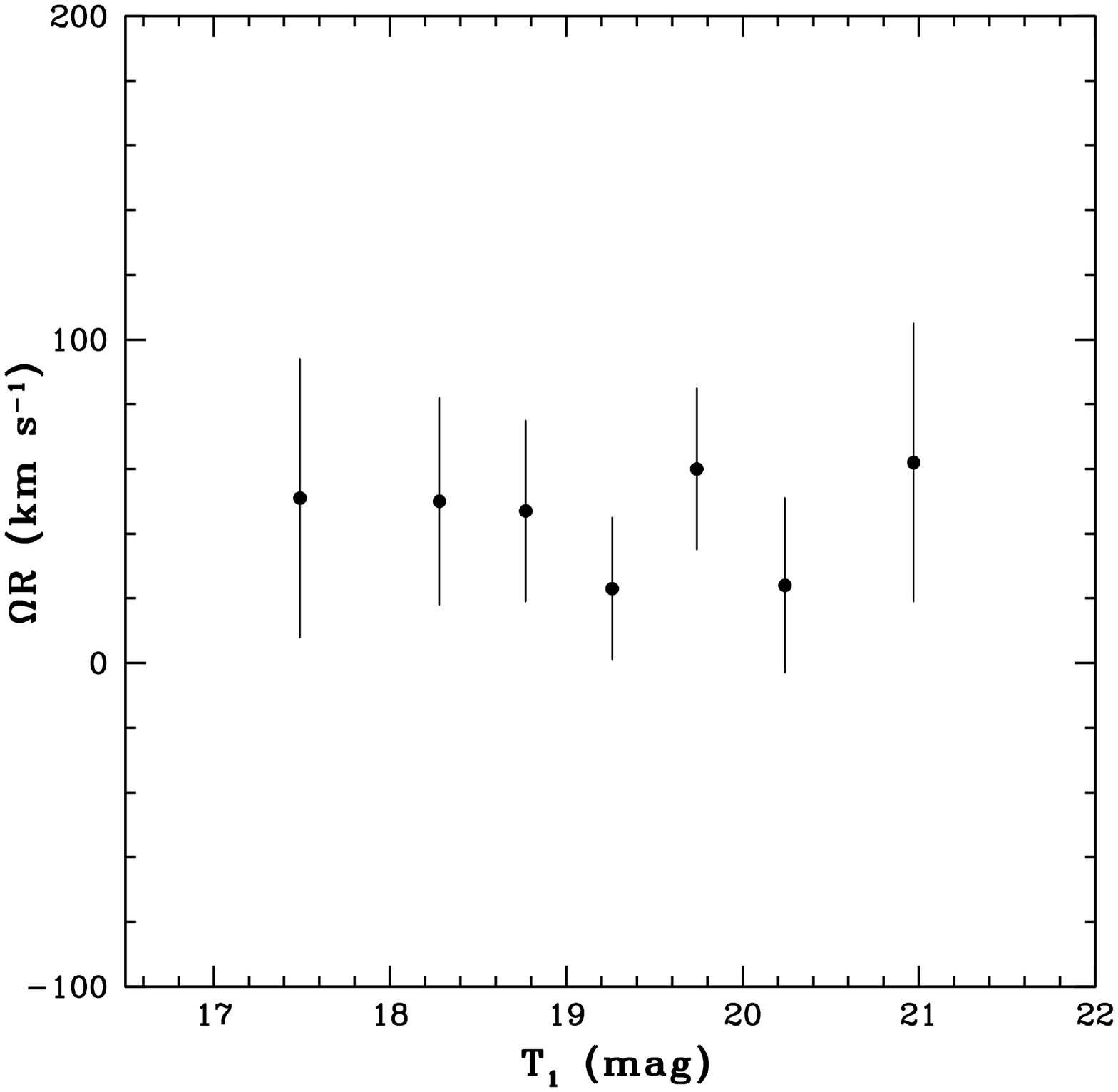,width=0.3\linewidth,clip=} \\
\epsfig{file=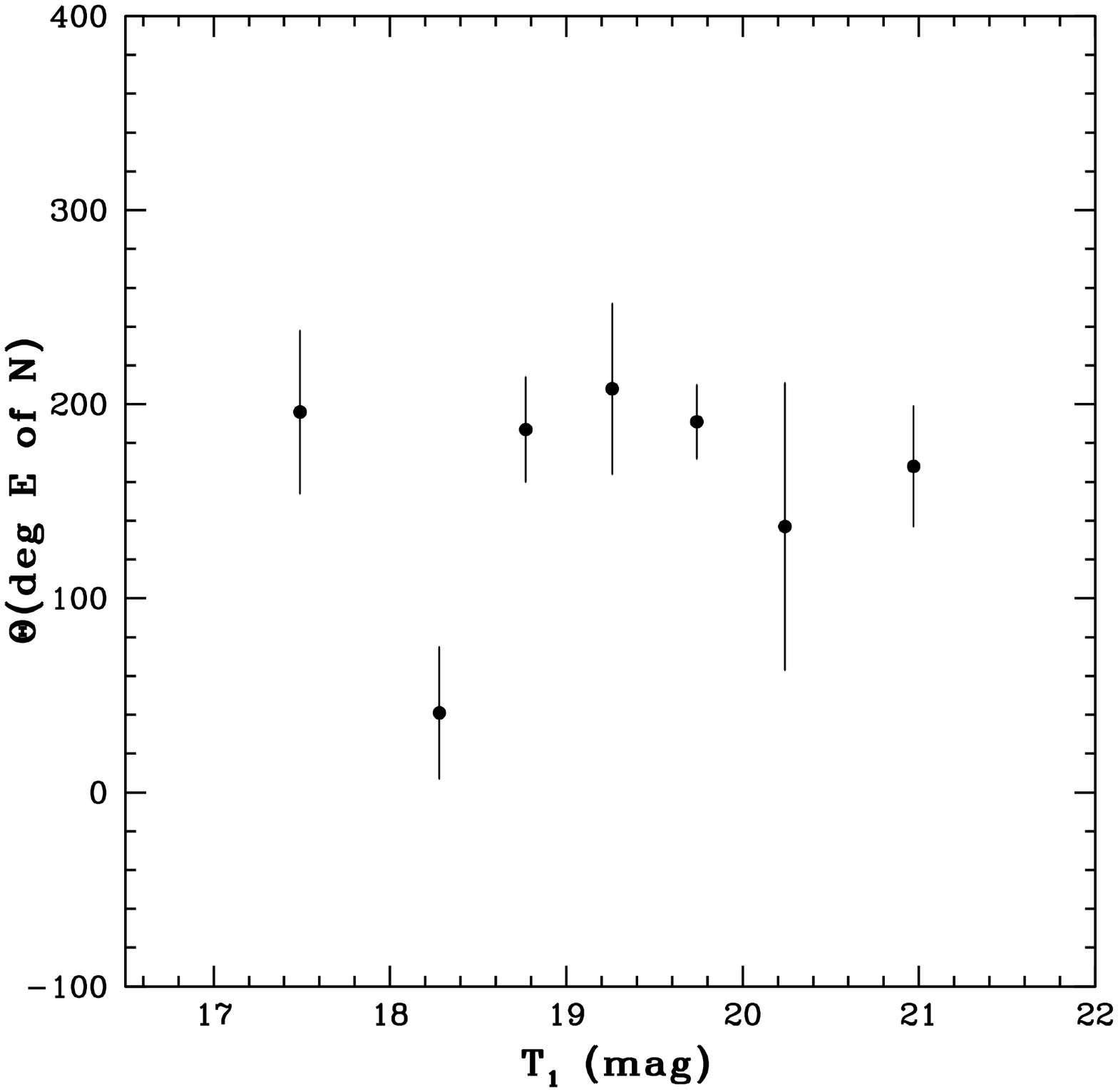,width=0.3\linewidth,clip=} &
\epsfig{file=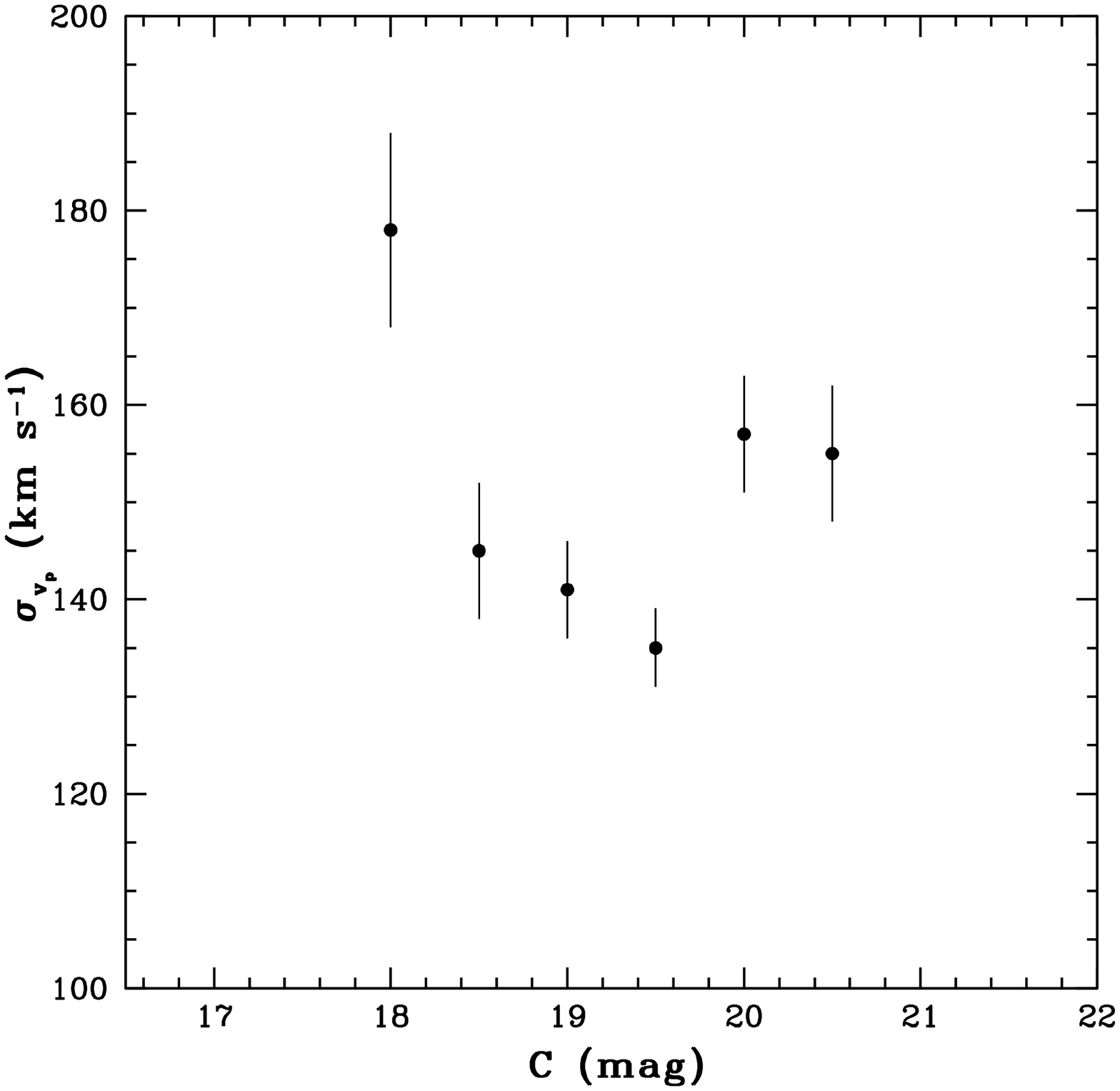,width=0.3\linewidth,clip=} \\
\end{tabular}
\caption{The $v_{sys}$ ({\it top left panel}), $\Omega R$({\it top right
    panel}), $\Theta_0$ ({\it bottom left panel}), and $\sigma_{v_p}$
  ({\it bottom right panel}) are plotted as a
  function of T$_1$ magnitude. }
\label{fig:T_kin}
\end{figure}

The kinematics of the GCS with respect to T$_1$ magnitude has
also been analyzed upon removal of GCs that have extreme colors 
(see Section~\ref{sec:phot_rad}). The GCs are binned in magnitudes between $16-18$,
$18-18.5$, $18.5-19$, $19.5-20$, $20.5-23$ mags, in order to keep the
number of objects per bin above a minimum of 25.  We plot the results
for the systemic velocity, 
rotation amplitude, rotation axis, and velocity dispersion in
Figure~\ref{fig:T_kin}.  We find there are no major differences in the
kinematic properties of the GCs with changing magnitude. It does
appear, however, that the GCs between
$18-18.5$ magnitudes have a significantly different rotation axis
($41\pm34^o$ E of N) than the remaining GCs which have an axis of
$\sim200^o$ E of N. 
We have examined the projected distribution of the GCs in this bin
compared to the other magnitude bins and see no distinct difference.
We also notice that the brightest GCs have a significantly larger velocity
dispersion than the remaining GCs, indicating there may be evidence for
varying kinematics as a function of magnitude.  A similar finding was
identified in NGC 1407 \citep{romanowsky09}; however, as they
indicated, it could be due to spatial bias of the brighter objects in
the sample being close to the center of the galaxy.   This is not as
obvious in the case of NGC 5128, as almost all GCs in this study are
centrally concentrated within 15\arcmin. 

\subsubsection{Kinematics of the Planetary Nebulae}
\label{sec:pn}

The PNe are a field star population at one of the latest
stages of stellar evolution and are easily identified by their emission of [OIII].
Follow-up spectroscopy can then be performed on the PNe to measure
their radial
velocities which has been done for 780 PNe in NGC 5128, extending out
to over 80\arcmin, and compiled in \cite{peng04a}. 

The kinematic properties of the 780 PNe are investigated in
\cite{woodley07} in the 
same manner as the GCs presented in this study .  They found the PNe had a 
rotation amplitude of $76\pm6$ km s$^{-1}$, a rotation axis of 
$170\pm5^o$ E of N, and a velocity dispersion of $118\pm13$ km s$^{-1}$.
The rotation amplitude of the PNe is a bit larger than the GCS as a
whole and the rotation axis of the PNe is quite uniform
with increasing radius \citep[see Table~5 in][]{woodley07}.  
The rotation axis is also, interestingly,
similar to the isophotal major axis of the galaxy, closely mimicing
the metal-rich GC subpopulation in the inner
15\arcmin.  The velocity dispersion of the PNe is
different than the GCs.  The PNe are dynamically colder in the inner
5\arcmin and their velocity dispersion continues to decline with increasing galactocentric radius.  

Declining velocity
dispersions of the PNe with  galactocentric radius have  been found for  NGC 3379
\citep{romanowsky03,douglas07}, NGC 821 \citep{romanowsky03}, NGC 4494
\citep{romanowsky03,napoliano09},  NGC 5128 \citep{peng04a,woodley07},
and  NGC 4697  \citep{mendez09}, which could
indicate a lack of dark matter in the system \citep{romanowsky03}.  However, 
GCSs studied within NGC 3379 \citep{pierce06} have shown an increasing
velocity dispersion, indicating there is likely to be some
amount of dark matter present.  
Recent work has shown that PNe may be on radial orbits 
\citep{dekel05,mamon05,abadi06}, which may also cause the declining
velocity dispersions.
Anisotropy may not necessarily be expected of the GCs because these
two populations trace different kinematical histories of their 
host galaxies, especially if the galaxy has undergone a major merging
event in the past.  Other studies have found that the GC populations have approximately isotropic orbits
\citep{zepf00,cote01,cote03,bridges06,schuberth09}.   However, separating the blue and
red GCs have shown there may be slight anisotropy as well, with M87
\citep{cote01} and NGC 4649 \citep{bridges06,hwang08} as examples.   In
the case of NGC 5128 studied here, we therefore suggest that the PNe
may be following {\it more} radial
orbits than the GCs.  

\section{Discussion}
\label{sec:discussion}

The kinematics of the GCSs that have been performed in other giant elliptical galaxies based on
current samples of $>100$ GCs includes: NGC 1407 \citep{romanowsky09},
an elliptical galaxy in the Eridanus A group at a distance of 20.9 Mpc
\citep{forbes06}; NGC 4649 (M60) \citep{bridges06,hwang08}, a luminous
giant elliptical galaxy in the Virgo cluster of galaxies at a distance
of      17.3      Mpc      \citep{mei07};     NGC      4486      (M87)
\citep{kisslerpatig98b,cohen00,cote01},  a   supermassive  cD  galaxy,
located at a  distance of $16.7\pm0.2$ Mpc \citep{mei07}  in the Virgo
cluster of galaxies;  NGC 4472 (M49) \citep{sharples98,zepf00,cote03},
a giant elliptical  galaxy and also the brightest  member of the Virgo
cluster  of  galaxies,  located  at  a distance  of  $16.4\pm0.2$  Mpc
\citep{mei07};  NGC 4636  \citep{schuberth06,lee09},  the southernmost  giant
elliptical galaxy  from the dynamical  center of the Virgo  cluster of
galaxies, located 14.7 Mpc away \citep{tonry01}; and NGC 1399, a giant
elliptical galaxy and also the  brightest galaxy in the Fornax cluster
of  galaxies, 19 Mpc  away \citep{richtler04,schuberth09}.   The main  findings of
these  studies  have  been   well  summarized  in  \cite{hwang08}  and
\cite{romanowsky09} and we do not repeat these here.

We find  that the kinematic  trends  of NGC  1407
\citep{romanowsky09}  are quite  similar  to NGC  5128  found in  this
study.  Both  systems have larger  rotation in the metal-rich  GCs (at
least  in  the inner-most  region  where  both datasets  have  strong
azimuthal coverage),  with the rotation  of the metal-rich  GCs around
the isophotal major axis.  Both systems have velocity dispersions that
appear  flat or  decreasing  in  the inner  regions  with a  potential
increase at larger radii.  They  are also both dominated by dispersion
over their entire  GC population.  Both NGC 5128 and  NGC 1407 are the
only two elliptical galaxies in a group environment with large samples
of GCs  available in  order to perform  a kinematic analysis.   With a
sample of 2 galaxies, it is a  far stretch to propose that the GCSs in
elliptical   galaxies  found  in   group  environments   have  similar
kinematics due to similar formation histories.  With this in mind,
if  these  galaxies hierarchically  built  up  the  majority of  their
stellar  content at  early  times, later  small  accretions and  minor
merging will not severely disrupt  the orbital and kinematic properties
in the  outer halo.   The kinematics from  the major episodes  of star
formation within these galaxies may still be intact.

\subsection{Implications for Galaxy Formation}

The              monolithic              collapse             scenario
\citep{eggen62,tinsley72,larson74,larson75,silk77,arimoto87}   suggests
the GCs in  massive galaxies should be old,  forming at high redshift.
The collapse  would be  very rapid indicating  that the spread  in old
ages should be small and the kinematic properties of the GCs should
be quite similar.  Any rotation that exists in the GCs would come from
tidal  torques from  nearby forming  galaxies  \citep{peebles69}.  The
expected amount of rotation would be small, as the angular momentum of
the initial  cloud would be small to  begin with in order  to form the
spheroidal shape of the  elliptical galaxy.   
This scenario is not supported by the  
trend towards  younger ages as metallicity  increases in the GCS \citep{peng04b,beasley08,woodley09}. 
Neither is it supported by the large  amounts  of rotation  
that  are seen  for  some  GCSs (for  the
metal-rich GCs in NGC 4636,  and both metal-rich and metal-poor GCs in
NGC 4486 and NGC 4649) or by the different rotational properties of some
metal-rich and metal-poor GCs (in NGC 4636, NGC 4486, and NGC 4472, for
example).  

The galaxy merging model \citep{schweizer87,ashman92} would naturally
suggest younger ages of the metal-rich GCs, which form, according to
this scenario, only in the merging events, while the metal-poor GCs
pre-exist in the progenitor galaxies.  This theory expects a multimodal
color distribution of the GCS.  Numerical simulations by
\cite{bekki02} have shown that the metal-poor GCs in this scenario are
spatially extended while the newly formed metal-rich GCs are more
centrally concentrated and more extended along the isophotal major
axis of the galaxy.  While the above properties are not disputed here,
this does not explain the old ages for the majority of metal-rich GCs
found in previous work on NGC 5128, unless all the major merging 
events took place at very early times ($\sim 10$ Gyr ago). 
Any existing angular momentum
would be transferred to the outer regions of the remnant galaxy from the merging
process \citep{bekki05}.   The more extended
metal-poor GC population should have a higher rotation signature than the metal-rich GCs,
particularly in the outer regions. We do not see this in the GCs of NGC 5128 or
in the GCs of the other giant elliptical galaxies studied, with the
exception of NGC 4472, which generally agrees with the galaxy merger model.  

A  hierarchical merging  scenario, combining the  multiphase
dissipational collapse model  \citep{forbes97} and accretion scenarios
\citep{cote98,cote00,cote02},    proposed      by
\cite{beasley02,beasley03} and \cite{strader05}, is  more  consistent  with  our
findings for  NGC 5128.     This naturally
explains the GC color bimodality  and also predicts old ages for both
metal-rich and metal-poor GCs (with the metal-poor GCs having slightly
older  ages).   The  old ages obtained  for the stellar
population  in the  halo  of NGC  5128, as
well as similar chemical enrichment between the metal-rich GCs and
stellar halo \citep{rejkuba05,harris08},  are
consistent with  this scenario as  well. This  hierarchical scenario also
supports the younger metal-rich GCs in NGC 5128 forming in either more
recent merging  events and/or  on a more  extended timescale.   In the
sparse environment of  NGC 5128, the formation of  the more metal-rich
GCs could be extended because  the density in the group environment is
lower than  that in clusters of  galaxies.  This is  supported by both
the range  of ages  for the metal-rich  GCs as well as the [$\alpha$/Fe]
values suggesting longer formation timescales \citep{woodley09},
compared  to  GCs in  other  giant  galaxies.   In a  typical  massive
collapse, the metal-poor GCs would not necessarily be expected to form
with large  amounts of rotation and  instead they should  have a large
velocity dispersion  \citep{hwang08}.  However, with  the accretion of
small  protogalaxies in  the early  Universe,  the GCs  may have  some
rotation as well  as orbital anisotropy.  The metal-rich  GCs, in this
formation scenario, can  vary in rotation, depending on  the degree of
the second collapse \citep{forbes97}.  We see evidence for this in NGC
5128 in both the old ages and kinematic results.

\cite{bekki05} have numerically simulated the kinematic
signatures of GCs from the merging of Milky Way-type disk galaxies  
with pre-existing old
metal-poor and metal-rich GCs.  To simplify their simulations,
they assumed that the 
initial GCSs in the progenitor galaxies were supported by dispersion
with no rotation.  However, we see rotation in the GCSs of
M31 \citep{perrett02} and the Milky Way \citep{harris01}.  They also
assumed the mergers were dissipationless, but  we do see some small 
fraction of young GCs in NGC 5128.  However, these numerical simulations allow us to compare the kinematic
properties that we see in the GCS of NGC 5128 to the generally
expected kinematics from two old merging spiral galaxies.  
Their results indicate that both the metal-rich and
metal-poor populations 
would have increasing rotational signatures with galactocentric
radius.  \cite{bekki05} also find the velocity dispersions of the
GCSs are flat or declining out to large radii caused by
their radial anisotropic orbits.  Lastly, they found a variety 
of kinematic alignments for the metal-rich and metal-poor GCs.   

In our kinematic study of NGC 5128, we do not see a strong
case for significant rotation in either metallicity subpopulation, and
also no significant increase with radius in the rotation signature.
We do, however,
see indications for flat and declining velocity dispersions in
the inner 20\arcmin\ of NGC 5128 for both metal-rich and metal-poor
GCs, and a different axis of rotation for the metal-rich and metal-poor GCs. 
The results found here are not
clearly consistent with the idea of the GCs forming their kinematic
signature from a 
major disk-disk merger, though this kind of origin is not ruled out.
In the case of other galaxies, NGC 1399 shows little to no 
rotation in either metal-rich or metal-poor GCs and
NGC 1407 also show little to moderate rotation signatures as well.
  
There is an important implication from the study of
\cite{bekki05} that should not be underplayed.  Their results suggest
that the initial 
kinematic signatures of GCs in the progenitor galaxies undergo orbital
mixing and it may, therefore, not be possible to trace the {\it original}
kinematics of the GCs \citep{hwang08}.  It may only be
possible to examine the GC kinematics from the most recent major
interaction.  As suggested by \cite{kisslerpatig98a}, it would then be very
difficult to use the current kinematic signature to trace the orbital
history of GCs.  In this respect, we can suggest that the {\it most recent}
major interaction in NGC 5128 was not likely a major disk-disk merger.
Again, it seems more likely that the galaxy formed hierarchically in
early times, with many infalling protogalaxies.

\section{Conclusions}
\label{sec:conclusions}
We   have obtained spectroscopy   with  LDSS-2/Magellan,
VIMOS/VLT, and Hydra/CTIO in order to measure the radial velocities of GCs in the nearby elliptical galaxy,
NGC 5128.  With these datasets, we  have remeasured radial velocities for 218 known GCs as
well as confirmed 155 new GCs.  There are now 605 GCs confirmed in NGC
5128 and  564 of  these GCs have  measured  radial velocities.  This is  the
second largest kinematic dataset of GCs in any galaxy after NGC 1399 \citep{schuberth09}.

The large sample of GCs with radial velocity measurements allows us to perform a
new kinematical analysis of the system.  We have investigated the
systemic velocity, the rotation amplitude,
the rotation axis, and the velocity dispersion for the entire GCS, as
well as the metal-rich and metal-poor GC subpopulations. We have also
analyzed these kinematic properties as a function of galactocentric
radius.   We find the
metal-poor GCs have a low rotation $26\pm15$ km s$^{-1}$ extending out to
45\arcmin, and does not have a well-defined rotation
axis. The motion of the metal-poor GCs is supported by dispersion with
a measured value of $149\pm4$ km s$^{-1}$.  It has a flat/declining velocity dispersion profile
extending to 20\arcmin, at which point it appears to increase.  The metal-rich GCs
have small to moderate rotation of $43\pm15$ km s$^{-1}$ out to 
40\arcmin\ around the isophotal major axis of the galaxy.  It
has similar velocity dispersion for the whole system ($156\pm4$ km
s$^{-1}$) compared to the metal-poor GCs.  The metal-rich velocity
dispersion profile decreases out to a galactocentric radius of
5\arcmin\ and then flattens.  This increase in the velocity dispersion
profiles in the metal-poor GC subpopulation may be a
result of the incomplete and spatially biased sample beyond
15\arcmin\ in the GCS, or caused by Milky Way halo
contamination.  Or perhaps it may be due to the presence of a substructure 
among the GCS which could artificially inflat the velocity dispersion 
in the outer halo, or it could also be an indicator of an extensive
dark matter halo.  
As a consequence of the low rotation amplitude, both the metal-poor and 
metal-rich subpopulations have low values of the rotation parameter of $0.17\pm0.09$ and $0.28\pm0.10$, respectively.

We have also analyzed the kinematics for the GCs that have estimated
ages and metallicities from the work of \cite{woodley09} and found that the
young ($<5$ Gyr) metal-rich GCs rotate on a different axis than the entire
metal-rich GC population.   If these GCs were formed from the
accretion of a gas-rich satellite galaxy, we could  expect their
kinematic signature to be different from the older GCs. 

We estimate the mass and mass-to-light ratio of NGC 5128 using its GCs.  
We find a mass of $(5.9\pm2.0)\times10^{11}$ M$_{\sun}$ out to
20\arcmin\ with M$/$L$_B = 16.4$ M$_{\sun}/$L$_{B\sun}$.  Our mass estimates
determined using the metal-rich GCs and the metal-poor GCs all agree
within uncertainties.  This mass estimate clearly puts NGC 5128 in the
class of giant elliptical galaxies, with a potentially large dark
matter halo.

\acknowledgements K.A.W., W.E.H, and G.L.H.H. thank NSERC  for  their financial
support. M.G. thanks Proyecto DI-36-09/R for financial support.   D.G. gratefully acknowledges support from the Chilean {\it Centro de Astrof{\'i}sica} FONDAP No. 15010003 and the Chilean Centro de Excelencia de Astrof{\'i}sica y Technolog{\'i}as Afines (CATA).   K.A.W. also thanks Dr. Peter Frinchaboy and Dr. Matthew
Walker for helpful discussions regarding data reduction of the Hydra/CTIO dataset.




\clearpage
\begin{deluxetable}{llll}
\tablecolumns{4}
\tabletypesize{\scriptsize}
\tablecaption{The Observation Data Summary\label{tab:fields}}
\tablewidth{0pt}
\tablehead{
\colhead{Instrument} & \colhead{R.A. (J2000)} & \colhead{Decl. (J2000)}&
\colhead{Time (hr)}  \\
}\startdata
LDSS-2& 13 26 18.31& -42 49 17.5 & 1.0\\
LDSS-2& 13 26 10.00& -42 53 40.0 & 0.7\\
LDSS-2& 13 26 10.00& -42 57 20.0 & 0.7\\
LDSS-2& 13 25 35.37& -42 51 27.6 & 0.7\\
LDSS-2& 13 25 37.00& -42 56 15.0 & 0.5\\
LDSS-2& 13 25 37.00& -43 09 00.0 & 0.6\\
LDSS-2& 13 25 16.00& -43 14 15.0 & 1.0\\
LDSS-2& 13 25 10.00& -43 18 30.0 & 1.0\\
LDSS-2& 13 24 44.00& -42 51 00.0 & 0.8\\
LDSS-2& 13 24 44.00& -42 53 30.0 & 1.0\\
LDSS-2& 13 24 57.00& -43 03 30.0 & 0.7\\
LDSS-2& 13 24 49.00& -43 08 00.0 & 0.7\\
LDSS-2& 13 24 49.00& -43 13 00.0 & 1.0\\
VIMOS & 13 25 46.03& -42 44 43.7 & 1.1 \\
VIMOS & 13 26 17.05& -43 00 30.4 & 1.7 \\
VIMOS & 13 24 39.04& -42 59 59.8 & 1.1 \\
VIMOS & 13 26 43.61& -43 15 52.0 & 3.3 \\
VIMOS & 13 24 31.02& -43 15 36.7 & 1.7 \\
Hydra & 13 25 16.91& -42 58 08.0 & 4.5 \\        
Hydra & 13 25 33.82& -43 02 49.6 & 4.5 \\
Hydra & 13 24 58.60& -42 58 25.7 & 3.5 \\ 
\enddata
\end{deluxetable}

\begin{deluxetable}{lllllll}
\tablecolumns{7}
\tabletypesize{\small}
\tablecaption{Radial Velocity Measurements of Previously Confirmed
  Globular Clusters in NGC 5128\label{tab:vel_knowns}}
\tablewidth{0pt}
\tablehead{
\colhead{GC ID} & \colhead{R.A.}& \colhead{Decl.} &\colhead{v$_{r,LDSS-2}$}
&\colhead{v$_{r,VIMOS}$} &\colhead{v$_{r,Hydra}$} &\colhead{v$_{r}$} \\
\colhead{ } & \colhead{(J2000)} &\colhead{(J2000)} & \colhead{(km
  s$^{-1}$)} & \colhead{(km s$^{-1}$)}& \colhead{(km s$^{-1}$)} & \colhead{(km s$^{-1}$)} \\
}\startdata
GC0001 & 13 25 01.16& -42 56 51.5& -        &503$\pm$16&546$\pm$39&   516$\pm$ 10 \\
GC0005 & 13 23 44.19& -43 11 11.8& -        &712$\pm$32& -         &  642$\pm$  1  \\
GC0007 & 13 23 54.52& -43 20 01.1& -        &273$\pm$52& -         &  274$\pm$ 49  \\
GC0009 & 13 23 58.58& -42 57 17.0& -        &305$\pm$38& -         &  590$\pm$144  \\
GC0010 & 13 23 58.76& -43 01 35.2& -        &482$\pm$31& -         &  496$\pm$ 21  \\
GC0011 & 13 23 59.51& -43 17 29.1& -        &617$\pm$24& -         &  652$\pm$ 37  \\
GC0018 & 13 24 10.97& -43 12 52.8& -        &496$\pm$17& -         &  439$\pm$118  \\
GC0020 & 13 24 18.92& -43 14 30.1& -        &686$\pm$21& -         &  749$\pm$ 32  \\
GC0021 & 13 24 21.40& -43 02 36.8& -        &604$\pm$21& -         &  596$\pm$ 17  \\
GC0022 & 13 24 23.72& -43 07 52.1& -        &599$\pm$20& -         &  613$\pm$ 16  \\
GC0023 & 13 24 23.98& -42 54 10.7& -        &606$\pm$33& -         &  582$\pm$ 81  \\
GC0024 & 13 24 24.15& -42 54 20.6& -        &574$\pm$14& -         &  616$\pm$ 25  \\
GC0028 & 13 24 28.44& -42 57 52.9& -        &516$\pm$50& -         &  558$\pm$ 97  \\
GC0031 & 13 24 29.73& -43 02 06.5& -        &559$\pm$19& -         &  595$\pm$202  \\
GC0032 & 13 24 31.35& -43 11 26.7&775$\pm$34& -        & -         &  734$\pm$ 25  \\      
GC0033 & 13 24 32.17& -43 10 56.9&813$\pm$40& -        & -         &  775$\pm$ 32  \\	
GC0037 & 13 24 36.87& -43 19 16.2& -        &505$\pm$49& -         &  570$\pm$ 31  \\
GC0038 & 13 24 37.75& -43 16 26.5& -        &  -        &213$\pm$10&  212$\pm$ 10  \\
GC0040 & 13 24 38.98& -43 20 06.4& -        &355$\pm$20& -         &  363$\pm$  1  \\
GC0041 & 13 24 40.39& -43 18 05.3& -        &754$\pm$18&597$\pm$16 &  725$\pm$  1  \\
GC0044 & 13 24 40.60& -43 13 18.1& -        &711$\pm$24& -         &  699$\pm$ 17  \\
GC0046 & 13 24 41.20& -43 01 45.6&483$\pm$49&  -       & -         &  518$\pm$ 18  \\ 	
GC0048 & 13 24 43.60& -42 53  7.3&550$\pm$48\tablenotemark{a}&-&-  &  509$\pm$ 15   \\	
GC0050 & 13 24 44.58& -43 02 47.3&731$\pm$44&  -       & -         &  718$\pm$ 16  \\	
GC0053 & 13 24 45.78& -43 02 24.5&444$\pm$58&558$\pm$24& -         & 503 $\pm$17  \\
GC0054 & 13 24 46.46& -43 04 11.6&736$\pm$58&-         &  -        & 712 $\pm$20   \\
GC0056 & 13 24 47.10& -43 06 01.7&550$\pm$53&-         &  -        & 528 $\pm$12   \\
GC0059 & 13 24 47.61& -43 10 48.5&343$\pm$39&-         &  -        & 344 $\pm$37   \\
GC0060 & 13 24 48.06& -43 08 14.2&804$\pm$46&-         &  -        & 786 $\pm$18   \\
GC0061 & 13 24 48.71& -42 52 35.5&512$\pm$172&-	      &509$\pm$63  & 509 $\pm$59   \\
GC0062 & 13 24 48.97& -42 57 48.4& -        &625$\pm$28& -         & 611 $\pm$17  \\
GC0063 & 13 24 49.38& -43 08 17.7&474$\pm$226&	-      &-          & 567 $\pm$76   \\	
GC0064 & 13 24 50.09& -43 07 36.2&588$\pm$41&	-      &-          & 594$\pm$ 36   \\	
GC0066 & 13 24 50.87& -43 01 22.9& -        &  -       &546$\pm$18\tablenotemark{b}&  550$\pm$10 \\
GC0067 & 13 24 51.49& -43 12 11.1&640$\pm$30&616$\pm$20& -         & 624 $\pm$13   \\	
GC0068 & 13 24 52.06& -43 04 32.7& -         &  -      &158$\pm$90 & 191 $\pm$27 \\
GC0070 & 13 24 53.29& -43 04 34.8&271$\pm$77&  -       & -         & 485 $\pm$23   \\	
GC0072 & 13 24 54.18& -42 54 50.4&540$\pm$46&534$\pm$32& -         & 564 $\pm$33  \\
GC0074 & 13 24 54.35& -42 53 24.8& -        &712$\pm$14&739$\pm$15 & 745 $\pm$ 7  \\
GC0075 & 13 24 54.49& -43 05 34.7& -        &733$\pm$20& -         & 695$\pm$ 45  \\
GC0076 & 13 24 54.55& -42 48 58.7&397$\pm$48& -        & -         & 439$\pm$26   \\		 
GC0077 & 13 24 54.73& -43 01 21.7& -        &783$\pm$18&765$\pm$14\tablenotemark{c}&  791$\pm$1 \\
GC0081 & 13 24 55.71& -43 20 39.1& -        &295$\pm$16& -         & 279$\pm$38  \\
GC0082 & 13 24 56.06& -42 54 29.6&513$\pm$54&459$\pm$17&535$\pm$30 & 529 $\pm$ 26  \\
GC0083 & 13 24 56.08& -43 10 16.4&692$\pm$33\tablenotemark{d}&606$\pm$20&-  &  687$\pm$31 \\
GC0086 & 13 24 57.44& -43 01 08.1&613$\pm$55& -        & -         & 685$\pm$ 9   \\ 	
GC0091 & 13 24 58.21& -42 56 10.0& -        &528$\pm$13& -         & 561$\pm$ 1  \\
GC0095 & 13 24 59.92& -43 09 08.6& -        &374$\pm$34& -         & 374$\pm$34 \\
GC0098 & 13 25 00.64& -43 05 30.3&468$\pm$64&  -       & -         & 503$\pm$21  \\	
GC0099 & 13 25 00.83& -43 11 10.6&941$\pm$66&919$\pm$38& -         & 941$\pm$66 \\
GC0103 & 13 25  1.60& -42 54 40.9&686$\pm$60&  -       & -         & 603$\pm$32  \\ 
GC0107 & 13 25 01.86& -42 52 27.8&635$\pm$73&401$\pm$47& -         & 635$\pm$73 \\
GC0108 & 13 25 02.76& -43 11 21.2&481$\pm$45&388$\pm$21& -         & 456$\pm$26 \\
GC0109 & 13 25 03.13& -42 56 25.1& -        &502$\pm$13& -         & 523$\pm$ 8 \\
GC0110 & 13 25 03.18& -43 03 02.5& -        &606$\pm$16&583$\pm$22 & 577$\pm$ 5 \\
GC0111 & 13 25 03.24& -42 57 40.5& -        &679$\pm$22& -         & 648$\pm$29 \\
GC0113 & 13 25 03.37& -42 50 46.2& -        &722$\pm$26& -         & 718$\pm$10 \\
GC0115 & 13 25 03.67& -42 51 21.7& -        &384$\pm$102&426$\pm$34& 425$\pm$34  \\
GC0117 & 13 25 04.48& -43 10 48.4& -        &648$\pm$24& -         & 626$\pm$22 \\
GC0119 & 13 25 04.81& -43 09 38.8&520$\pm$27\tablenotemark{e}&463$\pm$13& - & 481$\pm$10 \\
GC0120 & 13 25 05.02& -42 57 15.0& -        &674$\pm$ 9& -         & 677$\pm$7 \\
GC0121 & 13 25 05.29& -42 58 05.8& -        &807$\pm$18& -         & 787$\pm$19 \\
GC0122 & 13 25 05.46& -43 14 02.6&707$\pm$48&680$\pm$16& -         & 679 $\pm$13 \\
GC0123 & 13 25 05.72& -43 10 30.7&565$\pm$49&430$\pm$ 8&447$\pm$13\tablenotemark{f}  & 439$\pm$1 \\
GC0125 & 13 25 06.25& -43 15 11.6& -        & -      &588$\pm$18\tablenotemark{g}& 597$\pm$10 \\
GC0126 & 13 25 07.33& -43 08 29.6& -        &532$\pm$30& -         & 550$\pm$28 \\
GC0128 & 13 25 07.48& -43 12 29.4&503$\pm$68&616$\pm$59&           & 548$\pm$21 \\
GC0129 & 13 25 07.62& -43 01 15.2& -        &701$\pm$15& -         & 689$\pm$11 \\
GC0130 & 13 25 08.51& -43 02 57.4&326$\pm$50&384$\pm$14& -         & 361$\pm$12 \\
GC0135 & 13 25 09.54& -42 55 18.5& -        &405$\pm$31& -         & 448$\pm$23  \\
GC0137 & 13 25 10.25& -42 55 09.5& -        &547$\pm$11& -         & 576$\pm$12  \\
GC0138 & 13 25 10.27& -42 53 33.1& -        &414$\pm$ 9& -         & 416$\pm$ 8  \\
GC0143 & 13 25 11.17& -43 03 09.6&395$\pm$67&447$\pm$49& -         & 426$\pm$21  \\
GC0144 & 13 25 11.98& -43 04 19.3&865$\pm$77&612$\pm$11& -         & 622$\pm$18  \\
GC0145 & 13 25 12.11& -42 57 25.2& -        &487$\pm$12& -         & 490$\pm$10  \\
GC0146 & 13 25 12.21& -43 16 33.9&510$\pm$25\tablenotemark{h}&-&-  & 573$\pm$16   \\
GC0148 & 13 25 12.45& -43 14 07.4& -        &467$\pm$24&-          & 526$\pm$33  \\
GC0149 & 13 25 12.84& -42 56 59.8& -        &586$\pm$29&-          & 664$\pm$141  \\
GC0150 & 13 25 12.90& -43 07 59.1& -        &600$\pm$13&-          & 616$\pm$ 3  \\
GC0151 & 13 25 13.19& -43 16 35.6& -        &592$\pm$38&-          & 568$\pm$53  \\
GC0153 & 13 25 13.31& -42 52 12.4& -        &433$\pm$30&-          & 474$\pm$23  \\
GC0154 & 13 25 13.88& -43 07 32.5& -        &580$\pm$23&-          & 623$\pm$29  \\
GC0156 & 13 25 14.07& -43 00 51.3& -        & -        &577$\pm$55 & 577$\pm$55  \\
GC0158 & 13 25 14.24& -43 07 23.5& -        &592$\pm$12& -         & 616$\pm$18  \\
GC0160 & 13 25 15.12& -42 50 30.4& -        &704$\pm$26& -         & 716$\pm$55  \\
GC0164 & 13 25 15.93& -43 06 03.3& -        &550$\pm$23& -         & 570$\pm$32  \\
GC0165 & 13 25 16.06& -43 05 06.5& -        &734$\pm$18& -         & 707$\pm$19  \\
GC0166 & 13 25 16.08& -43 02 19.0& -        & -        &468$\pm$34 & 467$\pm$34  \\
GC0169 & 13 25 16.26& -42 50 53.3& -        &526$\pm$21& -         & 566$\pm$13  \\
GC0172 & 13 25 16.73& -42 50 18.4& -        &717$\pm$34& -         &711 $\pm$30   \\
GC0174 & 13 25 16.96& -43 09 28.0& -        &556$\pm$17& -         &561 $\pm$26  \\
GC0175 & 13 25 17.06& -43 02 44.6& -        &655$\pm$16& -         &655$\pm$16  \\
GC0177 & 13 25 17.33& -43 08 39.0& -        &469$\pm$23& -         & 485$\pm$52   \\
GC0180 & 13 25 18.27& -42 53 04.8& -        &379$\pm$17& -         & 384$\pm$16   \\
GC0181 & 13 25 18.44& -43 04 09.8& -        &599$\pm$25& -         & 618$\pm$108   \\
GC0182 & 13 25 18.52& -43 01 16.0& -        & -       &548$\pm$19  & 548$\pm$19    \\
GC0184 & 13 25 19.50& -43 02 28.4& -        &372$\pm$20& -         & 371$\pm$14   \\
GC0187 & 13 25 20.44& -42 54 08.5&920$\pm$88&801$\pm$57& -         & 855$\pm$73   \\
GC0188 & 13 25 20.72& -43 06 35.9& -        &888$\pm$27&-          & 817$\pm$47   \\
GC0189 & 13 25 21.29& -42 49 17.7& -        &452$\pm$17& -         & 450$\pm$15   \\
GC0192 & 13 25 22.35& -43 15 00.1&424$\pm$49&-     	& -        & 451$\pm$30   \\
GC0195 & 13 25 24.40& -43 07 58.9&587$\pm$64&-	       & -         & 542$\pm$34   \\
GC0197 & 13 25 25.49& -42 56 31.2&670$\pm$58&617$\pm$30&-          & 597$\pm$34   \\
GC0199 & 13 25 25.75& -43 05 16.5& -        &610$\pm$31&-          & 534$\pm$31   \\
GC0203 & 13 25 26.78& -42 52 39.9&503$\pm$62&479$\pm$35&-          & 508$\pm$40   \\
GC0204 & 13 25 26.82& -43 09 40.5&589$\pm$44&487$\pm$41&-          & 524$\pm$23   \\
GC0205 & 13 25 27.97& -43 04 02.2& -        &723$\pm$16& -         & 715$\pm$10   \\
GC0207 & 13 25 28.81& -43 04 21.6& -        & -        &553$\pm$155& 553$\pm$155  \\
GC0218 & 13 25 30.41& -43 11 49.6& -        &620$\pm$21& -         & 663$\pm$1   \\
GC0220 & 13 25 30.72& -42 48 13.4& -        &453$\pm$23& -         & 484$\pm$47   \\
GC0223 & 13 25 31.08& -43 04 17.0& -        &584$\pm$32& -         & 554$\pm$60   \\
GC0227 & 13 25 31.73& -43 19 22.6& -        &446$\pm$36& -         & 505$\pm$1   \\
GC0229 & 13 25 32.32& -43 07 17.1&546$\pm$52&-      	&457$\pm$26& 456$\pm$20   \\
GC0230 & 13 25 32.42& -42 58 50.2& -        &484$\pm$21&-          & 489$\pm$14   \\
GC0234 & 13 25 33.17& -42 59 03.2& - &-&536$\pm$26\tablenotemark{i}& 518$\pm$10   \\
GC0235 & 13 25 33.82& -43 02 49.60& -        & -       &608$\pm$31 & 607$\pm$31   \\
GC0236 & 13 25 33.93& -43 03 51.40& -        & -       &299$\pm$67 & 299$\pm$67   \\
GC0240 & 13 25 34.36& -42 51 05.9&297$\pm$50&-	       & -         & 270$\pm$27   \\	  
GC0241 & 13 25 34.64& -43 03 16.4& -        &  -       &585$\pm$97 &  585$\pm$97 \\
GC0242 & 13 25 34.64& -43 03 27.8& -         & -        &626$\pm$57&  662$\pm$  1 \\
GC0245 & 13 25 35.12& -42 56 45.3&347$\pm$38&297$\pm$30& -         &  333$\pm$ 14  \\
GC0246 & 13 25 35.16& -42 53 01.0& -        &545$\pm$25& -         &  593$\pm$ 43  \\
GC0250 & 13 25 35.64& -43 08 36.8& -        &557$\pm$46& -         &  542$\pm$ 56  \\
GC0257 & 13 25 38.13& -43 13 02.2& -        &510$\pm$36& -         &  558$\pm$ 43  \\
GC0258 & 13 25 38.43& -43 05 02.6& -        & -        &28$\pm$9   &  28$\pm$9\tablenotemark{j} \\
GC0259 & 13 25 38.45& -43 03 28.9& -        & -        &577$\pm$107&  574$\pm$106  \\
GC0262 & 13 25 39.17& -43 04 33.8& -        &406$\pm$24&355$\pm$140&  354$\pm$140  \\
GC0265 & 13 25 39.73& -42 55 59.2&733$\pm$44&752$\pm$15&  -        &  780$\pm$  1  \\
GC0267 & 13 25 40.09& -43 03 07.1& -        & -        &480$\pm$256&  480$\pm$256  \\
GC0268 & 13 25 40.56& -42 56 01.3&631$\pm$91&-	       &-          &  631$\pm$ 91  \\
GC0269 & 13 25 40.52& -43 07 17.9&511$\pm$9&-      	&440$\pm$47&  443$\pm$ 23  \\		 
GC0270 & 13 25 40.61& -43 21 13.6& -        &614$\pm$12&-          &  631$\pm$  9  \\
GC0272 & 13 25 40.90& -43 08 16.0& -        &429$\pm$43&-          &  446$\pm$ 31  \\
GC0273 & 13 25 41.36& -42 58 08.9& -        &828$\pm$16&-          &  774$\pm$ 35  \\
GC0274 & 13 25 41.63& -43 03 45.8& -        &701$\pm$43& -         & 701$\pm$43 \\
GC0276 & 13 25 42.09& -43 03 19.5& -        & -        &536$\pm$25 &  536$\pm$25   \\
GC0281 & 13 25 43.23& -42 58 37.4&682$\pm$62&-	  	& -        &  691$\pm$19  \\    
GC0282 & 13 25 43.40& -43 07 22.8& - &-&629$\pm$25\tablenotemark{k}&  613$\pm$ 6   \\
GC0283 & 13 25 43.43& -43 04 56.5& -        &319$\pm$19&313$\pm$24 &  316$\pm$14  \\
GC0284 & 13 25 43.80& -43 07 54.9&504$\pm$95 &462$\pm$51&-         &  534$\pm$58  \\
GC0285 & 13 25 43.90& -42 50 42.7&630$\pm$39&-	  	& -        &  579$\pm$20  \\	
GC0289 & 13 25 45.90& -42 57 20.2&192$\pm$59&-  	& -        &  214$\pm$20  \\	
GC0291 & 13 25 46.00& -42 56 53.0& -        &661$\pm$43& -         &  606$\pm$30  \\
GC0292 & 13 25 46.06& -43 08 24.5& -        &596$\pm$18& -         &  595$\pm$14  \\
GC0295 & 13 25 46.59& -42 57 03.0& -        & -   	&506$\pm$52&  471$\pm$10  \\
GC0296 & 13 25 46.68& -42 53 48.6& -        &661$\pm$27& -         &  633$\pm$108  \\
GC0297 & 13 25 46.92& -43 08 06.6&529$\pm$55&409$\pm$13\tablenotemark{l}&-&  417$\pm$12  \\
GC0302 & 13 25 48.46& -43 07 12.5&756$\pm$40&-  	&539$\pm$41&  650$\pm$28 \\	
GC0303 & 13 25 48.54& -42 57 41.2& -        &584$\pm$12& -         &  592$\pm$ 8 \\
GC0304 & 13 25 48.77& -43 11 38.7& -        &589$\pm$39& -         &  631$\pm$42 \\
GC0305 & 13 25 49.27& -43 02 20.4& -        &371$\pm$28& -         &  393$\pm$25 \\
GC0306 & 13 25 49.69& -42 54 49.3&795$\pm$49&721$\pm$11&752$\pm$17 &  738$\pm$ 6 \\
GC0307 & 13 25 49.73& -43 05 04.7& -        &579$\pm$16& -         &  477$\pm$54 \\
GC0308 & 13 25 49.82& -42 50 15.3&605$\pm$42&-	  	& -        &  557$\pm$17 \\	
GC0312 & 13 25 50.34& -43 04 08.2& -        &465$\pm$17&-          &  468$\pm$36 \\
GC0314 & 13 25 50.40& -42 58 02.3& -        &618$\pm$23&-          &  585$\pm$22 \\
GC0319 & 13 25 52.14& -42 58 30.2&379$\pm$51&-	  	&-         &  475$\pm$18 \\	
GC0320 & 13 25 52.74& -43 05 46.4& - &-&460$\pm$13\tablenotemark{m}&  461$\pm$ 1  \\
GC0321 & 13 25 52.78& -42 58 41.7& -        &295$\pm$14& -         &  348$\pm$22 \\
GC0322 & 13 25 52.88& -43 02 00.0& -        &453$\pm$22& -         &  478$\pm$13 \\
GC0324 & 13 25 53.37& -42 51 12.4& -        &457$\pm$31& -         &  470$\pm$24 \\
GC0325 & 13 25 53.50& -43 03 56.6& -        &739$\pm$10& -         &  728$\pm$13 \\
GC0326 & 13 25 53.57& -42 59 07.6& -        &540$\pm$12&548$\pm$10 &  577$\pm$89 \\
GC0327 & 13 25 53.75& -43 19 48.6& -        &329$\pm$ 7&-          &  335$\pm$6 \\
GC0329 & 13 25 54.39& -43 18 40.1& -        &540$\pm$16&-          &  673$\pm$ 1 \\
GC0330 & 13 25 54.58& -42 59 25.4& -        &639$\pm$12&672$\pm$10 & 673 $\pm$1   \\
GC0333 & 13 25 56.26& -43 01 32.9& -        &285$\pm$22&-          & 438 $\pm$80   \\
GC0334 & 13 25 56.59& -42 51 46.6& -        &401$\pm$18&-          & 403 $\pm$16   \\
GC0340 & 13 25 58.69& -43 07 11.0& -        &517$\pm$39&-          & 500 $\pm$19   \\
GC0341 & 13 25 58.91& -42 53 18.9& -        &331$\pm$34&-          & 410 $\pm$20   \\
GC0342 & 13 25 59.48& -42 55 30.7& -        &539$\pm$17&-          & 448 $\pm$97   \\
GC0350 & 13 26 01.00& -43 06 55.3& -        &474$\pm$39&-          & 517 $\pm$99   \\
GC0351 & 13 26 01.11& -42 55 13.5& -        &331$\pm$ 8&359$\pm$28 & 383 $\pm$12   \\
GC0353 & 13 26 01.83& -42 58 15.0& -        &472$\pm$15&-          & 458 $\pm$31   \\
GC0354 & 13 26 02.25& -43 08 55.6& -        &377$\pm$16&-          & 457 $\pm$38   \\
GC0356 & 13 26 02.79& -42 57 05.0& -        &326$\pm$93& -         &326$\pm$93\\
GC0357 & 13 26 02.85& -42 56 57.0& -        &694$\pm$10&720$\pm$16 & 702 $\pm$ 6  \\
GC0361 & 13 26 04.20& -42 55 44.7& -        &585$\pm$14& -         & 583 $\pm$10  \\
GC0362 & 13 26 04.61& -43 09 10.2& -        &441$\pm$38& -         & 264$\pm$131  \\
GC0365 & 13 26 05.41& -42 56 32.4&-&573$\pm$ 9&599$\pm$21\tablenotemark{n}&  594$\pm$1 \\
GC0368 & 13 26 06.55& -43 06 14.5& -        &432$\pm$26& -         & 439 $\pm$25  \\
GC0370 & 13 26 06.93& -42 57 35.1&740$\pm$53&649$\pm$10& -         & 652 $\pm$ 8  \\
GC0372 & 13 26 07.73& -42 52 00.3&704$\pm$38&705$\pm$18& -         & 702 $\pm$ 1  \\
GC0373 & 13 26 08.38& -42 59 18.9&620$\pm$49&590$\pm$20& -         & 586 $\pm$46  \\
GC0376 & 13 26 09.71& -42 50 29.5& -        &753$\pm$37&-          &  716$\pm$ 48  \\
GC0378 & 13 26 10.58& -42 53 42.7&637$\pm$42&578$\pm$11&-          &  612$\pm$  1  \\
GC0381 & 13 26 15.27& -42 48 29.4& -        &378$\pm$19&-          &  377$\pm$ 11  \\
GC0382 & 13 26 15.88& -42 55 00.9&443$\pm$68&	-  &555$\pm$107    &  587$\pm$  1  \\
GC0384 & 13 26 19.64& -43 03 18.6& -        & -   	&705$\pm$25&  690$\pm$ 22   \\
GC0388 & 13 26 21.11& -42 48 41.1&489$\pm$42&392$\pm$12& -         &  430$\pm$ 23  \\
GC0390 & 13 26 21.31& -42 57 19.1&431$\pm$43&-	  	& -        &  453$\pm$ 23  \\	
GC0391 & 13 26 21.99& -42 53 45.5&371$\pm$49&-	  	& -        &  459$\pm$ 23  \\	
GC0393 & 13 26 22.08& -43 09 10.7& -        &357$\pm$35& -         &  505$\pm$ 78  \\
GC0394 & 13 26 22.65& -42 46 49.8& -        &631$\pm$24& -         &  575$\pm$ 41  \\
GC0395 & 13 26 23.60& -43 03 43.9& -        &542$\pm$23& -         &  470$\pm$ 66  \\
GC0397 & 13 26 23.78& -42 54 01.1&431$\pm$46&-& 413$\pm$12\tablenotemark{o}&  405 $\pm$1  \\
GC0400 & 13 26 25.50& -42 57 06.2&505$\pm$152&387$\pm$22& -        &  406$\pm$18  \\
GC0403 & 13 26 33.55& -42 51 00.9& -        &585$\pm$17& -         &  534$\pm$33  \\
GC0404 & 13 26 38.49& -42 45 45.7& -        & -  	&690$\pm$26&  627$\pm$21  \\
GC0405 & 13 26 41.43& -43 11 25.0& -        &541$\pm$12& -         &  572$\pm$67  \\
GC0409 & 13 26 53.94& -43 19 17.7& -        &583$\pm$17& -         &  605$\pm$12  \\
GC0445 & 13 24 40.51& -43 01 56.7& -        & -  	&192$\pm$39& 325 $\pm$29  \\
GC0446 & 13 24 50.43& -43 04 51.0& -        &390$\pm$33&446$\pm$28 & 370 $\pm$20  \\
GC0448 & 13 24 53.64& -42 57 59.3& -        &634$\pm$28&655$\pm$84 & 633 $\pm$23  \\
GC0449 & 13 24 53.96& -43 00 43.6& -        & -  	&374$\pm$27& 362 $\pm$22 \\
GC0451 & 13 24 55.42& -42 59 48.5& -        & -         &627$\pm$52&  656$\pm$32  \\
GC0452 & 13 24 57.74& -42 58 51.8& -        &643$\pm$44& -         &  643$\pm$39  \\
GC0453 & 13 24 58.60& -42 58 16.6& -        &615$\pm$14&615$\pm$56 &  620$\pm$ 9  \\
GC0454 & 13 25 02.82& -43 02 04.4& -        &404$\pm$28&438$\pm$82 &  399$\pm$24  \\
GC0455 & 13 25 05.02& -43 01 33.6& -        &504$\pm$32& -         &  498$\pm$27  \\
GC0456 & 13 25 06.72& -42 59 17.1& -        & -   	&489$\pm$28&  443 $\pm$19  \\
GC0457 & 13 25 06.83& -42 57 32.2& -        & -   	&608$\pm$32&  587$\pm$25  \\
GC0458 & 13 25 07.68& -42 55 49.5& -        &525$\pm$37& -         &  533 $\pm$32  \\
GC0460 & 13 25 20.41& -43 06 09.2& -        &737$\pm$20&715$\pm$89 &  737 $\pm$17  \\
GC0463 & 13 25 27.28& -42 58 28.8& -        & -  	&265$\pm$59&  237 $\pm$30  \\
GC0464 & 13 25 27.45& -42 55 30.9& -        &458$\pm$40 & -        &  461 $\pm$35  \\
GC0471 & 13 25 42.10& -42 57 24.0& -        & -   	&311$\pm$28&  275 $\pm$24  \\
GC0478 & 13 25 49.23& -43 00 02.2& -        & -        &475$\pm$134&  465 $\pm$29  \\

\tablenotetext{a}{The weighted average of two radial velocity measurements with LDSS-2, $610\pm66$ km s$^{-1}$ and $485\pm69$ km s$^{-1}$.}
\tablenotetext{b}{The weighted average of three radial velocity
  measurements with Hydra, $563\pm47$ km s$^{-1}$, $555\pm26$ km
  s$^{-1}$, and $530\pm28$ km s$^{-1}$.}
\tablenotetext{c}{The weighted average of two radial velocity
  measurements with Hydra, $760\pm18$ km s$^{-1}$ and $772\pm23$ km
  s$^{-1}$.}
\tablenotetext{d}{The weighted average of two radial velocity measurements with LDSS-2, $691\pm44$ km s$^{-1}$ and $693\pm49$ km s$^{-1}$.}
\tablenotetext{e}{The weighted average of two radial velocity measurements with LDSS-2, $566\pm46$ km s$^{-1}$ and $495\pm34$ km s$^{-1}$.}
\tablenotetext{f}{The weighted average of two radial velocity
  measurements with Hydra, $457\pm17$ km s$^{-1}$ and $432\pm21$ km
  s$^{-1}$.}
\tablenotetext{g}{The weighted average of three radial velocity
  measurements with Hydra, $580\pm44$ km s$^{-1}$, $535\pm67$ km
  s$^{-1}$, and $594\pm20$ km s$^{-1}$.}
\tablenotetext{h}{The weighted average of two radial velocity measurements with LDSS-2, $465\pm51$ km s$^{-1}$ and $524\pm28$ km s$^{-1}$.}
\tablenotetext{i}{The weighted average of two radial velocity
  measurements with Hydra, $600\pm47$ km s$^{-1}$ and $508\pm31$ km
  s$^{-1}$.}
\tablenotetext{j}{This object was originally classified as a GC from
  a{\it Hubble Space Telescope} study by \cite{harris06}.  This is the
  first radial velocity measurement of this object, and it has a
  greater than 0.8 correlation peak for the recorded velocity,
  indicating it may be a star and not a genuine GC.}
\tablenotetext{k}{The weighted average of two radial velocity
  measurements with Hydra, $634\pm29$ km s$^{-1}$ and $615\pm50$ km
  s$^{-1}$.}
\tablenotetext{l}{The weighted average of two radial velocity
  measurements with VIMOS, $392\pm18$ km s$^{-1}$ and $427\pm19$ km
  s$^{-1}$.}
\tablenotetext{m}{The weighted average of three radial velocity
  measurements with Hydra, $458\pm23$ km s$^{-1}$, $461\pm29$ km
  s$^{-1}$, and $460\pm17$ km s$^{-1}$.}
\tablenotetext{n}{The weighted average of two radial velocity
  measurements with Hydra, $601\pm29$ km s$^{-1}$ and $596\pm30$ km
  s$^{-1}$.}
\tablenotetext{o}{The weighted average of three radial velocity
  measurements with Hydra, $417\pm40$ km s$^{-1}$, $413\pm14$ km
  s$^{-1}$, and $409\pm28$ km s$^{-1}$.}
\enddata
\end{deluxetable}

\begin{deluxetable}{lllllllllllll}
\tablecolumns{13}
\tabletypesize{\small}
\tablecaption{Radial Velocity Measurements of Newly Confirmed
  Globular Clusters in NGC 5128\label{tab:vel_new}}
\tablewidth{0pt}
\tablehead{
\colhead{GC ID} & \colhead{R.A.}& \colhead{Decl.} &\colhead{C}&\colhead{$\sigma_C$}&\colhead{M}&\colhead{$\sigma_M$}&\colhead{T$_1$}&\colhead{$\sigma_{T_1}$}&\colhead{v$_{r,LDSS-2}$}
&\colhead{v$_{r,VIMOS}$} &\colhead{v$_{r,Hydra}$} &\colhead{v$_{r}$} \\
\colhead{ } & \colhead{(J2000)} &\colhead{(J2000)} & \colhead{(mag)}& \colhead{(mag)}& \colhead{(mag)}& \colhead{(mag)}& \colhead{(mag)}& \colhead{(mag)}& \colhead{(km s$^{-1}$)}& \colhead{(km s$^{-1}$)}& \colhead{(km s$^{-1}$)}& \colhead{(km s$^{-1}$)} \\
}\startdata

GC0416 & 13 24 33.63 & -43 12 01.6 & 21.69 & 0.02 & 20.94& 0.01 & 20.12& 0.01 & 597$\pm$52  &- &543$\pm$10 & 546$\pm$  9\\
GC0417 & 13 24 42.39 & -42 51 49.1 & 20.50 & 0.02 & 19.82& 0.02 & 19.09& 0.01 & 681$\pm$35  &- &577$\pm$39 & 634$\pm$ 26\\
GC0418 & 13 24 44.77 & -43 06 33.3 & 22.01 & 0.06 & 20.86& 0.05 & 19.87& 0.03 & 487$\pm$102 &- &363$\pm$32 & 374$\pm$ 30  \\
GC0419 & 13 24 55.31 & -43 10 39.3 & 21.87 & 0.02 & 20.83& 0.02 & 19.89& 0.01 & 653$\pm$57  &- &658$\pm$32 & 656$\pm$ 27\\
GC0420 & 13 24 55.97 & -43 02 15.9 & 21.29 & 0.03 & 20.64& 0.02 & 19.87& 0.03 & 807$\pm$65  &763$\pm$40 &526$\pm$87& 735$\pm$28 \\
GC0421 & 13 24 59.58 & -43 06 40.8 & 22.62 & 0.03 & 21.49& 0.03 & 20.47& 0.01 & 841$\pm$124 &719$\pm$31 &-& 726$\pm$ 30\\
GC0422 & 13 25 03.28 & -43 08 14.4 & 21.11 & 0.03 & 20.56& 0.02 & 19.76& 0.01 & 666$\pm$140 &646$\pm$38 & 805$\pm$60 & 690$\pm$ 31\\
GC0423 & 13 25 04.28 & -43 17 10.8 & 22.36 & 0.04 & 21.80& 0.03 & 20.92& 0.02 & 577$\pm$73  &499$\pm$51 &-& 524$\pm$ 41\\
GC0424 & 13 25 06.87 & -43 02 40.2 & 20.65 & 0.01 & 19.93& 0.01 & 19.13& 0.01 & 247$\pm$49  &367$\pm$18 & 410$\pm$38 & 354$\pm$ 14\\
GC0425 & 13 25 09.60 & -43 04 37.9 & 21.65 & 0.02 & 20.94& 0.01 & 20.11& 0.01 & 658$\pm$34  &657$\pm$32 &-& 648$\pm$ 19\\
GC0426 & 13 25 14.18 & -43 04 46.6 & 21.23 & 0.03 & 20.67& 0.02 & 19.87& 0.02 & 705$\pm$120 &881$\pm$36 &-& 858$\pm$ 30\\
GC0427 & 13 25 21.49 & -43 19 29.6 & 19.59 & 0.01 & 19.23& 0.01 & 18.66& 0.01 & 163$\pm$57  &- &-& 163$\pm$ 57\\
GC0428 & 13 25 22.16 & -43 16 38.8 & 21.06 & 0.04 & 20.48& 0.04 & 19.76& 0.02 & 401$\pm$55  &- &466$\pm$96 & 417$\pm$ 47\\
GC0429 & 13 25 22.88 & -43 08 10.1 & 20.62 & 0.01 & 19.20& 0.01 & 18.14& 0.01 & 146$\pm$110 &-&- & 146$\pm$110\\
GC0430 & 13 25 24.38 & -42 57 16.8 &   -   &  -   & 19.25& 0.29 & 18.90& 0.01 & 503$\pm$75  &-&- & 503$\pm$ 75\\
GC0431 & 13 25 28.72 & -42 50 12.3 & 20.60 & 0.04 & 19.93& 0.04 & 19.36& 0.01 & 663$\pm$54  &593$\pm$40 & 687$\pm$38 &646$\pm$ 24\\
GC0432 & 13 25 34.08 & -43 10 45.7 & 19.87 & 0.31 & 20.81& 0.07 & 20.00& 0.05 & 396$\pm$84  &-&- & 396$\pm$ 84\\
GC0433 & 13 25 36.11 & -42 56 16.8 & 22.61 & 0.03 & 22.58& 0.03 & 22.00& 0.09 & 575$\pm$56  &-&- & 575$\pm$ 56\\
GC0434 & 13 25 37.85 & -42 56 28.0 & 20.20 & 0.01 & 20.59& 0.01 & 18.51& 0.00 & 505$\pm$51  &478$\pm$17 &-& 480$\pm$ 16\\
GC0435 & 13 25 38.53 & -42 57 19.9 & 21.32 & 0.02 & 20.30& 0.03 & 19.42& 0.01 & 706$\pm$63  &-& -& 706$\pm$ 63\\
GC0436 & 13 25 41.71 & -42 58 31.6 & 20.34 & 0.02 & 19.29& 0.01 & 18.52& 0.01 & 523$\pm$71  &440$\pm$15 & 478$\pm$18 & 456$\pm$ 10\\
GC0437 & 13 25 45.23 & -42 54 23.8 & 22.04 & 0.03 & 21.00& 0.04 & 20.14& 0.02 & 490$\pm$113 &-&- & 490$\pm$113\\
GC0438 & 13 25 45.91 & -42 51 41.2 & 21.34 & 0.02 & 21.55& 0.01 & 19.56& 0.01 & 837$\pm$59  &-&- & 837$\pm$ 59\\
GC0439 & 13 25 47.00 & -42 55 29.9 & 20.56 & 0.05 & 19.75& 0.03 & 18.89& 0.02 & 700$\pm$48  &-&- & 700$\pm$ 48\\
GC0440 & 13 25 50.59 & -42 51 40.1 & 20.19 & 0.02 & 19.60& 0.01 & 18.95& 0.02 & 622$\pm$50  &-&649$\pm$42 & 637$\pm$ 32\\
GC0441 & 13 25 54.48 & -42 57 27.6 & 20.97 & 0.02 & 19.91& 0.01 & 19.12& 0.01 & 698$\pm$67  &597$\pm$11&626$\pm$27 & 603$\pm$ 10\\
GC0442 & 13 26 09.37 & -42 53 17.5 & 21.09 & 0.03 & 20.38& 0.02 & 19.67& 0.02 & 615$\pm$81  &-&646$\pm$69 & 632$\pm$52\tablenotemark{a}\\
GC0443 & 13 26 12.11 & -42 49 03.5 & 19.87 & 0.01 & 19.31& 0.01 & 18.73& 0.01 & 225$\pm$35  &-&- & 225$\pm$ 35\\
GC0444 & 13 26 20.63 & -42 53 46.0 & 21.49 & 0.01 & 20.80& 0.01 & 20.15& 0.01 & 271$\pm$64  &-&- & 271$\pm$ 64\\
GC0480 & 13 23 50.49 & -43 11 43.9 & 20.96 & 0.03 & 21.87& 0.03 & 23.02& 0.02 &-& 646$\pm$61  &-&646$\pm$61\\ 
GC0481 & 13 23 54.18 & -43 21 56.6 & 21.89 & 0.05 & 22.81& 0.03 & 24.10& 0.03 &-& 512$\pm$84  &-&512$\pm$84\\ 
GC0482 & 13 24 03.98 & -43 17 20.3 & 21.73 & 0.03 & 22.55& 0.02 & 23.45& 0.03 &-& 599$\pm$105 &-&599$\pm$105\\ 
GC0483 & 13 24 09.86 & -43 22 39.0 & 20.89 & 0.05 & 21.90& 0.03 & 23.02& 0.03 &-& 756$\pm$37  &-&756$\pm$37\\ 
GC0484 & 13 24 10.98 & -43 12 15.8 & 20.53 & 0.02 & 21.47& 0.01 & 22.58& 0.01 &-& 498$\pm$22  &-&498$\pm$22\\ 
GC0485\tablenotemark{d} & 13 24 13.76 & -43 05 33.4 & 19.96 & 0.01 & 20.83& 0.01 & 21.84& 0.01 &-& 388$\pm$20  &-&388$\pm$20\\ 
GC0486 & 13 24 19.45 & -42 54 31.5 & 18.98 & 0.01 & 19.83& 0.01 & 20.77& 0.01 &-& $566\pm12$ & $503\pm25$ & 554$\pm$11  \\ 
GC0487 & 13 24 20.60 & -43 08 07.9 & 19.88 & 0.01 & 20.83& 0.01 & 22.01& 0.01 &-& $579\pm17$ & $553\pm53$ &  576$\pm$16\\ 
GC0488 & 13 24 27.23 & -42 57 25.7 & 20.99 & 0.03 & 21.92& 0.03 & 22.87& 0.02 &-& 559$\pm$35 &-& 559$\pm$35\\ 
GC0489 & 13 24 41.51 & -43 12 53.2 & 20.30 & 0.03 & 21.11& 0.02 & 21.76& 0.01 &-& 506$\pm$41 &-& 506$\pm$41\\ 
GC0490 & 13 24 47.40 & -43 09 59.8 & 20.20 & 0.01 & 21.14& 0.01 & 22.15& 0.01 &-& 568$\pm$22 &-& 568$\pm$22\\ 
GC0491 & 13 24 54.98 & -43 11 36.9 & 20.40 & 0.04 & 21.41& 0.03 & 22.46& 0.02 &-& 670$\pm$19 &-& 670$\pm$19\\ 
GC0492 & 13 24 56.61 & -43 12 23.6 & 20.74 & 0.03 & 21.42& 0.03 & 22.00& 0.02 &-& 713$\pm$28 &-& 713$\pm$28\\ 
GC0493 & 13 24 57.89 & -43 07 06.3 & 20.11 & 0.03 & 21.04& 0.02 & 21.97& 0.01 &-& 559$\pm$40 &-& 559$\pm$40\\ 
GC0494 & 13 24 58.30 & -43 18 33.6 & 20.20 & 0.02 & 20.95& 0.02 & 21.55& 0.01 &-& 676$\pm$42 &-& 676$\pm$42\\ 
GC0495 & 13 24 58.96 & -43 11 28.2 & 19.98 & 0.02 & 20.93& 0.01 & 21.99& 0.01 &-& $494\pm17$ & $619\pm47$ & 508$\pm$15\\ 
GC0496 & 13 24 59.80 & -42 57 33.2 & 19.68 & 0.01 & 20.47& 0.02 & 21.11& 0.01 &-& $470\pm33$ & $635\pm23$ & 581$\pm$18\\ 
GC0497 & 13 25 00.89 & -43 21 07.7 & 20.37 & 0.01 & 21.37& 0.01 & 22.48& 0.01 &-& $492\pm25$ & $437\pm44$ & 478$\pm$21\\ 
GC0498 & 13 25 01.36 & -43 05 46.5 & 20.38 & 0.04 & 21.39& 0.01 & 22.47& 0.01 &-& 472$\pm$36 &-& 472$\pm$36\\ 
GC0499 & 13 25 01.81 & -43 03 48.1 & 20.43 & 0.02 & 21.39& 0.02 & 22.49& 0.03 &-& 600$\pm$35 &-& 600$\pm$35\\ 
GC0500 & 13 25 03.60 & -43 01 40.6 & 20.42 & 0.03 & 21.15& 0.01 & 21.85& 0.03 &-& 386$\pm$52 &-& 386$\pm$52\\ 
GC0501 & 13 25 06.14 & -42 46 18.8 & 20.02 & 0.01 & 20.91& 0.01 & 21.97& 0.01 &-& 564$\pm$71 &-& 564$\pm$71\\ 
GC0502 & 13 25 06.68 & -43 08 58.2 & 20.98 & 0.06 & 22.11& 0.05 & 23.28& 0.04 &-& 240$\pm$53 &-& 240$\pm$53\\ 
GC0503 & 13 25 07.36 & -43 03 23.9 & 19.08 & 0.01 & 20.69& 0.01 & 21.18& 0.01 &-& 670$\pm$41 &-& 670$\pm$41\\ 
GC0504 & 13 25 08.94 & -43 08 53.7 & 20.04 & 0.04 & 20.70& 0.03 & 21.29& 0.01 &-& 678$\pm$26 &-& 678$\pm$26\\ 
GC0505 & 13 25 12.69 & -43 01 55.9 & 19.80 & 0.04 & 20.54& 0.03 & 21.57& 0.03 &-& 578$\pm$20 &-& 578$\pm$20\\ 
GC0506 & 13 25 12.77 & -43 10 43.5 & 20.42 & 0.04 & 21.44& 0.04 & 22.45& 0.02 &-& 280$\pm$29 &-& 280$\pm$29\\ 
GC0507 & 13 25 13.04 & -42 55 56.2 & 19.55 & 0.02 & 20.32& 0.01 & 20.94& 0.01 &-& $460\pm29$ & $478\pm91$ &  461$\pm$27\\ 
GC0508\tablenotemark{b} & 13 25 13.70 & -42 47 36.9 & 21.27 & 0.01 & 22.04& 0.01 & 22.80& 0.01 &-& 326$\pm$71 &-& 326$\pm$71\\ 
GC0509 & 13 25 13.70 & -42 40 32.4 & 20.73 & 0.01 & 21.45& 0.01 & 22.22& 0.01 &-& 306$\pm$35 &-& 306$\pm$35\\ 
GC0510 & 13 25 17.68 & -42 52 56.5 & 20.62 & 0.05 & 21.44& 0.06 & 22.12& 0.02 &-& 670$\pm$45 &-& 670$\pm$45\\ 
GC0511 & 13 25 18.29 & -42 47 52.4 & 20.81 & 0.02 & 21.66& 0.02 & 22.62& 0.02 &-& 487$\pm$39 &-& 487$\pm$39\\ 
GC0512 & 13 25 18.92 & -42 39 03.6 & 19.70 & 0.02 & 20.40& 0.02 & 21.10& 0.01 &-& $664\pm34$ & $441\pm82$ & 631$\pm$31\\ 
GC0513 & 13 25 20.11 & -43 22 21.8 & 20.33 & 0.01 & 21.11& 0.01 & 21.72& 0.01 &-& 844$\pm$54 &-& 844$\pm$54\\ 
GC0514 & 13 25 20.74 & -42 41 47.1 & 19.81 & 0.01 & 20.67& 0.01 & 21.56& 0.01 &-& 518$\pm$19 &-& 518$\pm$19\\ 
GC0515 & 13 25 23.08 & -43 06 20.9 & 20.65 & 0.02 & 21.12& 0.02 & 22.11& 0.02 &-& 591$\pm$29 &-& 591$\pm$29\\ 
GC0516 & 13 25 23.31 & -42 48 47.7 & 19.06 & 0.03 & 19.88& 0.04 & 20.93& 0.03 &-& $517\pm12$ & $581\pm28$ & 527$\pm$11   \\ 
GC0517 & 13 25 24.75 & -43 13 37.4 & 20.40 & 0.04 & 20.74& 0.04 & 21.50& 0.02 &-& 798$\pm$60 &-& 798$\pm$60\\ 
GC0518 & 13 25 25.36 & -42 49 27.5 & 21.05 & 0.03 & 21.75& 0.03 & 22.45& 0.03 &-& 589$\pm$63 &-& 589$\pm$63\\ 
GC0519 & 13 25 26.09 & -43 11 22.4 & 18.11 & 0.02 & 18.81& 0.02 & 19.50& 0.02 &-& $238\pm25$ & $250\pm29$ & 243$\pm$19   \\ 
GC0520 & 13 25 27.18 & -42 58 37.9 & 20.05 & 0.03 & 20.91& 0.02 & 21.91& 0.01 &-& 572$\pm$48 &-& 572$\pm$48\\ 
GC0521 & 13 25 28.81 & -42 53 15.9 & 19.82 & 0.01 & 20.80& 0.02 & 21.60& 0.01 &-& $160\pm74$ & $179\pm26$ & 176$\pm$24\\ 
GC0522 & 13 25 30.10 & -43 22 33.3 & 20.44 & 0.02 & 21.33& 0.01 & 22.26& 0.01 &-& 573$\pm$49&-&  573$\pm$48\\ 
GC0523\tablenotemark{b} & 13 25 33.10 & -43 09 21.6 & 20.40 & 0.03 & 21.16& 0.03 & 21.76& 0.03 &-& 351$\pm$69&-&  351$\pm$69\\ 
GC0524 & 13 25 33.94 & -42 51 39.4 & 20.85 & 0.02 & 22.22& 0.01 & 22.03& 0.02 &-& 527$\pm$61&-&  527$\pm$61\\ 
GC0525 & 13 25 37.17 & -42 43 23.7 & 20.77 & 0.04 & 21.81& 0.04 & 22.90& 0.02 &-& 551$\pm$41&-&  551$\pm$41\\ 
GC0526\tablenotemark{b} & 13 25 37.63 & -42 39 32.2 & 20.37 & 0.02 & 21.11& 0.02 & 21.93& 0.02 &-& $507\pm39$ & $519\pm45$ & 512$\pm$29\\ 
GC0527 & 13 25 39.41 & -42 58 24.1 & 19.84 & 0.01 & 20.44& 0.01 & 21.08& 0.01 &-& $262\pm74$ & $312\pm22$ & 301$\pm$19\\ 
GC0528 & 13 25 48.32 & -42 55 06.9 & 19.91 & 0.02 & 20.61& 0.01 & 21.60& 0.01 &-& 489$\pm$19 &-& 489$\pm$19\\ 
GC0529 & 13 25 50.54 & -43 08 02.9 & 20.30 & 0.02 & 21.01& 0.01 & 21.50& 0.01 &-& 720$\pm$80 &-& 720$\pm$80\\ 
GC0530 & 13 25 52.55 & -43 03 00.1 & 20.36 & 0.01 & 21.28& 0.02 & 22.08& 0.01 &-& 488$\pm$50 &-& 488$\pm$50\\ 
GC0531 & 13 25 54.76 & -43 01 40.5 & 20.33 & 0.03 & 21.43& 0.03 & 22.84& 0.01 &-& 341$\pm$23 &-& 341$\pm$23\\ 
GC0532 & 13 25 54.79 & -42 52 57.1 & 20.63 & 0.05 & 21.53& 0.03 & 22.62& 0.02 &-& 399$\pm$32 &-& 399$\pm$32\\ 
GC0533 & 13 25 55.09 & -43 10 04.8 & 20.00 & 0.01 & 20.88& 0.01 & 21.69& 0.01 &-& $350\pm26$ & $422\pm96$ &  354$\pm$25\\ 
GC0534 & 13 25 55.35 & -42 53 40.2 & 20.67 & 0.02 & 21.25& 0.01 & 21.81& 0.02 &-& 347$\pm$41 &-& 347$\pm$41\\ 
GC0535 & 13 25 55.80 & -42 51 39.8 & 21.72 & 0.06 & 22.62& 0.05 & 23.67& 0.04 &-& 557$\pm$93 &-& 557$\pm$93\\ 
GC0536 & 13 25 56.02 & -42 46 26.1 & 20.48 & 0.03 & 20.99& 0.02 & 21.45& 0.01 &-& 448$\pm$39 &-& 448$\pm$39\\ 
GC0537 & 13 25 56.14 & -43 02 53.4 & 20.49 & 0.02 & 21.36& 0.01 & 22.66& 0.01 &-& 398$\pm$30 &-& 398$\pm$30\\ 
GC0538 & 13 25 56.78 & -42 52 48.4 & 18.87 & 0.09 & 19.91& 0.05 & 20.89& 0.10 &-& $512\pm12$ & $508\pm47$ &  511$\pm$11\\ 
GC0539 & 13 25 57.20 & -42 47 44.6 & 19.84 & 0.01 & 20.60& 0.01 & 21.49& 0.01 &-& 544$\pm$33 &458$\pm$94& 534$\pm$31\tablenotemark{c}\\ 
GC0540 & 13 26 00.63 & -42 49 09.5 & 21.21 & 0.03 & 22.13& 0.02 & 23.15& 0.02 &-& 609$\pm$46 &-& 609$\pm$46\\ 
GC0541 & 13 26 01.33 & -43 02 34.9 & 20.52 & 0.01 & 21.36& 0.02 & 22.27& 0.01 &-& 471$\pm$30 &-& 471$\pm$30\\ 
GC0542 & 13 26 05.38 & -42 55 22.4 & 21.22 & 0.05 & 22.00& 0.04 & 23.06& 0.03 &-& 364$\pm$47 &-& 364$\pm$47\\ 
GC0543 & 13 26 08.51 & -43 04 45.1 & 20.28 & 0.01 & 21.02& 0.01 & 21.68& 0.01 &-& 466$\pm$33 &-& 466$\pm$33\\ 
GC0544 & 13 26 17.27 & -43 09 58.0 & 20.91 & 0.03 & 21.93& 0.02 & 23.12& 0.02 &-& 473$\pm$39 &-& 473$\pm$39\\ 
GC0545 & 13 26 21.34 & -42 49 59.3 & 20.50 & 0.02 & 21.20& 0.01 & 21.83& 0.01 &-& 439$\pm$28 &-& 439$\pm$28\\ 
GC0546 & 13 26 21.73 & -43 18 28.8 & 21.08 & 0.04 & 21.83& 0.03 & 22.46& 0.03 &-& 507$\pm$84 &-& 507$\pm$84\\ 
GC0547 & 13 26 23.07 & -43 22 39.1 & 20.88 & 0.04 & 21.62& 0.02 & 22.52& 0.03 &-& 521$\pm$57 &-& 521$\pm$57\\ 
GC0548 & 13 26 24.16 & -42 47 00.7 & 18.04 & 0.03 & 18.76& 0.02 & 19.52& 0.01 &-& $410\pm13$ & $339\pm33$ & 400$\pm$12\\ 
GC0549 & 13 26 28.90 & -43 13 24.7 & 20.55 & 0.04 & 21.56& 0.03 & 22.62& 0.02 &-& 459$\pm$31 &-& 459$\pm$31\\ 
GC0550 & 13 26 29.09 & -42 58 31.2 & 20.04 & 0.03 & 20.85& 0.02 & 21.59& 0.01 &-& 818$\pm$45 &-& 818$\pm$45\\ 
GC0551 & 13 26 30.44 & -43 03 15.3 & 20.39 & 0.06 & 21.28& 0.09 & 21.86& 0.03 &-& 349$\pm$91 &-& 349$\pm$91\\ 
GC0552 & 13 26 37.68 & -43 10 41.1 & 21.17 & 0.09 & 23.12& 0.04 & 24.84& 0.03 &-& 548$\pm$33 &-& 548$\pm$33\\ 
GC0553 & 13 26 37.70 & -42 58 53.5 & 18.65 & 0.01 & 19.46& 0.01 & 20.43& 0.01 &-& 276$\pm$22 &-& 276$\pm$22\\ 
GC0554 & 13 26 42.18 & -43 17 11.9 & 19.92 & 0.03 & 20.65& 0.02 & 21.16& 0.02 &-& $344\pm91$ & $429\pm40$&  415$\pm$36\\ 
GC0555 & 13 26 44.33 & -43 19 10.6 & 19.66 & 0.03 & 20.38& 0.02 & 20.92& 0.02 &-& 336$\pm$33 &-& 336$\pm$33\\ 
GC0556 & 13 26 45.96 & -43 11 41.7 & 20.03 & 0.04 & 20.82& 0.03 & 21.40& 0.02 &-& 563$\pm$39 &-& 563$\pm$39\\ 
GC0557 & 13 26 50.87 & -43 07 20.5 & 20.44 & 0.01 & 21.25& 0.01 & 21.96& 0.01 &-& 372$\pm$42 &-& 372$\pm$42\\ 
GC0558 & 13 26 55.84 & -43 20 11.3 & 18.99 & 0.02 & 19.77& 0.01 & 20.45& 0.01 &-& 616$\pm$25 &-& 616$\pm$25\\ 
GC0559 & 13 27 02.84 & -42 59 04.8 & 19.02 & 0.03 & 19.72& 0.03 & 20.31& 0.02 &-& 285$\pm$36 & 267$\pm$49 & 278$\pm$29\\ 
GC0560 & 13 23 26.11&-42 55 27.2 &19.52 & 0.01& 18.91 & 0.01 &18.27 & 0.01  &- &- &235$\pm$43	 &235$\pm$43	\\
GC0561 & 13 23 28.24&-42 47 42.3 &17.61 & 0.01& 17.00 & 0.01 &16.38 & 0.01  &- &- &181$\pm$68	 &181$\pm$68	\\
GC0562 & 13 23 28.43&-42 53 12.5 &20.10 & 0.01& 19.24 & 0.01 &18.49 & 0.01  &- &- &238$\pm$26	 &238$\pm$26	\\
GC0563 & 13 23 45.02&-43 01 11.5 &20.51 & 0.02& 19.48 & 0.02 &18.65 & 0.01  &- &- &533$\pm$35	 &533$\pm$35	\\
GC0564 & 13 23 47.42&-43 04 17.8 &22.22 & 0.03& 21.18 & 0.03 &20.23 & 0.02  &- &- &357$\pm$66	 &357$\pm$66	\\
GC0565 & 13 24 06.60&-42 52 58.6 &18.86 & 0.01& 17.75 & 0.01 &16.87 & 0.01  &- &- &285$\pm$29	 &285$\pm$29	\\
GC0566 & 13 24 09.78&-42 44 54.1 &19.75 & 0.01& 18.91 & 0.01 &18.21 & 0.01  &- &- &173$\pm$117 &173$\pm$117\\  
GC0567 & 13 24 19.71&-42 50 43.8 &21.55 & 0.03& 20.48 & 0.02 &19.55 & 0.01  &- &- &189$\pm$34	 &189$\pm$34	\\
GC0568 & 13 24 28.19&-43 01 36.4 &22.34 & 0.10& 21.15 & 0.09 &20.27 & 0.10  &- &- &211$\pm$16	 &211$\pm$16	\\
GC0569 & 13 24 29.52&-43 00 03.5 &20.99 & 0.04& 19.87 & 0.03 &19.02 & 0.04  &- &- &586$\pm$30	 &586$\pm$30	\\
GC0570 & 13 24 35.14&-42 57 59.6 &21.09 & 0.02& 20.38 & 0.02 &19.56 & 0.01  &- &- &297$\pm$21	 &297$\pm$21	\\
GC0571 & 13 24 36.96&-43 18 31.9 &21.72 & 0.04& 20.82 & 0.01 &19.88 & 0.01  &- &- &926$\pm$44	 &926$\pm$44	\\
GC0572 & 13 24 37.35&-43 06 31.4 &22.05 & 0.02& 21.08 & 0.02 &20.13 & 0.01  &- &- &183$\pm$49	 &183$\pm$49	\\
GC0573 & 13 24 40.81&-42 41 01.6 &20.07 & 0.01& 18.96 & 0.01 &18.06 & 0.01  &- &- &517$\pm$25\tablenotemark{d}&517$\pm$25\\	
GC0574 & 13 24 44.26&-42 58 54.4 &22.19 & 0.02& 21.25 & 0.02 &20.19 & 0.02  &- &- &512$\pm$30	 &512$\pm$30	\\
GC0575 & 13 24 47.81&-43 08 43.5 &20.64 & 0.01& 20.14 & 0.01 &19.52 & 0.01  &- &- &353$\pm$29	 &353$\pm$29	\\
GC0576 & 13 24 48.79&-43 16 17.3 &21.13 & 0.02& 20.43 & 0.02 &19.60 & 0.01  &- &- &426$\pm$41	 &426$\pm$41	\\
GC0577 & 13 24 54.82&-43 08 49.2 &20.16 & 0.01& 19.50 & 0.02 &18.79 & 0.01  &- &- &197$\pm$29	 &197$\pm$29	\\
GC0578 & 13 24 55.37&-42 58 15.1 &22.24 & 0.01& 21.24 & 0.02 &20.29 & 0.01  &- &- &385$\pm$27	 &385$\pm$27	\\
GC0579 & 13 24 56.53&-42 39 31.4 &19.98 & 0.01& 18.70 & 0.00 &17.76 & 0.01  &- &- &158$\pm$50	 &158$\pm$50	\\
GC0580 & 13 24 57.09&-42 38 55.0 &19.60 & 0.01& 19.08 & 0.00 &18.48 & 0.01  &- &- &170$\pm$45	 &170$\pm$45	\\
GC0581 & 13 24 58.64&-42 58 05.6 &21.44 & 0.05& 20.92 & 0.04 &20.22 & 0.03  &- &- &524$\pm$74	 &524$\pm$74	\\
GC0582 & 13 25 01.21&-43 04 01.6 &22.02 & 0.02& 20.87 & 0.02 &19.90 & 0.02  &- &- &539$\pm$28	 &539$\pm$28	\\
GC0583 & 13 25 02.19&-42 56 51.5 &21.79 & 0.02& 21.00 & 0.02 &20.18 & 0.01  &- &- &602$\pm$84	 &602$\pm$84	\\
GC0584 & 13 25 09.06&-43 10 02.0 &21.11 & 0.01& 20.22 & 0.01 &19.30 & 0.01  &- &- &552$\pm$26	 &552$\pm$26	\\
GC0585 & 13 25 14.06&-43 02 42.7 &21.20 & 0.03& 20.56 & 0.02 &19.85 & 0.02  &- &- &510$\pm$82	 &510$\pm$82	\\
GC0586 & 13 25 16.53&-43 06 12.1 &22.07 & 0.02& 20.95 & 0.01 &20.01 & 0.01  &- &- &679$\pm$45	 &679$\pm$45	\\
GC0587 & 13 25 20.05&-43 03 10.0 &21.55 & 0.02& 20.46 & 0.02 &19.52 & 0.02  &- &- &278$\pm$37	 &278$\pm$37	\\
GC0588 & 13 25 30.15&-42 54 00.9 &21.23 & 0.02& 20.52 & 0.03 &19.87 & 0.01  &- &- &1051$\pm$52   &1051$\pm$52  \\	
GC0589 & 13 25 38.28&-42 54 53.7 &21.65 & 0.01& 20.57 & 0.02 &19.76 & 0.01  &- &- &686$\pm$98	 &686$\pm$98	\\
GC0590 & 13 25 46.96&-42 52 34.0 &21.08 & 0.02& 20.18 & 0.01 &19.43 & 0.01  &- &- &355$\pm$66	 &355$\pm$66	\\
GC0591 & 13 25 50.07&-42 49 58.1 &21.13 & 0.05& 20.60 & 0.04 &20.02 & 0.03  &- &- &702$\pm$42	 &702$\pm$42	\\
GC0592 & 13 25 51.94&-43 12 16.6 &19.58 & 0.01& 19.07 & 0.01 &18.45 & 0.01  &- &- &187$\pm$41	 &187$\pm$41	\\
GC0593 & 13 25 52.07&-42 55 28.3 &21.41 & 0.02& 20.54 & 0.02 &19.84 & 0.01  &- &- &223$\pm$26	 &223$\pm$26	\\
GC0594 & 13 25 55.06&-43 00 41.7 &21.30 & 0.01& 20.58 & 0.01 &19.92 & 0.01  &- &- &630$\pm$87	 &630$\pm$87	\\
GC0595 & 13 25 57.71&-43 05 13.3 &21.46 & 0.04& 20.55 & 0.03 &19.68 & 0.03  &- &- &410$\pm$37	 &410$\pm$37	\\
GC0596 & 13 25 59.36&-42 48 52.1 &21.70 & 0.05& 20.86 & 0.04 &20.07 & 0.03  &- &- &319$\pm$19	 &319$\pm$19	\\
GC0597 & 13 26 06.56&-43 07 08.2 &21.95 & 0.02& 20.92 & 0.01 &20.03 & 0.01  &- &- &240$\pm$30	 &240$\pm$30	\\
GC0598 & 13 26 09.52&-43 08 52.4 &21.45 & 0.02& 20.42 & 0.02 &19.51 & 0.01  &- &- &483$\pm$67	 &483$\pm$67	\\
GC0599 & 13 26 10.53&-43 01 05.9 &21.66 & 0.02& 20.55 & 0.02 &19.77 & 0.01  &- &- &311$\pm$81	 &311$\pm$81	\\
GC0600 & 13 26 11.71&-43 16 05.6 &20.73 & 0.02& 19.82 & 0.02 &18.93 & 0.01  &- &- &355$\pm$40	 &355$\pm$40	\\
GC0601 & 13 26 13.30&-43 04 40.8 &22.33 & 0.05& 21.29 & 0.03 &20.33 & 0.03  &- &- &184$\pm$43	 &184$\pm$43	\\
GC0602 & 13 26 17.77&-42 46 14.1 &22.03 & 0.04& 20.61 & 0.02 &19.55 & 0.01  &- &- &178$\pm$51	 &178$\pm$51	\\
GC0603 & 13 26 21.12&-43 02 59.4 &21.55 & 0.02& 20.70 & 0.03 &19.85 & 0.01  &- &- &506$\pm$56	 &506$\pm$56	\\
GC0604 & 13 26 30.75&-42 45 14.9 &21.49 & 0.05& 20.90 & 0.04 &19.87 & 0.02  &- &- &222$\pm$34	 &222$\pm$34	\\
GC0605 & 13 26 54.04&-43 05 06.2 &20.49 & 0.04& 19.85 & 0.03 &19.11 & 0.02  &- &- &948$\pm$64	 &948$\pm$64	\\

\enddata                                                                
\tablenotetext{a}{A second strong correlation peak was found at 182$\pm$31 km s$^{-1}$.}
\tablenotetext{b}{Structural parameter measurements indicate the
  object may be a star.}
\tablenotetext{c}{A second strong correlation peak was found at 103$\pm$106 km s$^{-1}$.}
\tablenotetext{d}{The weighted velocity of two measurements with Hydra, 539$\pm$50 km s$^{-1}$ and 510$\pm$29 km s$^{-1}$.}

\end{deluxetable}

\begin{deluxetable}{lllll}
\tablecolumns{5}    
\tabletypesize{\scriptsize}
\tablecaption{Gaussian Fits for the Radial Velocity Distributions\label{tab:vel_fits}}
\tablewidth{0pt}    
\tablehead{
\colhead{Group} & \colhead{R}& \colhead{mean} &\colhead{sigma} &\colhead{$\chi^2_{red}$}  \\
\colhead{ } & \colhead{(arcmin)} &\colhead{(km s$^{-1}$)}
&\colhead{(km s$^{-1}$)} & \colhead{ }\\
}\startdata
All GCs & 0-5   & $516.1\pm13.8$ & $149.6\pm10.0$& 1.09 \\
All GCs & 5-10  & $524.9\pm10.3$ & $142.6\pm7.4$ & 0.87 \\
All GCs & 10-15 & $525.3\pm14.3$ & $152.1\pm10.3$& 1.15 \\
All GCs & 15-20 & $461.0\pm24.2$ & $177.7\pm17.8$& 1.80 \\
All GCs & 20-45 & $520.4\pm25.2$ & $207.9\pm18.6$& 1.16 \\
MP GCs  & 0-5   & $493.3\pm21.7$ & $147.1\pm16.1$& 0.88 \\
MP GCs  & 5-10  & $521.0\pm15.2$ & $141.4\pm11.0$& 1.44 \\
MP GCs  & 10-15 & $506.3\pm18.8$ & $152.1\pm13.5$& 1.31 \\
MP GCs  & 15-20 & $446.0\pm33.5$ & $169.5\pm24.4$& 1.37 \\
MP GCs  & 20-45 & $475.1\pm31.5$ & $195.1\pm23.6$& 1.42 \\
MR GCs  & 0-5   & $529.6\pm22.5$ & $166.8\pm16.6$& 1.27 \\
MR GCs  & 5-10  & $527.3\pm14.1$ & $144.0\pm10.1$& 0.84 \\
MR GCs  & 10-15 & $550.0\pm21.3$ & $148.1\pm15.4$& 0.67 \\
MR GCs  & 15-20 & $474.4\pm34.4$ & $183.8\pm25.7$& 1.33 \\
MR GCs  & 20-45 & $569.5\pm38.1$ & $204.1\pm27.7$& 1.74 \\
\enddata
\end{deluxetable}

\begin{deluxetable}{lllllllll}  
\tablecolumns{9}            
\tabletypesize{\small}      
\tablecaption{The Mass of NGC 5128 \label{tab:mass}}
\tablewidth{0pt}
\tablehead{
\colhead{GCs} & \colhead{R}& \colhead{N} &\colhead{$v_{sys}$} &
\colhead{$\Omega R$} &\colhead{$\Theta_o$}  &\colhead{M$_{p}$}  &\colhead{M$_{r}$} &\colhead{M$_{t}$} \\
\colhead{ } & \colhead{(arcmin)} &\colhead{ } &\colhead{(km s$^{-1}$)}
& \colhead{(km s$^{-1}$)} & \colhead{(deg. E of N)}  & \colhead{($\times 10^{11}$ M$_{\sun}$)}& \colhead{($\times 10^{11}$ M$_{\sun}$)}& \colhead{($\times 10^{11}$ M$_{\sun}$)}  \\
}\startdata
All GCs & 0-5  & 120 & $514\pm14$ & $14\pm21$ & $175\pm80$& $1.900\pm0.663$  & $0.003\pm0.009$  &  $1.902\pm0.663$  \\
All GCs & 0-10 & 317 & $523\pm8$ & $33\pm13$ & $194\pm20$ & $4.520\pm1.496$  & $0.028\pm0.022$  &  $4.548\pm1.496$  \\
All GCs & 0-15 & 432 & $524\pm7$ & $36\pm11$ & $190\pm14$ & $6.257\pm1.999$  & $0.050\pm0.030$  &  $6.307\pm2.000$  \\
All GCs & 0-20 & 487 & $515\pm7$ & $36\pm11$ & $189\pm15$ & $9.260\pm2.881$  & $0.067\pm0.041$  &  $9.327\pm2.881$  \\
All GCs & 0-45 & 549 & $516\pm7$ & $33\pm11$ & $185\pm15$ & $17.45\pm4.950$  & $0.119\pm0.079$  &  $17.56\pm4.949$  \\
All GCs & 5-10 & 197 & $528\pm10$ & $43\pm16$ & $198\pm19$& $2.427\pm0.869$  & $0.047\pm0.035$  &  $2.472\pm0.869$  \\
All GCs & 5-15 & 312 & $527\pm8$ & $44\pm13$ & $192\pm14$ & $4.128\pm1.456$  & $0.075\pm0.044$  &  $4.203\pm1.457$  \\
All GCs & 5-20 & 367 & $515\pm8$ & $44\pm13$ & $191\pm14$ & $5.832\pm2.038$  & $0.099\pm0.059$  &  $5.931\pm2.038$  \\
All GCs & 5-45 & 429 & $517\pm8$ & $39\pm12$ & $186\pm14$ & $10.58\pm3.525$  & $0.166\pm0.102$  &  $10.75\pm3.527$  \\
MP GCs  & 5-20 & 180 & $508\pm11$ & $46\pm18$ & $206\pm18$& $4.879\pm3.643$  & $0.109\pm0.085$  &  $4.987\pm3.644$  \\
MP GCs  & 5-45 & 216 & $508\pm11$ & $31\pm18$ & $192\pm24$& $9.942\pm7.153$  & $0.105\pm0.122$  &  $10.05\pm7.154$   \\
MR GCs  & 5-20 & 184 & $523\pm11$ & $50\pm17$ & $174\pm18$& $6.157\pm2.564$  & $0.127\pm0.086$  &  $6.284\pm2.566$   \\
MR GCs  & 5-45 & 214 & $525\pm11$ & $45\pm16$ & $180\pm18$& $9.466\pm3.766$  & $0.206\pm0.146$  &  $9.672\pm3.769$   \\
\enddata
\end{deluxetable}
  
\begin{deluxetable}{llllllllll}  
\tablecolumns{10}            
\tabletypesize{\small}      
\tablecaption{The Kinematics of the Globular Cluster System of NGC 5128 \label{tab:kin}}
\tablewidth{0pt}
\tablehead{
\colhead{GCs} & \colhead{R}&  \colhead{R$_{avg}$}& \colhead{N} &\colhead{$v_{sys}$} &
\colhead{$\Omega R$} &\colhead{$\Theta_o$} &\colhead{$\Theta_{o}$-$\Theta_{major}$} &\colhead{$\sigma_{v_p}$}  &\colhead{$\Omega R/\sigma_{v_p}$}  \\
\colhead{ } & \colhead{(arcmin)}& \colhead{(arcmin)} &\colhead{ } &\colhead{(km s$^{-1}$)}
& \colhead{(km s$^{-1}$)} & \colhead{(deg. E of N)} & \colhead{(deg. E of N)}  & \colhead{(km s$^{-1}$)}& \colhead{ }  \\
}\startdata
All GCs & 0-5   & 3.49 &120 & $514\pm14$& $ 14\pm21$ & $175\pm80$& -40 & $149\pm5 $&$0.09\pm0.14$\\
All GCs & 5-10  & 7.31 &197 & $528\pm10$& $ 43\pm16$ & $198\pm19$& -17 & $147\pm3 $&$0.29\pm0.12$ \\
All GCs & 10-15 &11.96 &115 & $524\pm14$& $ 49\pm24$ & $184\pm21$& -31 & $151\pm5 $&$0.32\pm0.17$ \\
All GCs & 15-20 &17.32 & 55 & $446\pm24$& $ 43\pm38$ & $168\pm43$& -47 & $151\pm5$ &$0.28\pm0.22$ \\
All GCs & 20-45 &26.83 & 62 & $526\pm22$& $ 23\pm37$ & $ 55\pm87$& -160 & $156\pm7 $&$0.15\pm0.25$ \\
All GCs & 0-45  &10.82 &549 & $517\pm7$ & $ 33\pm10$ & $185\pm15$& -30 & $150\pm2 $&$0.22\pm0.07$ \\       
MP GCs  & 0-5   & 3.78 & 47 & $496\pm21$& $ 43\pm29$ & $ 93\pm42$& -122 & $144\pm8 $&$0.30\pm0.22$ \\
MP GCs  & 5-10  & 7.41 & 89 & $524\pm16$& $ 46\pm24$ & $228\pm27$& 13 & $147\pm6 $&$0.31\pm0.17$ \\
MP GCs  & 10-15 &11.97 & 66 & $511\pm19$& $ 47\pm33$ & $182\pm30$& -33 & $150\pm7 $&$0.31\pm0.23$ \\
MP GCs  & 15-20 &17.48 & 25 & $426\pm31$& $ 25\pm43$ &$137\pm114$& -78 & $128\pm11$&$0.19\pm0.42$ \\
MP GCs  & 20-45 &28.50 & 36 & $507\pm28$& $ 67\pm50$ & $ 45\pm37$& -170 & $171\pm14$&$0.39\pm0.31$ \\
MP GCs  & 0-45  &11.90 &263 & $506\pm9$ & $ 26\pm15$ & $175\pm28$& -40 & $149\pm4 $&$0.17\pm0.09$ \\       
MR GCs  & 0-5   & 3.49 & 57 & $527\pm22$& $ 65\pm34$ & $244\pm27$& 29 & $178\pm11 $&$0.37\pm0.21$ \\
MR GCs  & 5-10  & 7.23 &107 & $527\pm14$& $ 51\pm20$ & $178\pm22$& -37 & $149\pm5 $&$0.34\pm0.15$ \\
MR GCs  & 10-15 &11.96 & 49 & $542\pm22$& $ 46\pm36$ & $189\pm34$& -26 & $148\pm7 $&$0.31\pm0.26$ \\
MR GCs  & 15-45 &20.47 & 54 & $506\pm24$& $ 33\pm41$ & $168\pm56$& -47 & $162\pm9 $&$0.20\pm0.21$ \\
MR GCs  & 0-45  &10.20 &267 & $526\pm10$& $ 43\pm15$ & $196\pm17$& -19 & $156\pm4 $&$0.28\pm0.10$  \\       
\enddata
\end{deluxetable}

\begin{deluxetable}{llllll}  
\tablecolumns{6}            
\tabletypesize{\small}      
\tablecaption{The Kinematics of Globular Clusters with Ages,                    
  Metallicities, and [$\alpha$/Fe]\tablenotemark{a}\label{tab:age_kin}}
\tablewidth{0pt}
\tablehead{
\colhead{Group} & \colhead{N}& \colhead{R$_{avg}$} &\colhead{$v_{sys}$} & \colhead{$\Omega R$} &\colhead{$\Theta_o$} \\
\colhead{ } & \colhead{ } &\colhead{(arcmin)} &\colhead{(km s$^{-1}$)} & \colhead{(km s$^{-1}$)} & \colhead{(deg. E of N)}  \\
}\startdata
All GCs                                   & 72&5.7 &550$\pm$18 &25$\pm$26 & 110$\pm$61 \\        
GCs  Age $\geq 8$ Gyr                     & 49&5.8 &543$\pm$23 &27$\pm$36 & 107$\pm$65 \\
GCs  $5 <$ Age $ < 8$ Gyr                 & 10&4.9 &577$\pm$42 &53$\pm$78 & 235$\pm$54 \\        
GCs  Age $ < 5$ Gyr                       & 13&5.7 &562$\pm$51 &58$\pm$92 &  80$\pm$84 \\
GCs  Age $\geq 8$ Gyr $\&$ [Z/H]$ > -1$   & 23&5.9 &574$\pm$32 &17$\pm$50 & 265$\pm$143\\  
GCs  Age $\geq 8$ Gyr $\&$ [Z/H] $\leq -1$& 26&5.7 &515$\pm$34 &55$\pm$48 & 112$\pm$51 \\         
\enddata
\tablenotetext{a} {The ages, metallicities and [$\alpha$/Fe] used to
eclassify these globular clusters are from \cite{woodley09}.}
\end{deluxetable}

                  
\clearpage

\end{document}